\documentstyle[12pt,axodraw,graphicx]{article}
\begin{document}
\baselineskip 0.6cm

\def\simgt{\mathrel{\lower2.5pt\vbox{\lineskip=0pt\baselineskip=0pt
           \hbox{$>$}\hbox{$\sim$}}}}
\def\simlt{\mathrel{\lower2.5pt\vbox{\lineskip=0pt\baselineskip=0pt
           \hbox{$<$}\hbox{$\sim$}}}}
\def\lrD{\stackrel{\leftrightarrow}{\cal D}}

\begin{titlepage}

\begin{flushright}
UCB-PTH-05/12 \\
LBNL-57456
\end{flushright}

\vskip 1.8cm

\begin{center}

{\Large \bf 
The Supersymmetric Fine-Tuning Problem \\ and TeV-Scale Exotic Scalars
}

\vskip 1.0cm

{\large
Yasunori Nomura and Brock Tweedie
}

\vskip 0.4cm

{\it Department of Physics, University of California,
           Berkeley, CA 94720} \\
{\it Theoretical Physics Group, Lawrence Berkeley National Laboratory,
           Berkeley, CA 94720} \\

\vskip 1.2cm

\abstract{
A general framework is presented for supersymmetric theories that do not 
suffer from fine-tuning in electroweak symmetry breaking.  Supersymmetry is 
dynamically broken at a scale $\Lambda \approx (10\!\sim\!100)~{\rm TeV}$, 
which is transmitted to the supersymmetric standard model sector through 
standard model gauge interactions.  The dynamical supersymmetry breaking 
sector possesses an approximate global $SU(5)$ symmetry, whose $SU(3) 
\times SU(2) \times U(1)$ subgroup is explicitly gauged and identified 
as the standard model gauge group.  This $SU(5)$ symmetry is dynamically 
broken at the scale $\Lambda$, leading to pseudo-Goldstone boson states, 
which we call xyons.  We perform a detailed estimate for the xyon mass 
and find that it is naturally in the multi-TeV region.  We study general 
properties of xyons, including their lifetime, and study their collider 
signatures.  A generic signature is highly ionizing tracks caused by 
stable charged bound states of xyons, which may be observed at the LHC. 
We also consider cosmology in our scenario and find that a consistent 
picture can be obtained.  Our framework is general and does not depend 
on the detailed structure of the Higgs sector, nor on the mechanism of 
gaugino mass generation. 
}

\end{center}
\end{titlepage}

\section{Introduction}
\label{sec:intro}

Weak-scale supersymmetry has long been the leading candidate for physics 
beyond the standard model.  It stabilizes the Higgs potential against 
potentially huge radiative corrections, giving a consistent theory of 
electroweak symmetry breaking.  Combined with the idea of dynamical 
supersymmetry breaking (DSB), a large hierarchy between the weak and 
the Planck scales is explained by a dimensional transmutation associated 
with DSB gauge interactions~\cite{Witten:1981nf}.  The framework also 
leads to an elegant picture of gauge coupling unification at a scale of 
$M_X \simeq 10^{16}~{\rm GeV}$~\cite{Dimopoulos:1981zb}.  In addition, 
weak-scale supersymmetry has a virtue that it is relatively easier to 
evade constraints from precision electroweak measurements, compared 
with other candidates for physics beyond the standard model.

Despite these impressive successes, the most naive supersymmetric 
extension of the standard model suffers from problems.  First of all, 
generic superparticle masses of order the weak scale lead to too 
large flavor changing neutral currents.  This requires some organizing 
principle for the spectrum of superparticles.  The most promising 
one is ``universality'' --- the superparticles having the same 
standard model gauge quantum numbers are highly degenerate in mass. 
With this spectrum, contributions to flavor changing neutral currents 
from superparticle loops are canceled to high degree, evading severe 
experimental constraints.  Another problem comes from the fact that 
LEP~II did not discover any superparticles or the Higgs boson.  In most 
supersymmetric theories, this leads to severe fine-tuning of order 
a few percent to reproduce the correct scale for electroweak symmetry 
breaking.  This problem is called the supersymmetric fine-tuning 
problem.

In this paper we first present a general discussion on how to evade the 
above problems of generic supersymmetric theories while preserving their 
successful features.  We find that, requiring the absence of fine-tuning 
in electroweak symmetry breaking, the naturalness upperbound on the squark 
masses are at around $(660\!\sim\!850)~{\rm GeV}$ quite independently 
of the details of the Higgs sector.  The upperbound on the slepton 
masses is similarly given at about $(310\!\sim\!400)~{\rm GeV}$. 
We then argue that consideration along these lines naturally leads 
to a class of theories which predicts the existence of exotic scalar 
particles with mass in the multi-TeV region.  In these theories the 
DSB sector possesses a global ``unified'' symmetry, of which the $SU(3) 
\times SU(2) \times U(1)$ subgroup is explicitly gauged and identified 
as the standard model gauge group.  The DSB sector spontaneously breaks 
this global symmetry, as well as supersymmetry, at a dynamical scale 
of $(10\!\sim\!100)~{\rm TeV}$.  The exotic scalar particles then arise 
as composite pseudo-Goldstone bosons in the DSB sector and have the 
same gauge quantum numbers as those of grand unified gauge bosons --- 
$({\bf 3}, {\bf 2})_{-5/6}$ in the simplest case.  We call these 
particles {\it xyons}.

The scenario we consider leads to certain characteristic features for 
the superparticle spectrum, and we illustrate them by presenting the 
supersymmetry breaking masses for the gauginos, squarks and sleptons at 
a representative set of parameter points.  We also estimate the mass of 
xyons in generic situations, using naive scaling arguments, and calculate 
it in terms of fundamental parameters in a class of calculable theories 
formulated in holographic higher dimensional spacetime.  We find that 
the mass of xyons is naturally in the multi-TeV region, so that it 
is within the reach of the LHC in some of the parameter region and that 
the discovery potential of xyons at the Very Large Hadron Collider (VLHC) 
is very high.  We discuss general properties of xyons, especially their 
lifetime, and find that they generically lead to a distinctive signature 
of stable massive charged particles.  We also discuss other experimental 
signatures as well as cosmology in our scenario.

The organization of the paper is as follows.  In the next section we 
discuss the supersymmetric fine-tuning problem and present a general 
argument connecting the naturalness of electroweak symmetry breaking 
with certain properties of the supersymmetry breaking sector.  We then 
present our explicit framework in section~\ref{sec:framework} and study 
its general consequences in section~\ref{sec:main}.  In particular, we 
study properties of xyons both in the context of generic 4D theories 
and the holographic theories in 5D which were constructed earlier in 
Ref.~\cite{Nomura:2004is}.  We also discuss experimental signals and 
cosmology of our scenario in section~\ref{sec:exp}.  Finally, in 
section~\ref{sec:concl} we discuss the generality of our framework, 
particularly emphasizing that the framework is independent of the 
details of the Higgs sector or the mechanism of gaugino mass generation 
--- for example, our argument does not depend on whether the gaugino 
masses are Majorana or Dirac type.

\section{The Supersymmetric Fine-Tuning Problem}
\label{sec:fine-tuning}

Our argument starts from carefully identifying the sources of the 
supersymmetric fine-tuning problem.  A part of this argument also 
appears in~\cite{Chacko:2005ra}.  In the minimal supersymmetric 
standard model (MSSM), the minimization of the tree-level Higgs 
potential gives the equation $M_Z^2/2 \simeq -m_h^2 - |\mu|^2$, 
where $m_h^2$ is the soft supersymmetry-breaking mass squared for 
the up-type Higgs boson and $\mu$ the supersymmetric mass for the 
two Higgs doublets.  The $Z$-boson mass, $M_Z$, appears on the 
left-hand-side because the quartic coupling of the MSSM Higgs 
potential is given by $\lambda = (g^2+g'^2)/8$ at tree level. 
This relation is in general violated at radiative level, or if we 
extend the Higgs sector such that there are additional sources for 
the Higgs quartic coupling.  In these cases the above equation is 
modified to
\begin{equation}
  \frac{M_{\rm Higgs}^2}{2} \simeq -m_h^2 - |\mu|^2,
\label{eq:ewsb-cond}
\end{equation}
where $M_{\rm Higgs}$ is the mass of the physical Higgs boson. 
This equation is valid for moderately large values for $\tan\beta 
\equiv \langle h \rangle/\langle \bar{h} \rangle$, e.g. $\tan\beta 
\simgt 2$, where $h$ and $\bar{h}$ are the up-type and down-type 
Higgs doublets, respectively.  For smaller values of $\tan\beta$, 
correction terms of order $1/\tan^2\!\beta$ should be included 
in the right-hand-side.%
\footnote{Appropriate correction terms are given by $\{ m_h^2 
- m_{\bar{h}}^2 + 2(\mu B)\tan\beta \}/(\tan^2\!\beta+1)$, where 
$\mu B$ is the holomorphic supersymmetry-breaking mass for $h$ 
and $\bar{h}$, which is generically of order $1/\tan\beta$. 
Note that for small $\tan\beta$, this equation deviates from 
the famous relation, $M_{\rm Higgs}^2/2 = (m_{\bar{h}}^2 - m_h^2 
\tan^2\!\beta)/(\tan^2\!\beta-1) - |\mu|^2$.  Our equation is 
obtained by looking along the direction of the vacuum expectation value 
in the field space of $h$ and $\bar{h}$.  This is generically valid 
as long as the charged Higgs boson mass is somewhat larger than the 
lightest Higgs boson mass, $M_{\rm Higgs}$, which we expect to be 
the case to evade the constraint from $b \rightarrow s \gamma$.}

In the MSSM, $M_{\rm Higgs}$ is smaller than about $130~{\rm GeV}$. 
This implies that if we want to avoid significant fine-tuning 
between the two unrelated parameters $m_h^2$ and $\mu$, each term 
in the right-hand-side of Eq.~(\ref{eq:ewsb-cond}) cannot be much 
larger than about $(160\!\sim\!210~{\rm GeV})^2$.%
\footnote{The numbers we provide correspond to the requirement that 
each term in the equation determining the weak scale must not be larger 
than a factor $(3\!\sim\!5)$ times the sum of all the terms, i.e. the 
level of a cancellation must not be severer than $(20\!\sim\!33)\%$.}
The size of $|\mu|$ is constrained to be $|\mu| \simgt 100~{\rm GeV}$ 
by the chargino search, but it still allows $|\mu| \simlt 210~{\rm GeV}$. 
For $m_h^2$, there are several contributions to this quantity.  In 
particular, loops of the top quark and squarks give the contribution
\begin{equation}
  \delta m_h^2 \simeq -\frac{3y_t^2}{4\pi^2}\, m_{\tilde{t}}^2\, 
    \ln\Biggl( \frac{M_{\rm mess}}{m_{\tilde{t}}} \Biggr),
\label{eq:corr-Higgs}
\end{equation}
where $y_t$ is the top Yukawa coupling, and $m_{\tilde{t}}$ represents 
the masses of the two top squarks, which we have taken to be equal 
for simplicity.  Here, $M_{\rm mess}$ is the scale at which superparticle 
masses are generated (or at which supersymmetry breaking is mediated 
to the supersymmetric standard model sector), which we call the messenger 
scale.  The absence of unnatural cancellations then implies that the 
values of $|\delta m_h^2|$ should not be much larger than the bound 
on $m_h^2$ itself, $(160\!\sim\!210~{\rm GeV})^2$.

Assuming flavor universality for the squark and slepton masses, the 
current mass bound for the top squarks is roughly $m_{\tilde{t}} \simgt 
300~{\rm GeV}$~\cite{Eidelman:2004wy}.  Now, imagine that the messenger 
scale is very high, of order the Planck scale, $M_{\rm mess} \simeq 
M_{\rm Pl}$, as in the case of the minimal supergravity scenario. 
In this case the logarithm $\ln(M_{\rm mess}/m_{\tilde{t}})$ is very large 
($\simeq 35$), and Eq.~(\ref{eq:corr-Higgs}) gives $-\delta m_h^2$ as large 
as $(500~{\rm GeV})^2$ even for $m_{\tilde{t}}^2 \simeq (300~{\rm GeV})^2$. 
(In fact, the value of $m_{\tilde{t}}^2$ is constrained to be even 
larger in conventional supersymmetric theories, see discussions below.) 
This, therefore, requires a severe cancellation between the two terms in 
Eq.~(\ref{eq:ewsb-cond}) at a level of a few percent.  In fact, because 
of the large logarithm, a precise analysis requires summing up the 
leading logarithms using renormalization group equations.  However, 
the result is essentially unchanged, and a fine-tuning of order a few 
percent is still required~\cite{Chankowski:1998xv}.  This argument 
suggests that smaller values of $M_{\rm mess}$ are preferred from the 
naturalness point of view. 

Suppose now that $M_{\rm mess}$ is small and close to the electroweak 
scale, $M_{\rm mess} \simeq (10\!\sim\!100)~{\rm TeV}$.  In this case 
the logarithm in Eq.~(\ref{eq:corr-Higgs}) can be quite small, and of 
a factor of a few.  In the MSSM, however, this still does not allow 
us to eliminate the fine-tuning.  The obstacle arises essentially 
from the conflict between the LEP~II bound on the Higgs-boson mass, 
$M_{\rm Higgs} \simgt 114~{\rm GeV}$~\cite{unknown:2001xx}, and the 
tree-level MSSM prediction, $M_{\rm Higgs} \leq M_Z$, which requires 
a significant radiative correction to $M_{\rm Higgs}$.  Such a large 
correction arises in the MSSM only when the top squarks are heavy. 
For natural values of the top-squark mass parameters, the LEP~II bound, 
$M_{\rm Higgs} \simgt 114~{\rm GeV}$, requires $m_{\tilde{t}}$ to 
be larger than about $(800\!\sim\!1000)~{\rm GeV}$~\cite{Carena:1995wu}. 
This in turn requires cancellations among various terms in the 
right-hand-side of Eq.~(\ref{eq:ewsb-cond}) at a level of 
$(4\!\sim\!6)\%$, at best.%
\footnote{It is customary to define the amount of fine-tuning by the 
sensitivity of the weak scale, $M_Z$, to fractional changes of fundamental 
parameters $a_i$ of the theory~\cite{Barbieri:1987fn}: $\Delta^{-1} 
\equiv \min_i \{ (a_i/M_Z^2) (\partial M_Z^2/\partial a_i) \}^{-1}$. 
In the case that supersymmetry breaking is mediated by standard model 
gauge interactions, $a_i$'s are proportional to square roots of the 
scalar masses squared, $a_i \propto (m_{\tilde{f}}^2)^{1/2}$, so that 
the fine-tuning parameter is a factor 2 smaller than the amount of 
cancellations quoted in the text.  For the case discussed here, for 
example, $\Delta^{-1} \simlt (2\!\sim\!3)\%$ (see e.g.~\cite{Chacko:2005ra} 
for details).}

A lower bound on $m_{\tilde{t}}$ can also come from the pattern of 
superparticle masses.  For very small values of $M_{\rm mess}$, the 
most natural mechanism giving flavor-universal superparticle masses 
is to mediate supersymmetry breaking through standard model gauge 
interactions, $SU(3)_C \times SU(2)_L \times U(1)_Y$ (321).  This 
implies that the DSB sector, or some sector that feels the primary 
supersymmetry breaking, is charged under 321 gauge interactions and 
thus contributes to the evolution of the 321 gauge couplings above 
$M_{\rm mess}$.  This contribution, therefore, destroys the successful 
supersymmetric prediction for the low-energy gauge couplings unless 
it is somehow universal for $SU(3)_C$, $SU(2)_L$ and $U(1)_Y$.%
\footnote{We take the $SU(5)$-normalization for $U(1)_Y$ throughout 
the paper.}
The simplest possibility to preserve the prediction is then that the DSB 
sector, or the corresponding sector, respects a global $SU(5)$ symmetry 
that contains 321 as a subgroup, so that this sector does not affect the 
differential evolution of the 321 gauge couplings.  In this case 
the ratio of the top squark mass to the right-handed selectron mass is 
given by
\begin{equation}
  \frac{m_{\tilde{t}}^2}{m_{\tilde{e}}^2} 
    \simeq \frac{(4/3)g_3^4+\delta}{(3/5)g_1^4} 
    \simeq (7\!\sim\!8)^2,
\label{eq:te-ratio}
\end{equation}
where $g_3$ and $g_1$ are the $SU(3)_C$ and $U(1)_Y$ gauge couplings, 
renormalized at a scale of order $M_{\rm mess}$, and $\delta$ represents 
the small contributions from $SU(2)_L$ and $U(1)_Y$.%
\footnote{In fact, Eq.~(\ref{eq:te-ratio}) applies to quite large classes 
of theories in which supersymmetry breaking is mediated by standard model 
gauge interactions.  For example, Eq.~(\ref{eq:te-ratio}) applies to 
the minimal model of gauge mediation~\cite{Dine:1994vc} even if the 
messenger sector does not possess a global $SU(5)$ symmetry.  In general, 
Eq.~(\ref{eq:te-ratio}) applies to any gauge mediation models in which 
the messenger sector respects an approximate $SU(5)$ symmetry at the 
unification scale and in which supersymmetry breaking effects in the 
messenger sector are not very large.}
The non-discovery of the right-handed selectron at LEP~II pushes up its 
mass to be above $\simeq 100~{\rm GeV}$~\cite{Eidelman:2004wy}.  This leads 
to $m_{\tilde{t}}$ as large as, at least, $700~{\rm GeV}$, which in turn 
leads to $-\delta m_h^2$ larger than about $(300~{\rm GeV})^2$ even for 
$\ln(M_{\rm mess}/m_{\tilde{t}})$ as small as a factor of a few.  For 
$M_{\rm Higgs} \simlt 130~{\rm GeV}$, for example, this alone requires 
some cancellation at a level of $10\%$.

How can we avoid this unpleasant situation?  Barring a possibility of 
accidental cancellations, the necessary condition is to have an additional 
contribution to $M_{\rm Higgs}$, i.e. an additional source for the 
Higgs quartic coupling other than the $SU(2)_L \times U(1)_Y$ $D$-terms 
in the MSSM.  Such a contribution may arise, for example, from the 
superpotential coupling of the Higgs doublets to some other field, such 
as the singlet field in the next-to-minimal supersymmetric standard model 
(NMSSM)-type theories~\cite{Masip:1998jc,Harnik:2003rs,Birkedal:2004zx}, 
or from the $D$-term of additional gauge interactions other than 321 
of the MSSM~\cite{Cvetic:1997ky,Batra:2003nj}.  Is this sufficient to 
eliminate the fine-tuning?  Equation~(\ref{eq:ewsb-cond}) implies 
that for arbitrarily large $M_{\rm Higgs}$, the naturalness requirement 
on the sizes of $|m_h^2|$ and $|\mu|^2$ becomes arbitrarily weaker.%
\footnote{Note that Eq.~(\ref{eq:ewsb-cond}) applies not only to 
the MSSM but also to more general supersymmetric extensions of the 
standard model. Our argument is also independent of the origin of 
the supersymmetric Higgs mass $\mu$.  For example, $\mu$ may arise 
from the expectation value of a singlet field at the weak scale.}
However, if we make $M_{\rm Higgs}$ very large, we lose one of the 
virtues of supersymmetric theories: precision electroweak constraints 
can be satisfied without unnatural cancellations among various 
contributions.  If we want to keep this virtue, i.e. if we want to 
fit to the data just by decoupling the contributions from new physics, 
the mass of the Higgs boson is bounded as~\cite{unknown:2003ih}
\begin{equation}
  M_{\rm Higgs} \simlt 250~{\rm GeV}.
\label{eq:MH-bound}
\end{equation}
We then find from Eq.~(\ref{eq:ewsb-cond}) that the values of $|m_h^2|$, 
and thus $|\delta m_h^2|$, should not be much larger than about 
$(310\!\sim\!400~{\rm GeV})^2$.  This still places strong constraints 
on the sizes of $m_{\tilde{t}}^2$ and $\ln(M_{\rm mess}/m_{\tilde{t}})$. 
We find that {\it even with an additional contribution to $M_{\rm Higgs}$ 
other than that in the MSSM, $M_{\rm mess}$ should still be much smaller 
than the Planck scale and $m_{\tilde{t}}^2$ is still subject to upperbounds 
that depend on the values of $M_{\rm mess}$}.  This is the key observation 
leading to the framework discussed in the next section.

\section{Dynamical Breaking of Supersymmetry and $SU(5)$}
\label{sec:framework}

\begin{figure}[t]
\begin{center} 
\begin{picture}(290,100)(5,125)
  \Text(60,145)[t]{\large (a)}
  \Photon(5,190)(32,190){3.5}{3}   \Line(5,190)(32,190)
  \Photon(88,190)(115,190){3.5}{3} \Line(88,190)(115,190)
  \Text(5,198)[b]{$\lambda$} \Text(115,198)[b]{$\lambda$} 
  \GOval(60,190)(23,28)(0){0.85} 
  \Text(61,190)[]{\large SUSY}  \Line(40,185)(80,195)
  \Text(240,145)[t]{\large (b)}
  \DashLine(185,188)(213,188){3} \Text(185,196)[b]{$\tilde{f}$} 
  \DashLine(267,188)(295,188){3} \Text(295,196)[b]{$\tilde{f}$} 
  \PhotonArc(238,188)(25,130,180){3}{2.5} \CArc(238,188)(25,130,180)
  \PhotonArc(242,188)(25,0,50){3}{2.5}    \CArc(242,188)(25,0,50)
  \Text(207,203)[bl]{$\lambda$} \Text(274,203)[br]{$\lambda$}
  \CArc(240,201)(30,204,336) \Text(240,167)[t]{$f$}
  \GOval(240,208)(13,18)(0){0.85} 
  \Text(241,208)[]{\small SUSY}  \Line(225,204)(255,212)
\end{picture}
\caption{Examples of the diagrams that give (a)~gaugino masses and 
 (b)~sfermion masses, where $\lambda$, $\tilde{f}$ and $f$ represent 
 gauginos, sfermions and fermions, respectively.}
\label{fig:gauge-med}
\end{center}
\end{figure}
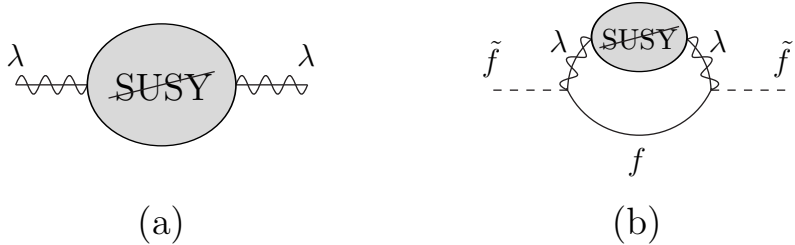

We here specify our framework more explicitly.  Because of small 
values of $M_{\rm mess}$, we assume that supersymmetry breaking in 
the DSB sector is transmitted to the supersymmetric standard model 
sector through 321 gauge interactions.  The gaugino and sfermion 
masses are then generated by the diagrams in Fig.~\ref{fig:gauge-med}. 
(The diagram giving the gaugino masses may be different; see discussion 
in section~\ref{sec:concl}.)
Here, the gray disks at the centers represent the contributions from 
the DSB sector, or some sector that feels primary supersymmetry breaking 
such as the messenger sector in gauge mediation models~\cite{Dine:1994vc,%
Dine:1981gu}.  We call this sector, collectively, the supersymmetry 
breaking sector. 

There is an immediate consequence of mediating supersymmetry breaking 
by 321 gauge interactions.  Suppose that the supersymmetry breaking 
sector carries the Dynkin index $\hat{b}$ under $SU(3)_C$, $SU(2)_L$ 
and $U(1)_Y$, so that it contributes to the beta-function coefficients 
for the 321 gauge couplings by $\hat{b}$ (the quantity $\hat{b}$ 
should be universal for $SU(3)_C$, $SU(2)_L$ and $U(1)_Y$ in order 
not to destroy gauge coupling unification).  The requirement that the 
321 gauge couplings do not hit the Landau pole below the unification 
scale then gives a constraint $\hat{b} \simlt 5$, where we have taken 
$M_{\rm mess} = O(10\!\sim\!100~{\rm TeV})$.  Now, the masses of 
the gauginos $\tilde{M}$ and the sfermions $\tilde{m}$ are generated 
at $M_{\rm mess}$ as threshold effects, through the diagrams in 
Fig.~\ref{fig:gauge-med}.  Therefore, their values are bounded as 
$\tilde{M} \simlt (g^2/16\pi^2) \hat{b} M_{\rm mess}$ and $\tilde{m}^2 
\simlt (g^2/16\pi^2)^2 C \hat{b} M_{\rm mess}^2$, where $g$ and $C$ 
represent the standard model gauge coupling and a Casimir factor. This 
gives a lower bound on the mediation scale $M_{\rm mess} \simgt 20~{\rm 
TeV}$, and thus the size of the logarithm $\ln(M_{\rm mess}/m_{\tilde{t}}) 
\simgt 3.5$.  Using Eq.~(\ref{eq:corr-Higgs}), we then find a rough 
upperbound on $m_{\tilde{t}}$ (and on generic squark masses 
$m_{\tilde{q}}$):
\begin{equation}
  m_{\tilde{q}} \sim m_{\tilde{t}} \simlt (660\!\sim\!850) 
    \left( \frac{M_{\rm Higgs}}{250~{\rm GeV}} \right){\rm GeV},
\label{eq:bound-squark}
\end{equation}
where the first relation comes from flavor universality. (In a realistic 
theory the lightest top-squark mass is somewhat smaller than the other 
squark masses by the top-Yukawa and left-right mixing effects.)  Here, 
we have explicitly shown the dependence of the bound on the physical 
Higgs-boson mass, $M_{\rm Higgs}$, coming from Eq.~(\ref{eq:ewsb-cond}), 
to make it clear that the bound becomes tighter for smaller values of 
$M_{\rm Higgs}$.

The naturalness bound derived above gives an immediate tension with the 
unified mass relation of Eq.~(\ref{eq:te-ratio}), with the LEP~II bound 
of $m_{\tilde{e}} \simgt 100~{\rm GeV}$.  We find that the unified mass 
relation is not compatible with the naturalness bound unless the mass 
of the physical Higgs boson is rather large, $M_{\rm Higgs} \simgt 
200~{\rm GeV}$, which is consistent with the precision electroweak data 
only if the top-quark mass lies in the upper edge of the latest world 
average $m_t = 178.0 \pm 4.3~{\rm GeV}$~\cite{Azzi:2004rc}.  For $M_{\rm 
Higgs} \simgt 200~{\rm GeV}$, having the right-handed selectron mass as 
small as its experimental lower bound, $m_{\tilde{e}} \simeq 100~{\rm GeV}$, 
allows the required amount of cancellations to be reduced to the level 
of $30\%$ even with the unified mass relation of Eq.~(\ref{eq:te-ratio}). 
Predictions of such a scenario, then, would be a heavy Higgs boson, 
$M_{\rm Higgs} \simeq (200\!\sim\!250)~{\rm GeV}$, a heavy top quark, 
$m_t \simeq (180\!\sim\!182)~{\rm GeV}$, light right-handed sleptons, 
$m_{\tilde{e}} \simeq 100~{\rm GeV}$, and sfermions with masses given 
by the unified mass relations: $m_{\tilde{q}} \simeq m_{\tilde{u}} 
\simeq m_{\tilde{d}} \simeq 750~{\rm GeV}$ and $m_{\tilde{l}} \simeq 
200~{\rm GeV}$.  However, the viability of this scenario depends crucially 
on the mass of the top quark.  For example, if the top-quark mass is 
within the $1\sigma$ region of the recently reported CDF Run~II value 
$m_t = 173.5 \pm 4.1~{\rm GeV}$~\cite{CDF}, the possibility of having 
the unified mass relation without fine-tuning disappears.  We thus 
conclude that {\it unless the top-quark mass is in the upper edge of 
the experimentally allowed range, the unified mass relation of 
Eq.~(\ref{eq:te-ratio}) is not compatible with the requirement from 
naturalness}.  In the rest of the paper, we only consider the case 
in which the unified mass relation is violated.

We can also obtain a naturalness upperbound on the masses of the sleptons 
in a similar way.  Mediations of supersymmetry breaking by 321 gauge 
interactions imply that the masses of the doublet sleptons are the 
same as $m_h^2$ before taking into account the effects from the Yukawa 
couplings.  This implies that these masses should not be much larger 
than $m_h^2$, giving
\begin{equation}
  m_{\tilde{l}} \simlt (310\!\sim\!400) 
    \left( \frac{M_{\rm Higgs}}{250~{\rm GeV}} \right){\rm GeV}.
\label{eq:bound-slepton}
\end{equation}
The bound on the singlet slepton masses is more indirect, but we expect 
that it is not much different from Eq.~(\ref{eq:bound-slepton}).  This 
is because the difference of the doublet and singlet slepton masses 
arises only from the $SU(2)_L$ contribution and the factor 2 difference 
of hypercharges.  These effects are not much larger than the values 
in Eq.~(\ref{eq:bound-slepton}) themselves and also work in opposite 
directions.  It is important here to notice that the bounds in 
Eqs.~(\ref{eq:bound-squark},~\ref{eq:bound-slepton}) rely only on 
$M_{\rm mess} \simgt 20~{\rm TeV}$.  {\it They do not depend on any 
details of the supersymmetry breaking mechanism or on the origin 
of the additional contribution to the physical Higgs-boson mass.}

Now, let us consider properties of the supersymmetry breaking sector. 
The basic requirements on this sector are (i) it contributes to the 
evolution of the 321 gauge couplings universally, but (ii) it generates 
superparticle masses at $M_{\rm mess}$ that do not obey simple $SU(5)$ 
relations such as the one given in Eq.~(\ref{eq:te-ratio}).  It is, 
of course, possible that these requirements are satisfied simply 
because the matter content of the supersymmetry breaking sector fills 
out complete $SU(5)$ multiplets but the couplings in this sector do 
not respect $SU(5)$ at all.%
\footnote{An example of such theories is given by models of gauge 
mediation with non-minimal messenger sectors~\cite{Chacko:2005ra,%
Agashe:1997kn}.}
In this case, however, we should regard the appearance of the complete 
$SU(5)$ multiplets somewhat accidental.%
\footnote{A class of ``natural'' models, however, arises if 
the messenger fields of gauge mediation live in the bulk of higher 
dimensional unified theories realized at the scale of $1/R \simeq 
(10^{15}\!\sim\!10^{16})~{\rm GeV}$, since these fields then have 
the 321 quantum numbers coming from an $SU(5)$ multiplet but do not 
obey any $SU(5)$ relations~\cite{Hall:2001pg}.}
Moreover, if the supersymmetry breaking sector is strongly coupled over 
a wide energy interval above the weak scale, we expect non-universal 
corrections to the evolution of the 321 gauge couplings at higher loops, 
which are not necessarily suppressed.  Therefore, here we consider an 
alternative possibility that the supersymmetry breaking sector in fact 
possesses a global $SU(5)$ symmetry, of which the $SU(3) \times SU(2) 
\times U(1)$ subgroup is explicitly gauged and identified as the 
standard model gauge group.  In this case the global $SU(5)$ symmetry 
ensures that the contribution to the 321 gauge coupling evolution from 
the supersymmetry breaking sector is universal.  On the other hand, 
the effect of this $SU(5)$ symmetry should be absent in the spectrum of 
superparticles, as otherwise it would lead to the unified mass relations 
such as the one in Eq.~(\ref{eq:te-ratio}).  How can this happen? 

The simplest possibility to realize the situation described above is 
to consider that the DSB sector possesses a global $SU(5)$ symmetry, 
of which the 321 subgroup is gauged and identified as the standard model 
gauge group, and to assume that the dynamics of this sector breaks 
not only supersymmetry but also the global $SU(5)$ symmetry to the 
321 subgroup at the dynamical scale $\Lambda \approx M_{\rm mess} 
\simeq (10\!\sim\!100)~{\rm TeV}$.  In this case the gray disks in 
Fig.~\ref{fig:gauge-med} represent the DSB sector itself, and not some 
other sector feeling the primary supersymmetry breaking more directly 
than the supersymmetric standard model sector.  This setup also has an 
advantage that the lowest possible values of $M_{\rm mess}$ are obtained, 
as the mediation of supersymmetry breaking is most direct.  Because of 
the spontaneous breakdown of $SU(5)$ at the scale $\Lambda$, where the 
superparticle masses are generated, the resulting superparticle spectrum 
does not respect unified mass relations.  A class of theories implementing 
this mechanism was first constructed in Ref.~\cite{Nomura:2004is} 
in holographic higher dimensional spacetime, and was used 
in~\cite{Chacko:2005ra} to ameliorate fine-tuning in the context 
of specific models.  Here, we will work in a more general context: 
we allow an arbitrary form of the Higgs potential and/or an arbitrary 
origin for the additional contribution to $M_{\rm Higgs}$.  The main 
conclusions of our analysis also do not depend on the mechanism of 
gaugino mass generation, as will be discussed in detail in 
section~\ref{sec:concl}.

\section{TeV-Scale Exotic Scalars}
\label{sec:main}

The framework described in the previous section has one general consequence. 
Because of the spontaneous breakdown of the global $SU(5)$ group to 
the 321 subgroup, the spectrum in the DSB sector contains light scalar 
particles, whose 321 gauge quantum numbers are the same as the unified 
gauge bosons (XY gauge bosons): $({\bf 3}, {\bf 2})_{-5/6}$.  The original 
global $SU(5)$ symmetry is also explicitly broken by the gauging of 
the 321 subgroup.  Therefore, these particles are pseudo-Goldstone 
bosons and obtain masses at loop level through 321 gauge interactions. 
Since both $SU(5)$ and supersymmetry are broken at the scale $\Lambda 
\approx M_{\rm mess}$, their masses squared are generically of order 
$(g^2 C/16\pi^2) M_{\rm mess}^2$.  We call these light scalar particles 
{\it xyons}.  Note that the superpartners of xyons obtain masses of order 
$M_{\rm mess}$ because of the supersymmetry breaking at the scale $\Lambda$. 

It is possible that the DSB sector has a global symmetry larger than 
$SU(5)$, in which case the 321 quantum numbers of xyons are more 
complicated.  For example, if the global group of the DSB sector is 
$SO(10)$ and spontaneously broken to the $SU(4)_C \times SU(2)_L \times 
SU(2)_R$ subgroup at the scale $\Lambda$, as is the case considered 
in Ref.~\cite{Nomura:2004it}, the 321 quantum numbers of xyons are given 
by $({\bf 3}, {\bf 2})_{-5/6} + ({\bf 3}, {\bf 2})_{1/6}$.

\subsection{Xyon mass --- estimate}
\label{subsec:xyon-mass-1}

We here estimate the mass of xyons using naive scaling arguments. 
Since the xyon mass is generated at one loop through 321 gauge 
interactions, the mass squared $m_\varphi^2$ for xyons is roughly 
given by
\begin{equation}
  m_\varphi^2 \simeq \sum_{i=1,2,3} 
    \frac{g_i^2 C_i^\varphi}{16\pi^2} M_\rho^2,
\label{eq:xyon-mass}
\end{equation}
where $g_i$ are the 321 gauge couplings with $i=1,2,3$ representing 
$U(1)_Y$, $SU(2)_L$ and $SU(3)_C$, and $C_i^\varphi$ are the group 
theoretical factors: $(C_1^\varphi, C_2^\varphi, C_3^\varphi) = 
(5/12, 3/4, 4/3)$ for $SU(5)$ xyons.  The parameter $M_\rho$ is 
defined to be the mass scale for the resonances in the DSB sector 
($M_\rho \approx \Lambda \approx M_{\rm mess}$).  The xyon mass is 
essentially determined by the quantity $M_\rho$. 

Since the DSB sector generates the superparticle masses through 321 gauge 
interactions, the size of $M_\rho$ is related to these masses.  The 
gaugino masses are generated by the diagram in Fig.~\ref{fig:gauge-med}a, 
and given by
\begin{equation}
  M_i \simeq g_i^2 \frac{\hat{b}}{16\pi^2}\, 
    (\hat{\zeta}_i M_\rho),
\label{eq:gaugino-masses}
\end{equation}
where $\hat{\zeta}_i$ are parameters of $O(1)$ which can take different 
values for $i=1,2,3$ reflecting the spontaneous breakdown of the global 
$SU(5)$ symmetry.  This expression can be obtained using large-$N$ 
scaling, by identifying the effective number of ``colors'' in the DSB 
sector as $\hat{b}$, the contribution of the DSB sector to the evolution 
of the 321 gauge couplings, which is appropriate in the present context. 
The scalar masses are similarly given by
\begin{equation}
  m_{\tilde{f}}^2 \simeq 2\!\! \sum_{i=1,2,3} 
    \frac{g_i^4 C_i^{\tilde{f}}}{16\pi^2} 
    \frac{\hat{b}}{16\pi^2}\, 
    (\hat{\zeta}_i M_\rho)^2,
\label{eq:scalar-masses}
\end{equation}
where $\tilde{f} = \tilde{q}, \tilde{u}, \tilde{d}, \tilde{l}, \tilde{e}$ 
represents the MSSM squarks and sleptons, and $C_i^{\tilde{f}}$ are 
the group theoretical factors given by $(C_1^{\tilde{f}}, C_2^{\tilde{f}}, 
C_3^{\tilde{f}}) = (1/60,3/4,4/3)$, $(4/15,0,4/3)$, $(1/15,0,4/3)$, 
$(3/20,3/4,0)$ and $(3/5,0,0)$ for $\tilde{f} = \tilde{q}, \tilde{u}, 
\tilde{d}, \tilde{l}$ and $\tilde{e}$, respectively (the equation 
also applies to the Higgs fields with $C_i^{h} = C_i^{\bar{h}} = 
C_i^{\tilde{l}}$).  Here, the overall factor of $2$ has been inserted 
so that the expression smoothly matches to the general gauge mediation 
result. 

The size of $M_\rho$ is determined by the superparticle masses through 
Eqs.~(\ref{eq:gaugino-masses},~\ref{eq:scalar-masses}).  Let us now 
assume that the masses of the squarks are dominated by the $SU(3)_C$ 
contributions, which is a natural assumption given the bound in 
Eq.~(\ref{eq:bound-slepton}).  The gluino mass $M_{\tilde{g}} = M_3$ 
is given by
\begin{equation}
  M_{\tilde{g}} \simeq \frac{g_3^2\, \hat{b}}{16\pi^2}\, 
    (\hat{\zeta}_3 M_\rho),
\label{eq:gluino-masses}
\end{equation}
while the squark mass squared $m_{\tilde{q}}^2 \sim m_{\tilde{t}}^2$ 
is given by
\begin{equation}
  m_{\tilde{q}}^2 \simeq \frac{g_3^4\, \hat{b}}{96\pi^4}\, 
    (\hat{\zeta}_3 M_\rho)^2.
\label{eq:squark-masses}
\end{equation}
We thus find that the ``size'' of the DSB gauge group $\hat{b}$ is 
given by 
\begin{equation}
  \hat{b} \simeq \frac{8}{3} \frac{M_{\tilde{g}}^2}{m_{\tilde{q}}^2}.
\label{eq:hat-b}
\end{equation}
Note that $M_{\tilde{g}}$ and $m_{\tilde{q}}$ are the renormalized 
masses at the scale $\approx M_{\rm mess}$.  The corresponding 
equation in terms of the pole masses contains an extra factor of 
$(g_3(M_{\rm mess})/g_3(M_{\tilde{g}}))^4$ in the right-hand-side.

The xyon mass is also dominated by the $SU(3)_C$ contribution.  From 
Eq.~(\ref{eq:xyon-mass}), we find
\begin{equation}
  m_\varphi^2 \simeq \frac{g_3^2}{12\pi^2} M_\rho^2.
\label{eq:xyon-mass-2}
\end{equation}
The equations~(\ref{eq:squark-masses},~\ref{eq:xyon-mass-2}) then 
tell us
\begin{equation}
  m_\varphi^2 \simeq \frac{8\pi^2}{g_3^2} 
    \frac{1}{\hat{b}\, \hat{\zeta}_3^2} m_{\tilde{q}}^2 
  \simeq 
    (4 m_{\tilde{q}})^2 \Biggl( \frac{5}{\hat{b}} \Biggr)
    \Biggl( \frac{1}{\hat{\zeta}_3} \Biggr)^2,
\label{eq:xyon-squark-ratio}
\end{equation}
where we have used $g_3 = g_3(M_{\rm mess}) \simeq 1$.  This implies 
that if the ``size'' of the DSB sector $\hat{b}$ is close to its maximum 
value $5$, which is in fact naturally the case in the holographic 
theories discussed in the next subsection, the xyon mass is in the 
multi-TeV region. For $\hat{b} \simeq 5$ and $\hat{\zeta}_3 \simeq 1$, 
Eqs.~(\ref{eq:bound-squark},~\ref{eq:xyon-squark-ratio}) gives
\begin{equation}
  m_\varphi \simlt 3~{\rm TeV},
\end{equation}
which is encouraging for the search for xyons at the LHC.%
\footnote{It should be noted that the estimate given here is very rough. 
Since the mass of xyons is quadratically divergent in the low-energy 
effective theory below $M_\rho$, its precise value depends on the details 
of the DSB sector.  In the low-energy theory, this uncertainty can manifest 
in the fact that the $M_\rho$'s appearing in Eq.~(\ref{eq:xyon-mass}) 
and in Eqs.~(\ref{eq:gaugino-masses},~\ref{eq:scalar-masses}) are not 
generically equal.  The numbers for the xyon mass in this subsection, 
therefore, should be regarded only as a rough guide.  The xyon mass is 
calculated in the next subsection in an explicit ultraviolet theory.}

The value of $\hat{\zeta}_3$ may be somewhat suppressed to give squark 
masses smaller than their ``unified theoretic'' values, i.e. the values 
given by the unified mass relations as in Eq.~(\ref{eq:te-ratio}). 
To get some feeling about this potential suppression, we can take the 
ratio between the squark mass given in Eq.~(\ref{eq:squark-masses}) and 
the right-handed selectron mass given by Eq.~(\ref{eq:scalar-masses}), 
which leads to
\begin{equation}
  \frac{\hat{\zeta}_3}{\hat{\zeta}_1} \simeq 
    \sqrt{\frac{9}{20}} \frac{g_1^2 m_{\tilde{q}}}{g_3^2 m_{\tilde{e}}}
    \simeq 0.15 \frac{m_{\tilde{q}}}{m_{\tilde{e}}}.
\label{eq:zeta-ratio}
\end{equation}
We then find that for values of squark and slepton masses that satisfy the 
naturalness bounds of Eqs.~(\ref{eq:bound-squark},~\ref{eq:bound-slepton}) 
and the experimental constraints, the ratio of the $\hat{\zeta}$ parameters 
is naturally in a range $0.2 \simlt \hat{\zeta}_3/\hat{\zeta}_1 \simlt 0.8$. 
If we rewrite Eq.~(\ref{eq:xyon-squark-ratio}) in terms of $m_{\tilde{e}}$, 
using Eq.~(\ref{eq:zeta-ratio}), we obtain
\begin{equation}
  m_\varphi^2 
    \simeq (27 m_{\tilde{e}})^2 \Biggl( \frac{5}{\hat{b}} \Biggr)
    \Biggl( \frac{1}{\hat{\zeta}_1} \Biggr)^2.
\label{eq:xyon-slepton-ratio}
\end{equation}
This implies that if the violation of the unified mass relations is 
entirely due to a suppression of squark masses by $\hat{\zeta}_3 < 1$, 
and not due to an enhancement of slepton masses by $\hat{\zeta}_1 > 1$, 
the mass of xyons is larger than about $3~{\rm TeV}$, in which case 
the discovery of xyons at the LHC may be difficult.  It is, however, 
plausible that the violation of the unified mass relations is a combined 
effect of both $\hat{\zeta}_3 < 1$ and $\hat{\zeta}_1 > 1$.  In this 
case it is possible that xyons will in fact be discovered at the LHC.

\subsection{Holographic theories in warped space}
\label{subsec:holographic}

We here consider a class of calculable theories that naturally realizes 
the framework discussed in section~\ref{sec:framework}.  Suppose that the 
size (the number of ``colors'') $\tilde{N}$ of the DSB gauge group is large: 
$\tilde{N} \gg 1$.  In this case, the DSB sector produces a large number 
of hadronic resonances at the scale $\Lambda$, whose interaction strengths 
are suppressed by powers of $1/\sqrt{\tilde{N}}$~\cite{'tHooft:1973jz}. 
This suggests that the theory may have an equivalent but different 
description based on these weakly coupled resonances.  In fact, the 
gauge/gravity duality suggests that under certain conditions these 
resonances are identified as the Kaluza-Klein (KK) towers in some 
higher dimensional theory and that we can formulate the theory in 
higher dimensional spacetime compactified to four dimensions. 

To be more explicit, let us assume that the gauge coupling $\tilde{g}$ 
of the DSB gauge interaction is nearly conformal above $\Lambda$, 
i.e. it evolves very slowly over a wide energy interval between 
$\Lambda$ and a high scale of order the unification scale $M_X 
\simeq 10^{16}~{\rm GeV}$, and that it takes a value $\tilde{g}^2 
\tilde{N}/16\pi^2 \gg 1$.  The AdS/CFT correspondence then suggests 
that we may formulate our theory in 5D anti-de~Sitter (AdS) space 
truncated by two branes~\cite{Arkani-Hamed:2000ds}.  The metric of 
this spacetime is given by
\begin{equation}
  d s^2 \equiv G_{MN} dx^M dx^N 
    = e^{-2k|y|} \eta_{\mu\nu} dx^\mu dx^\nu + dy^2,
\label{eq:metric}
\end{equation}
where $y$ is the coordinate for the extra dimension and $k$ denotes the 
inverse curvature radius of the AdS space.  The two branes are located at 
$y=0$ (the UV brane) and $y=\pi R$ (the IR brane). This is the spacetime 
considered in Ref.~\cite{Randall:1999ee}, in which the large hierarchy 
between the weak and the Planck scales is generated by the AdS warp factor. 
The scales are chosen such that the scales on the UV and IR branes are 
roughly the 4D Planck scale and the scale $\Lambda$, respectively: 
$k \sim M_5 \sim M_* \sim M_{\rm Pl}$ and $kR \sim 10$ (the 4D Planck 
scale is given by $M_{\rm Pl}^2 \simeq M_5^3/k$).  Here, $M_5$ is the 
5D Planck scale, and $M_*$ the 5D cutoff scale, which is taken to be 
somewhat (typically a factor of a few) larger than $k$.  With this choice 
of scales, the characteristic mass scale for the KK towers, which are 
5D manifestations of the resonances in the DSB sector, is given by 
$\pi k e^{-\pi kR} \sim \Lambda \approx (10\!\sim\!100)~{\rm TeV}$. 

Strictly speaking, to have a ``dual'' higher dimensional description 
considered here, the ``original'' 4D theory must possess certain 
non-trivial properties such as the existence of mass gaps between 
the resonances with spins $\leq 2$ and those with higher spins. 
However, once we have the picture of 5D warped space and construct 
a theory on this space, we can forget about the ``original'' 4D picture 
for all practical purposes and work out all physical consequences 
using higher dimensional effective field theories.  This viewpoint was 
particularly emphasized in~\cite{Nomura:2004zs}, which we follow here. 
Note that the xyon mass calculated below is dominated by the momentum 
region around $\Lambda$, so that the precise geometry in the ultraviolet 
region is not important --- it can deviate from the pure AdS without 
changing the essential result. 

\begin{figure}[t]
\begin{center}
\begin{picture}(270,175)(0,-30)
  \Line(0,5)(270,5) \Text(135,120)[b]{\large $SU(5)$}
  \Line(30,-10)(30,120) \Text(30,130)[b]{\large $321$} 
  \Text(30,-18)[t]{\small $y=0$ (UV)}
  \Text(12,111)[r]{$Q,U,D$} \Text(22,97)[r]{$L,E$ $(N)$}
  \CArc(66,20)(68,127,146) \Line(26,75)(22,69) \Line(26,75)(19,72)
  \Text(-5,63)[r]{Yukawa} \Text(10,49)[r]{couplings}
  \Line(240,-10)(240,120) \Text(240,130)[b]{\large $321$}
  \Text(240,-18)[t]{\small $y=\pi R$ (IR)}
  \CArc(298,36)(67,127,143) \Line(244,76)(247,83) \Line(244,76)(250,80)
  \Text(278,98)[]{SUSY}  \Line(260,93)(294,103)
  \Text(273,86)[tl]{\footnotesize $F_Z \neq 0$}
  \Text(135,63)[b]{$A_\mu^{321}, A_\mu^{\rm XY}, H_{\bf 5}, H_{{\bf 5}^*}$}
  \Photon(72,38)(127,38){1}{4}  \Line(127,38)(122,41) \Line(127,38)(122,35)
  \Photon(143,27)(198,27){1}{4} \Line(143,27)(148,30) \Line(143,27)(148,24)
\end{picture}
\caption{The overall picture of the theory.}
\label{fig:theory}
\end{center}
\end{figure}
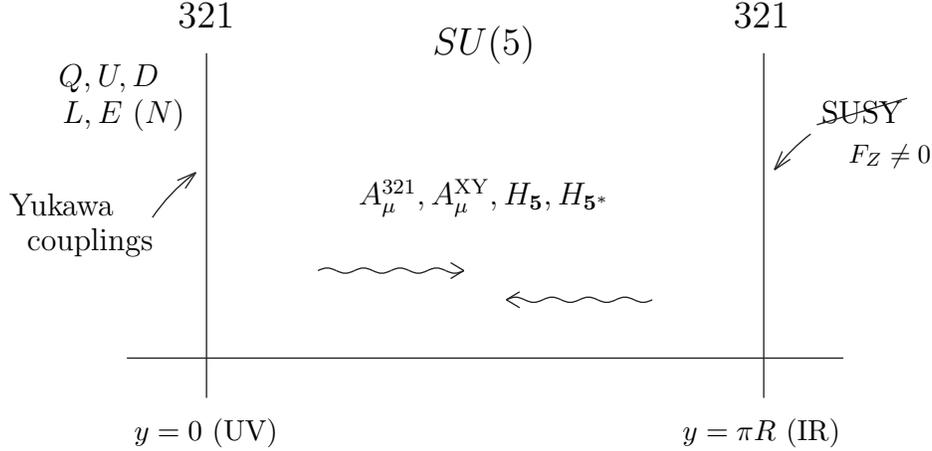

Since the global symmetry of the DSB sector corresponds to the gauge 
symmetry in the 5D picture, the minimal theory in 5D warped space has a 
gauge group $SU(5)$ in the bulk.  We thus consider a supersymmetric $SU(5)$ 
gauge theory on the truncated 5D warped spacetime, Eq.~(\ref{eq:metric}). 
The bulk $SU(5)$ symmetry is broken to the 321 subgroup at the UV brane, 
reflecting the fact that only the 321 subgroup is gauged in the 4D picture 
(at least at scales below $M_X$).  At the IR brane, both $SU(5)$ and 
supersymmetry are broken, reflecting the fact that these symmetries are 
both broken by the dynamics of the DSB sector; specifically, the $SU(5)$ 
symmetry is also broken to the 321 subgroup on this brane.  This class 
of theories was first constructed in Ref.~\cite{Nomura:2004is}, and 
we refer the readers there for further details.  The locations of 
the matter and Higgs fields are somewhat model dependent; the only 
restriction is that, to preserve the successful supersymmetric prediction 
for gauge coupling unification, the wavefunctions for the zero modes 
of these fields are either localized to the UV brane or conformally 
flat~\cite{Goldberger:2002pc} (for earlier work see~\cite{Pomarol:2000hp}). 
Here we simply put three generations of the quark and lepton superfields 
$Q, U, D, L$ and $E$ (and the right-handed neutrinos $N$) on the 
UV brane, and locate the Higgs fields, ${\bf 5} + {\bf 5}^*$ of 
$SU(5)$, in the bulk.  The case of bulk matter will be discussed in 
subsection~\ref{subsec:xyon-decay}.  The Yukawa couplings are located 
on the UV brane.  This setup leads to fully realistic phenomenologies 
--- proton decay is sufficiently suppressed and small neutrino masses 
are naturally obtained through the seesaw mechanism.  The overall 
picture of the theory is depicted in Fig.~\ref{fig:theory}.

How do xyons arise in this theory?  According to the general discussion 
at the beginning of this section, the theory must produce relatively 
light scalar particles that have the same 321 gauge quantum numbers as 
the XY gauge bosons.  To see this explicitly, we here adopt a simple 
mechanism of breaking $SU(5)$ by boundary conditions both at the 
UV and IR branes.%
\footnote{The $SU(5)$ breaking on the UV and/or IR branes can be due 
to vacuum expectation values (VEVs) of $SU(5)$-breaking Higgs fields 
located on the UV and/or IR branes.  This does not change any of the 
essential physics discussed here, as long as the VEV on the IR brane is 
sufficiently larger than the local AdS curvature scale and the VEV on 
the UV brane is of order $k$ or larger.  In the case that the UV-brane 
VEV is smaller, the estimate on the xyon mass could be affected.  We will 
comment on this case at the end of subsection~\ref{subsec:xyon-mass-2}.}
The boundary conditions on the 5D gauge multiplet are given by 
\begin{equation}
  \pmatrix{V \cr \Sigma}(x^\mu,-y) 
  = \pmatrix{ \hat{P} V \hat{P}^{-1} \cr 
             -\hat{P} \Sigma \hat{P}^{-1}}(x^\mu,y), 
\qquad
  \pmatrix{V \cr \Sigma}(x^\mu,-y') 
  = \pmatrix{ \hat{P} V \hat{P}^{-1} \cr 
             -\hat{P} \Sigma \hat{P}^{-1}}(x^\mu,y'), 
\label{eq:bc-g}
\end{equation}
where $y' = y - \pi R$, and $\hat{P}$ is a $5 \times 5$ matrix acting 
on gauge space: $\hat{P} = {\rm diag}(+,+,+,-,-)$.  Here, we have 
represented the 5D gauge multiplet in terms of a 4D $N=1$ vector 
superfield $V(A_\mu, \lambda)$ and a 4D $N=1$ chiral superfield 
$\Sigma(\chi+iA_5, \lambda')$, both of which are in the adjoint 
representation of $SU(5)$, and given the above boundary conditions in 
the orbifold picture.  The boundary conditions for the Higgs multiplets 
are given similarly.  Using notation where a bulk hypermultiplet is 
represented by two 4D $N=1$ chiral superfields $\Phi(\phi,\psi)$ and 
$\Phi^c(\phi^c,\psi^c)$ with opposite quantum numbers, the two Higgs 
hypermultiplets $\{ H, H^c \}$ and $\{ \bar{H}, \bar{H}^c \}$ obey 
boundary conditions similar to Eq.~(\ref{eq:bc-g}), i.e. $\hat{P}$ 
acting on the $SU(5)$ fundamental indices, with $V$ and $\Sigma$ 
replaced by $H$ and $H^c$ ($\bar{H}$ and $\bar{H}^c$), respectively.% 
\footnote{The boundary conditions for the Higgs multiplets on the IR 
brane can be reversed without destroying the successes of the model.}

The spectrum of the model in the supersymmetric limit is then given 
as follows~\cite{Nomura:2004is}.  For the gauge sector, the KK spectrum 
of the gauge tower, $m_n$, is approximately given by 
\begin{equation}
  \left\{ \begin{array}{ll} 
    V^{321}: & m_0 = 0, \\
    \{ V^{321}, \Sigma^{321} \}: 
      & m_n \simeq (n-\frac{1}{4})\pi k',
  \end{array} \right.
\qquad
  \left\{ \begin{array}{ll} 
    \Sigma^{\rm XY}: & m_0 = 0, \\
    \{ V^{\rm XY}, \Sigma^{\rm XY} \}: 
      & m_n \simeq (n+\frac{1}{4})\pi k',
  \end{array} \right.
\label{eq:spectrum-gauge}
\end{equation}
where $n = 1,2,\cdots$, and $k' \equiv k e^{-\pi kR} \approx 
(10\!\sim\!100)~{\rm TeV}$ is the rescaled AdS curvature scale. 
An important point is that the zero modes consist of not only the 321 
component of $V$, $V^{321}$, but also the $SU(5)/321$ (XY) component 
of $\Sigma$, $\Sigma^{\rm XY}$ (this exotic state, however, does 
not affect the gauge coupling prediction nor lead to rapid proton 
decay~\cite{Nomura:2004is}).  The spectrum for the Higgs sector is 
model dependent; it depends on the choice of boundary conditions, the 
shape of the zero-mode wavefunctions, and a potential supersymmetric 
mass term on the IR brane. 

After supersymmetry breaking, some of the zero-mode states obtain masses. 
Supersymmetry breaking is generically parameterized by an $F$-term 
vacuum expectation value (VEV) of a singlet chiral superfield $Z$ 
localized on the IR brane~\cite{Gherghetta:2000qt,Goldberger:2002pc}. 
The 321 gauginos, $\lambda^{321}$, then obtain masses through the 
following operators on the IR brane:
\begin{equation}
  S_{321} = \int\!d^4x \int_0^{\pi R}\!\!dy\,
    2 \delta(y-\pi R) \!\!\sum_{i=1,2,3} \biggl[ 
      -\int\!d^2\theta\, \frac{\zeta_i}{M_*} Z \, 
      {\rm Tr}[ {\cal W}_i^\alpha {\cal W}_{i \alpha} ] 
      + {\rm h.c.} \biggr],
\label{eq:gaugino-mass}
\end{equation}
where ${\cal W}_{i \alpha} \equiv -(1/8)\bar{\cal D}^2(e^{-2V_i} 
{\cal D}_\alpha e^{2V_i})$ represent field-strength superfields, and 
$i=1,2,3$ denotes $U(1)_Y$, $SU(2)_L$ and $SU(3)_C$, respectively.
Note that the three coefficients $\zeta_i$ are in general not equal, 
reflecting the $SU(5)$ breaking on the IR brane, so that the 321 
gaugino masses do not obey any unified relations in general.  This 
is the 5D manifestation of the spontaneous $SU(5)$ breaking in the 
DSB sector. 

Supersymmetry breaking also gives masses to zero modes in 
$\Sigma^{\rm XY}$ through the following operators on the IR brane:
\begin{eqnarray}
  S_{\rm XY} &=& \int\!d^4x \int_0^{\pi R}\!\!dy\,
    2 \delta(y-\pi R) \biggl[ \biggl\{ e^{-2\pi kR} 
    \int\!d^4\theta\, \frac{\eta}{2 M_*} Z^\dagger \,
    {\rm Tr}[ {\cal P}[{\cal A}] {\cal P}[{\cal A}]] + {\rm h.c.} \biggr\} 
\nonumber\\
  && \qquad\qquad\qquad
  {} -e^{-2\pi kR} \int\!d^4\theta\, \frac{\rho}{4 M_*^2} Z^\dagger Z \,
    {\rm Tr}[ {\cal P}[{\cal A}] {\cal P}[{\cal A}]] \biggr],
\label{eq:XY-mass}
\end{eqnarray}
where ${\cal A} \equiv e^{-2V}\!(\partial_y e^{2V}) - \sqrt{2}\, e^{-2V} 
\Sigma^\dagger e^{2V} - \sqrt{2}\, \Sigma$.%
\footnote{The expression for ${\cal A}$ in Ref.~\cite{Nomura:2004is} 
is valid only in the Wess-Zumino gauge.  The expression here applies 
in arbitrary gauges~\cite{Hebecker:2001ke}.}
Here, the trace is over the $SU(5)$ space and ${\cal P}[{\cal X}]$ 
is a projection operator: with ${\cal X}$ an adjoint of $SU(5)$, 
${\cal P}[{\cal X}]$ extracts the $({\bf 3}, {\bf 2})_{-5/6} + ({\bf 3}^*, 
{\bf 2})_{5/6}$ component of ${\cal X}$ under the decomposition to 321. 
The coefficients $\eta$ and $\rho$ are dimensionless parameters.  Apparent 
singularities arising from taking the thin-wall limit for the IR brane are 
absorbed by appropriately redefining the coefficients of IR-brane operators. 
At the leading order 
\begin{equation}
  \rho = -8 g_5^2 |\eta|^2 \delta(0) + \rho',
\label{eq:rho-prime}
\end{equation}
where $g_5$ is the 5D gauge coupling~\cite{Nomura:2004is}.  In the 
fundamental theory, these singularities are smoothed out by the 
effects of brane thickness, of order $M_*^{-1}$.

The operators in Eq.~(\ref{eq:XY-mass}) give masses of order $\pi k'$ 
to the zero modes of $\lambda'^{\rm XY}$ and $\chi^{\rm XY}$ at tree 
level, but not to $A_5^{\rm XY}$.  In fact, we find that 5D gauge 
invariance forbids any local operator giving a mass to $A_5^{\rm XY}$. 
The mass of $A_5^{\rm XY}$ is then generated only at loop level, and 
so is significantly smaller than those of the other XY states such as 
$\lambda'^{\rm XY}$ and $\chi^{\rm XY}$.  We thus find that the 
light states implied by the general argument are the zero modes of 
$A_5^{\rm XY}$ --- {\it xyons in our theory arise from the extra 
dimensional component of the grand unified gauge bosons.}%
\footnote{In general, if the $SU(5)$ breaking on the IR brane is 
caused by the VEV of a Higgs field localized on the IR brane, xyons 
are mixtures of the XY component of the brane Higgs field and the 
extra dimensional component of the grand unified gauge bosons.}

\subsection{Superparticle and xyon masses}
\label{subsec:xyon-mass-2}

We are now ready to calculate the mass of xyons $\varphi$, or the zero 
modes of $A_5^{\rm XY}$, as well as those of the MSSM superparticles, 
in the 5D theory.  These masses depend on the unknown coefficients 
$\zeta_i$, $\eta$ and $\rho'$ in Eqs.~(\ref{eq:gaugino-mass},%
~\ref{eq:XY-mass},~\ref{eq:rho-prime}) and the $F$-term VEV of $Z$ 
through the following combinations:
\begin{equation}
  M_{\lambda,i} \equiv \frac{\zeta_i F_Z}{M_*}, \qquad
  M_{\lambda,X} \equiv \frac{\eta F_Z^*}{M_*}, \qquad
  M_{\chi,X}^2 \equiv \frac{\rho' |F_Z|^2}{M_*^2},
\label{eq:susy-br-para-1}
\end{equation}
where $i=1,2,3$ and $F_Z$ is the $F$-term VEV of $Z$ defined by 
$\langle Z \rangle = - e^{-\pi kR} F_Z \theta^2$.  The natural sizes 
for these parameters are estimated using naive dimensional analysis 
as $\zeta_i \sim 1/4\pi$, $\eta \sim 1/4\pi$, $\rho' \sim 1$ 
and $F_Z \sim M_*^2/4\pi$~\cite{Chacko:1999hg}.  We thus define 
dimensionless parameters 
\begin{equation}
  r_{\lambda,i} \equiv \frac{M_{\lambda,i}}{M_*/16\pi^2}, \qquad
  r_{\lambda,X} \equiv \frac{M_{\lambda,X}}{M_*/16\pi^2}, \qquad
  r_{\chi,X} \equiv \frac{M_{\chi,X}^2}{M_*^2/16\pi^2},
\label{eq:susy-br-para-2}
\end{equation}
which are all expected to be $O(1)$: $r_{\lambda,i} \sim r_{\lambda,X} 
\sim r_{\chi,X} \sim 1$.  The masses of xyons and superparticles depend 
on these parameters as well as other model parameters, specifically 
$M_*/\pi k$ and $k' = k e^{-\pi kR}$. 

The calculation of the masses is performed following the procedure 
of Ref.~\cite{Nomura:2003qb}.  We first obtain the 5D action with all 
the couplings renormalized at a scale $k'$ measured in terms of the 4D 
metric $\eta_{\mu\nu}$.  The relevant ones are the bulk and brane 
gauge couplings, $g_5$, $\tilde{g}_{0,i}$ and $\tilde{g}_{\pi,i}$:
\begin{equation}
  S = -{1\over 4}\int\!d^4x \int_0^{\pi R}\!\!dy \, \sqrt{-G} 
    \biggl[ \frac{1}{g_5^2} F_{MN} F^{MN} 
    + 2 \delta(y) \frac{1}{\tilde{g}_{0,i}^2} {F^i}_{\mu\nu} {F^i}^{\mu\nu} 
    + 2 \delta(y - \pi R) \frac{1}{\tilde{g}_{\pi,i}^2} 
    {F^i}_{\mu\nu} {F^i}^{\mu\nu} \biggr].
\label{eq:gen-kin}
\end{equation}
Assuming that the sizes of these couplings are determined by 
naive dimensional analysis at the appropriate cutoff scale (see 
e.g.~\cite{Goldberger:2002pc}), the values of the UV-brane couplings 
evaluated at $k'$ are given by
\begin{equation}
  \frac{1}{\tilde{g}_{0,i}^2} \simeq 
    \frac{b^{\rm MSSM}_i}{8\pi^2} \ln\left(\frac{k}{k'}\right),
\label{eq:UV-couplings}
\end{equation}
where $(b^{\rm MSSM}_1, b^{\rm MSSM}_2, b^{\rm MSSM}_3) = (33/5, 1, -3)$ 
are the MSSM beta-function coefficients, and $k$ should be identified 
as the unification scale: $k \approx M_X \simeq 10^{16}~{\rm GeV}$. 
The values of the IR-brane couplings are determined by naive dimensional 
analysis as $1/\tilde{g}_{\pi,i}^2 \sim C_i/16\pi^2$, where $C_i$ are 
group theoretical factors of order one.  Because there is no logarithmic 
enhancement for the IR-brane terms, we can safely neglect these terms 
and set $1/\tilde{g}_{\pi,i}^2 = 0$ in our calculation.  (Neglecting the 
IR-brane terms yields errors of order $\pi k/M_* \simeq (20\!\sim\!50)\%$ 
for the xyon and sfermion squared-masses, but errors of this order are not 
very important for our discussion here.)  Setting $1/\tilde{g}_{\pi,i}^2 
= 0$, the 4D gauge couplings $g_i$ at the scale $k'$ are given by
\begin{equation}
  \frac{1}{g_i^2} = \frac{\pi R}{g_5^2} + \frac{1}{\tilde{g}_{0,i}^2}. 
\label{eq:4D-couplings}
\end{equation}
This determines the bulk gauge coupling, $g_5$, in terms of the 4D 
gauge couplings, $g_i$, and the UV-brane couplings evaluated at the 
scale $k'$, given by Eq.~(\ref{eq:UV-couplings}).

The masses of the gauginos and sfermions are calculated as 
in~\cite{Nomura:2003qb}, but now the gaugino mass parameters at the IR 
brane for $SU(3)_C$, $SU(2)_L$ and $U(1)_Y$ are not necessarily equal: 
$M_{\lambda,1} \neq M_{\lambda,2} \neq M_{\lambda,3}$.  The results of 
the calculation are summarized in Appendix~A.  In Fig.~\ref{fig:sparticles} 
we illustrate the behavior of the gaugino and sfermion masses as functions 
of $r_{\lambda,i}$ for $k' = 10~{\rm TeV}$ and $M_*/\pi k' = 3$. 
\begin{figure}[t]
  \center{\includegraphics[width=.6\textwidth]{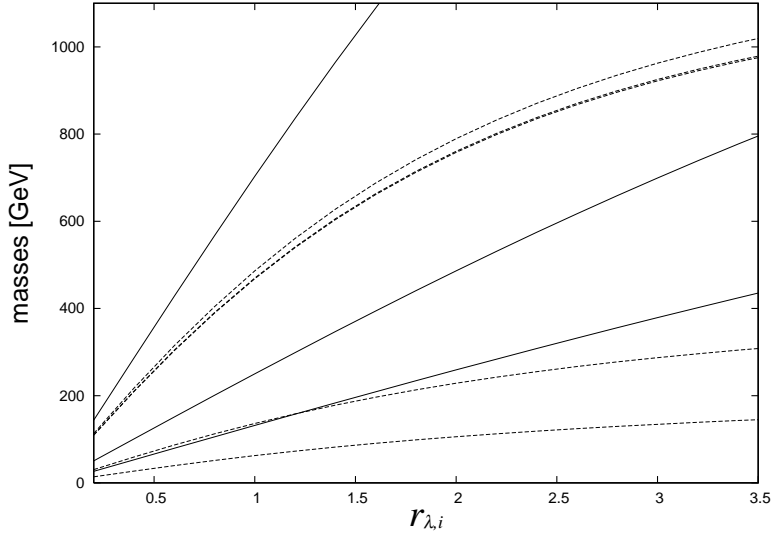}}
\caption{Masses of the MSSM gauginos (solid, with $M_3$, $M_2$ and 
 $M_1$ from above) and the MSSM scalars (dashed, with $m_{\tilde{q}}$, 
 $m_{\tilde{u}}$, and $m_{\tilde{d}}$ closely spaced and $m_{\tilde{l}}$ 
 and $m_{\tilde{e}}$ below).  The horizontal axis represents 
 dimensionless parameters $r_{\lambda,i}$ as explained in the text.}
\label{fig:sparticles}
\end{figure}
Here, the masses are the running masses but evolved down to the scale 
of superparticle masses, so that they are close to the pole masses.
The solid lines represent the gaugino masses for $SU(3)_C$, $SU(2)_L$ 
and $U(1)_Y$ from above, and the horizontal axis represents $r_{\lambda,3}$, 
$r_{\lambda,2}$, and $r_{\lambda,1}$, respectively, for each gaugino-mass 
line.  The dashed lines represent the sfermion masses ($m_{\tilde{q}}$, 
$m_{\tilde{u}}$, $m_{\tilde{d}}$, $m_{\tilde{l}}$ and $m_{\tilde{e}}$ 
from above) with $m_{\tilde{q}}$, $m_{\tilde{u}}$, and $m_{\tilde{d}}$ 
closely spaced and $m_{\tilde{l}}$ and $m_{\tilde{e}}$ below. 
To draw the sfermion-mass lines, we have taken $r_{\lambda,1} = 
r_{\lambda,2} = r_{\lambda,3}$ for simplicity; however, we can use 
Fig.~\ref{fig:sparticles} in the general case of $r_{\lambda,1} \neq 
r_{\lambda,2} \neq r_{\lambda,3}$ to obtain approximate values for the 
sfermion masses by identifying the horizontal axis to be $r_{\lambda,3}$ 
for $\{ m_{\tilde{q}}, m_{\tilde{u}}, m_{\tilde{d}} \}$, $r_{\lambda,2}$ 
for $m_{\tilde{l}}$, and $r_{\lambda,1}$ for $m_{\tilde{e}}$.  For 
different values of $k'$ the masses scale linearly in $k'$.  Taking 
different values of $M_*/\pi k'$ results in rescaling the horizontal 
axis, as the effects of $M_*/\pi k' \rightarrow \alpha (M_*/\pi k')$ is 
the same as those of $r_{\lambda,i} \rightarrow \alpha r_{\lambda,i}$ 
with a fixed value of $k'$. 

It may be useful here to present the approximate formulae for the gaugino 
and sfermion masses, derived in~\cite{Nomura:2004zs}.  These formulae 
are obtained by working out the correspondence relations between the 
4D and 5D theories, which are given by 
\begin{equation}
  M_\rho \approx \pi k', \qquad
  \frac{\hat{b}}{8\pi^2} \approx \frac{1}{g_5^2 k}, \qquad
  \hat{\zeta}_i \approx \frac{2 g_5^2 F_Z}{\pi M_*} \zeta_i.
\label{eq:corresp}
\end{equation}
Using these relations in Eqs.~(\ref{eq:gaugino-masses},%
~\ref{eq:scalar-masses}), we obtain the formulae for the gaugino masses
\begin{equation}
  M_i \simeq g_i^2 M_{\lambda,i}',
\label{eq:gaugino-masses-5D}
\end{equation}
and the sfermion masses
\begin{equation}
  m_{\tilde{f}}^2 \simeq \sum_{i=1,2,3} 
    \frac{g_i^4 C_i^{\tilde{f}}}{4 \pi^2}\, 
    (g_5^2 k) M_{\lambda,i}'^2,
\label{eq:scalar-masses-5D}
\end{equation}
where $M'_{\lambda,i} \equiv M_{\lambda,i}e^{-\pi kR} = 
(\zeta_i F_Z/M_*)(k'/k)$.  The equations~(\ref{eq:gaugino-masses-5D},%
~\ref{eq:scalar-masses-5D}) well reproduce the numerical results 
for the superparticle masses in the parameter region $\hat{\zeta}_i 
\simeq g_5^2 F_Z \zeta_i /M_* \simlt O(1)$.

The parameter region leading to the desired pattern for the 
superparticle masses can be read off from Fig.~\ref{fig:sparticles}, 
or Eqs.~(\ref{eq:gaugino-masses-5D},~\ref{eq:scalar-masses-5D}). 
For example, we find that the desired pattern is obtained by taking 
\begin{equation}
  \{ r_{\lambda,1}, r_{\lambda,2}, r_{\lambda,3} \} = \{ 2.5, 0.9, 1.0 \},
  \qquad M_*/\pi k = 3, \qquad k' = 10~{\rm TeV},
\label{eq:sample}
\end{equation}
which gives the soft supersymmetry breaking masses 
\begin{equation}
  M_1 \approx 320~{\rm GeV}, \qquad
  M_2 \approx 230~{\rm GeV}, \qquad
  M_3 \approx 700~{\rm GeV},
\label{eq:gaugino}
\end{equation}
for the gauginos and 
\begin{eqnarray}
 && m_{\tilde{q}}^2 \approx (480~{\rm GeV})^2, \qquad
    m_{\tilde{u}}^2 \approx (470~{\rm GeV})^2, \qquad
    m_{\tilde{d}}^2 \approx (470~{\rm GeV})^2,
\label{eq:squark}
\\
 && m_{\tilde{l}}^2 \approx (140~{\rm GeV})^2, \qquad
    m_{\tilde{e}}^2 \approx (120~{\rm GeV})^2,
\label{eq:slepton}
\end{eqnarray}
for the squarks and sleptons.

The spectrum in Eqs.~(\ref{eq:gaugino}~--~\ref{eq:slepton}) has the 
following characteristic features. 
\begin{enumerate}
\item[(A)]
Taking into account small mass splittings among the generations arising 
from the Yukawa couplings, the lightest among the MSSM gauginos, 
squarks and sleptons is the stau $\tilde{\tau}$, which is most likely 
the next-to-lightest supersymmetric particle (NLSP).  (The lightest 
supersymmetric particle (LSP) is the gravitino, whose mass is of order 
$m_{3/2} \approx \Lambda^2/M_{\rm Pl} \simeq (0.1\!\sim\!10)~{\rm eV}$ 
unless there is some other source of supersymmetry breaking than the 
DSB sector considered.)
\item[(B)]
The 5D theory with small tree-level UV-brane gauge kinetic terms 
corresponds in 4D to a theory in which the standard model gauge couplings 
are large at the unification scale $M_X \simeq 10^{16}~{\rm GeV}$ (see 
Ref.~\cite{Goldberger:2002pc}).  This implies $\hat{b} \simeq 5$, and 
we find from Eqs.~(\ref{eq:gaugino-masses},~\ref{eq:scalar-masses}), 
or Eqs.~(\ref{eq:gaugino-masses-5D},~\ref{eq:scalar-masses-5D}), that 
$M_3/m_{\tilde{q},\tilde{u},\tilde{d}} \approx (3\hat{b}/8)^{1/2} \simeq 
1.4$ and $M_2/m_{\tilde{l}} \approx (2\hat{b}/3)^{1/2} \simeq 1.8$. 
(The corresponding relation for $U(1)_Y$, $M_1/m_{\tilde{e}} \approx 
(5\hat{b}/6)^{1/2} \simeq 2.0$, does not hold well because the value 
of $\zeta_1$, or $\hat{\zeta}_1$, $r_{\lambda,i}$, is outside the 
region where the approximate mass formulae apply.) 
\item[(C)]
The masses of the squarks are close with each other.  They are larger 
than the slepton masses but only by a factor of a few.  In particular, 
the ratio of the squark to the slepton masses is smaller than that 
given by the unified mass relations.
\item[(D)]
The Higgsino mass parameter $\mu$ inferred from Eq.~(\ref{eq:ewsb-cond}) 
is generically small, $|\mu| \simlt 200~{\rm GeV}$ for $\tan\beta 
\simgt 2$, if we require the amount of fine-tuning to be smaller than 
$20\%$.  Therefore, if the Higgs sector has a certain resemblance to 
that of the MSSM, the two lightest neutralinos and the lighter chargino 
tend to have closer masses.  The value of $|\mu|$, however, can be 
larger if $\tan\beta$ is smaller, due to a larger value of $y_t$. 
(Note that $\tan\beta$ as small as $\simeq 1.2$ is possible in the 
present theory because the evolution of the top Yukawa coupling is 
strongly asymptotically free due to larger values of the $SU(3)_C$ 
gauge coupling at high energies.)  In general, $\mu$ is bounded as 
$|\mu| \simlt (310\!\sim\!400)~{\rm GeV}$ (see Eq.~(\ref{eq:ewsb-cond}) 
and the discussion below Eq.~(\ref{eq:MH-bound})), so that there are 
at least two neutralinos and a chargino with masses below $\simeq 
400~{\rm GeV}$.
\end{enumerate}
In fact, these features are somewhat generic in our scenario because 
they arise mainly from the gauge mediated nature of supersymmetry breaking 
and the relations between the gaugino mass parameters $r_{\lambda,2}, 
r_{\lambda,3} \simlt r_{\lambda,1}$, which comes from the requirement 
of reducing the fine-tuning and evading the experimental constraint 
$m_{\tilde{e}} \simgt 100~{\rm GeV}$.  It is, however, important 
to notice that they are also subject to some model dependencies. 
For instance, the mass ratios of gauginos to squarks and sleptons 
given in (B) become smaller if $\tilde{b}$ is smaller, which can 
be the case if the $SU(5)$ breaking on the UV brane is due to the 
Higgs mechanism.  The ratios may even be completely different if 
the mechanism of gaugino mass generation is different (see discussion 
in section~\ref{sec:concl}).  Also, the feature of (D) could depend 
on the detailed structure of the Higgs sector, which we do not specify 
explicitly in the present work.  In fact, the structure of the Higgs 
sector in our scenario is expected to be (much) richer than that of 
the MSSM. 

Here we note that the particular parameter point above has been chosen 
for illustrative purposes.  Experimental constraints will realistically 
reduce the parameter space of our scenario, especially that coming from 
the $b \rightarrow s\gamma$ process.  Determining the precise constraints 
on our parameter space is a complicated problem, mostly because of the 
theoretical uncertainties involved, and so we do not pursue it here. 
However, there is a large parametric freedom in our framework.  For 
example, if it turns out that the above parameter point is problematic, 
we can increase the value of $k'$ somewhat and/or change the ratios 
between $r_{\lambda,i}$'s, still keeping the fine-tuning small. 
Some of the constraints may also depend on the structure of the Higgs 
sector, which we have not specified in detail.  The constraints are also 
different if the gauginos are Dirac fermions as in the model discussed 
in section~\ref{sec:concl}. 

We now turn to the xyon mass.  The mass of xyons is obtained by 
calculating 5D one-loop diagrams.  The details of the calculation 
are given in Appendix~A.  The xyon mass squared, $m_\varphi^2$, is 
given in terms of the parameters $r_{\lambda,i}$, $r_{\lambda,X}$, 
$r_{\chi,X}$, $M_*/\pi k$ and $k'$.  In general, $m_\varphi^2$ 
depends on the complex phases of the IR-brane supersymmetry breaking 
parameters $M_{\lambda,i}$ and $M_{\lambda,X}$ (i.e. the phases of 
$r_{\lambda,i}$ and $r_{\lambda,X}$), but as is discussed in Appendix~A 
the dependence of $m_\varphi^2$ on these phases is small.  We thus 
take $r_{\lambda,i}$ and $r_{\lambda,X}$ to be real in the analysis 
here (the other parameters, $r_{\chi,X}$, $M_*/\pi k$ and $k'$, are 
always real). The analysis in Appendix~A also tells us the following. 
(i) The squared mass for xyons, $m_\varphi^2$, is positive for most of 
the parameter region, which is crucial for the model to be viable. (ii) 
In a natural parameter region $r_{\lambda,i} \sim r_{\lambda,X} \sim 
r_{\chi,X} \sim O(1)$ the value of $m_\varphi^2$ depends practically 
only on $r_{\lambda,X}$, $M_*/\pi k$ and $k'$. (iii) The effect of 
$M_*/\pi k \rightarrow \alpha (M_*/\pi k')$ on $m_\varphi^2$ is the 
same as that of $r_{\lambda,X} \rightarrow \alpha r_{\lambda,X}$ 
with a fixed value of $k'$. (iv) The xyon mass, $(m_\varphi^2)^{1/2}$, 
scales almost linearly with $k'$. 

\begin{figure}[t]
  \center{\includegraphics[width=.6\textwidth]{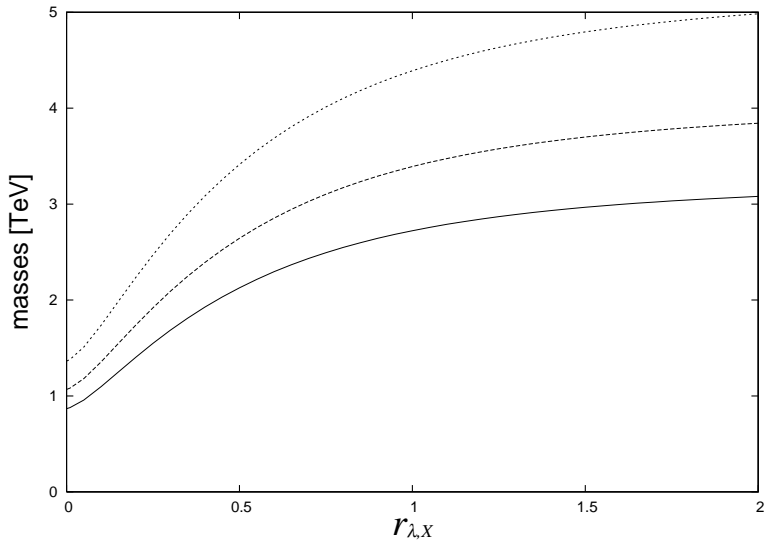}}
\caption{The mass of xyons as a function of $r_{\lambda,X}$ for 
 $\{ r_{\lambda,1}, r_{\lambda,2}, r_{\lambda,3} \} = \{ 2.5, 0.9, 
 1.0 \}$, $r_{\chi,X} = 1$ and $M_*/\pi k = 3$.  The three curves 
 correspond to three different values of $k'$: $k' = 8~{\rm TeV}$ 
 (solid) $10~{\rm TeV}$ (dashed) and $13~{\rm TeV}$ (dotted).}
\label{fig:xyon}
\end{figure}

The features (ii), (iii) and (iv) described above allows us to 
represent $m_\varphi^2$ as a function only of $r_{\lambda,X}$ with 
fixed values of the other parameters, since the dependence on the 
other parameters is either very weak or trivially reproduced.  In 
Fig.~\ref{fig:xyon} we plot the mass of xyons, $(m_\varphi^2)^{1/2}$ 
as a function of $r_{\lambda,X}$ for $k' = 8, 10$ and $13~{\rm TeV}$ 
with the fixed values of $\{ r_{\lambda,1}, r_{\lambda,2}, r_{\lambda,3} 
\} = \{ 2.5, 0.9, 1.0 \}$, $r_{\chi,X} = 1$ and $M_*/\pi k = 3$. 
As discussed before (see Eq.~(\ref{eq:sample}) and discussions around), 
these values give the superparticle masses desired for electroweak 
symmetry breaking.  The figure then shows that for natural values 
of $r_{\lambda,X} \sim O(1)$, the mass of xyons lies in a range 
$1~{\rm TeV} \simlt m_\varphi \simlt 5~{\rm TeV}$.  While the IR-brane 
operators neglected in our analysis can cause an error of order 
$\pi k/M_* \simeq 30\%$ in the xyon mass, we still conclude that 
{\it for natural values of the model parameters, $r_{\lambda,i} \sim 
r_{\lambda,X} \sim r_{\chi,X} \sim O(1)$ and $M_*/\pi k$ a factor of 
a few, the mass of xyons is expected to lie in the multi-TeV region}.%
\footnote{In the model of Ref.~\cite{Chacko:2005ra} where supersymmetry 
breaking is mediated to the Higgs sector through singlet fields, the 
scale of the IR brane tends to be higher, $k' = O(100~{\rm TeV})$, leading 
to $m_\varphi = O(10~{\rm TeV})$.  This corresponds to the region where 
$r_{\lambda,i}$, or $\zeta_i$, are smaller than the naive values, 
$r_{\lambda,i} = O(0.1)$.  It is possible, however, to modify the 
Higgs sector to allow more direct mediation so that the naive values 
for the couplings are accommodated and that the xyon mass is lowered 
to the multi-TeV region.}
This is encouraging for the discovery of xyons at the LHC.  As we will 
discuss in subsection~\ref{subsec:collider}, the reach of the LHC in 
the xyon mass is about $(2.0\!\sim\!2.2)~{\rm TeV}$.  Therefore, for 
$k' = 8~{\rm TeV}$ ($10~{\rm TeV}$), xyons may be discovered at the LHC 
if $r_{\lambda,i} \simlt 0.6$ ($0.4$).  It is also interesting that the 
mass of xyons saturates at larger values of $r_{\lambda,X}$.  For example, 
the mass is bounded for $k' = 8, 10, 13~{\rm TeV}$ by $\approx 3, 4, 
5~{\rm TeV}$, respectively.  The discovery of xyons at the VLHC will, 
therefore, be quite promising. 

A few comments are in order.  We have calculated so far the mass of 
xyons in the theory where $SU(5)$ is broken by boundary conditions 
at the UV brane.  This corresponds in the 4D picture to the case with 
$\hat{b} \simeq 5$, as the successful prediction for the low-energy gauge 
couplings requires the theory to be strongly coupled at the unification 
scale (the tree-level UV-brane gauge kinetic terms to be small).  The 
breaking of $SU(5)$ on the UV brane, however, may also be due to the 
Higgs mechanism.  Then, if the VEV of the $SU(5)$-breaking Higgs is 
sufficiently smaller than the cutoff scale $M_*$, we can have a sizable 
tree-level UV-brane gauge kinetic term without destroying the successful 
prediction.  This corresponds in 4D to having small 321 gauge couplings 
at the unification scale and thus $\hat{b}$ smaller than $\simeq 5$. 
This in turn raises the mass of xyons by a factor of $(5/\hat{b})^{1/2}$ 
compared with the values given in Fig.~\ref{fig:xyon} (see e.g. 
Eqs.~(\ref{eq:xyon-squark-ratio},~\ref{eq:xyon-slepton-ratio})). 

Finally, the masses of the superpartners of xyons are calculated 
using equations given in Appendix~A.  These masses are plotted in 
Fig.~\ref{fig:lambda-chi} for $M_*/\pi k = 3$ and $k' = 10~{\rm TeV}$. 
\begin{figure}[t]
  \center{\includegraphics[width=.6\textwidth]{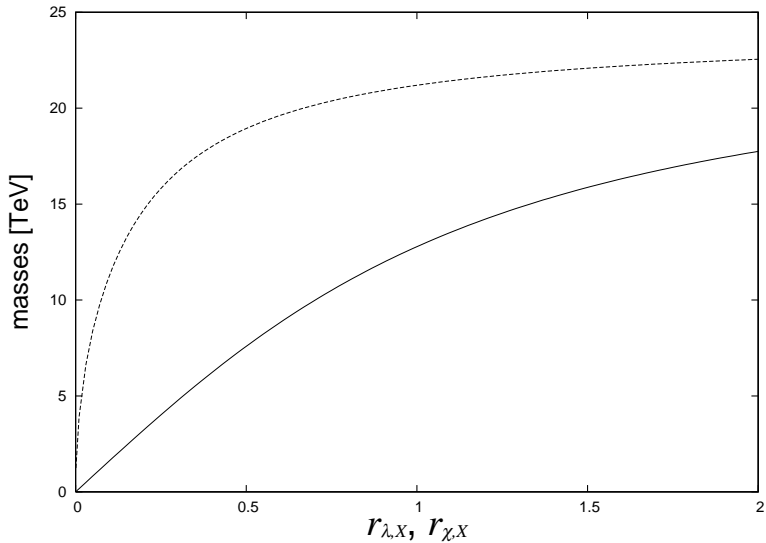}}
\caption{Masses of the superpartners of xyons, $\lambda'^{\rm XY}$ 
 (solid) and $\chi^{\rm XY}$ (dashed), for $M_*/\pi k = 3$ and 
 $k' = 10~{\rm TeV}$.  The horizontal axis represents $r_{\lambda,X}$ 
 and $r_{\chi,X}$ for $\lambda'^{\rm XY}$ and $\chi^{\rm XY}$, 
 respectively.}
\label{fig:lambda-chi}
\end{figure}
Solid and dashed lines represent the fermionic and bosonic superpartners 
of xyons, $\lambda'^{\rm XY}$ and $\chi^{\rm XY}$, respectively, and 
the horizontal axis corresponds to $r_{\lambda,X}$ and $r_{\chi,X}$ 
for $\lambda'^{\rm XY}$ and $\chi^{\rm XY}$.  These particles are out 
of the reach of the LHC, so we need a collider with larger energies 
to discover them.  We find from the figure that the fermionic partner 
$\lambda'^{\rm XY}$, which may be called the {\it xyino}, is within the 
reach of the VLHC with center-of-mass energies of $(50\!\sim\!200)~{\rm 
TeV}$ for $r_{\lambda,X} \simlt 1$.  However, the discovery of the 
scalar partner $\chi^{\rm XY}$, or {\it sxyon}, may be difficult 
even at the VLHC with highest possible energies.

\subsection{Xyon decay}
\label{subsec:xyon-decay}

Experimental properties of xyons depend strongly on their lifetime 
and/or decay modes.  We here study xyon decay in general theories, 
including the holographic theories discussed in the previous two 
subsections. 

Let us first consider the case where the DSB gauge interaction is 
asymptotically free (this is {\it not} the case in the holographic 
theories discussed before).  In this case xyons $\varphi$ arise 
as composite particles at the dynamical scale $\Lambda \approx 
(10\!\sim\!100)~{\rm TeV}$.  Now, let us introduce a composite chiral 
superfield $\Sigma$ that contains xyons as the imaginary part of the 
lowest component, $\Sigma = (\chi+i\varphi)+\cdots$.  Since xyons are 
(pseudo-)Goldstone bosons, the dynamics of the DSB sector respect the 
shift symmetry $\Sigma \rightarrow \Sigma + i\epsilon$ where $\epsilon$ 
is a constant.  Now, imagine that xyons are composed of $n$ constituent 
fields, collectively denoted as ${\cal Q}$, so that $\Sigma \approx 
{\cal Q}^n/\Lambda^{n-1}$ ($n \geq 2$).  Then the direct decay of 
xyons into the MSSM particles can be caused by interactions of the 
form ${\cal L} \approx \int\!d^4\theta\, {\cal Q}^n U^\dagger Q/M_*^n$, 
$\int\!d^4\theta\, {\cal Q}^n Q^\dagger E/M_*^n$ and $\int\!d^4\theta\, 
{\cal Q}^n L^\dagger D/M_*^n$.  Here, $M_*$ is the fundamental scale of 
the theory, expected to be of order the unification or Planck scale, and 
we have supplied powers of $M_*$ by dimensional analysis.  The dimension 
of ${\cal Q}$ is counted as $1$ (the canonical dimension), since the 
DSB gauge interaction is weak at the scale $M_*$ by assumption.  After 
confinement at the scale $\Lambda$, the interactions listed above lead 
to the effective operators ${\cal L} \sim \int\!d^4\theta\, (\Sigma 
+ \Sigma^\dagger) U^\dagger Q$, $\int\!d^4\theta\, (\Sigma + \Sigma^\dagger) 
Q^\dagger E$ and $\int\!d^4\theta\, (\Sigma + \Sigma^\dagger) L^\dagger D$ 
with coefficients of order $\Lambda^{n-1}/M_*^n$.  This gives the decay 
rate $\Gamma = O(m_\varphi^3 (\Lambda^{n-1}/M_*^n)^2 (m_q/m_\varphi)^2)$, 
where $m_q$ is the largest mass of the final state fermions, leading to 
the lifetime of order $10^{-23}(M_*/\Lambda)^{2n}~{\rm sec}$ (here we 
have used $m_q = m_t$).%
\footnote{For light final-state fermions, the decay into scalar 
superpartners can be faster with the rate given by $\Gamma = 
O(m_\varphi^3 (\Lambda^{n-1}/M_*^n)^2 \{(m_1^2-m_2^2)/m_\varphi^2\}^2)$, 
where $m_1$ and $m_2$ are the masses of the two final-state scalars.}
If these are the dominant decay modes, therefore, the lifetime of 
xyons are much longer than the age of the universe, $\tau_\varphi 
\gg 10^{10}~{\rm years}$.

It is possible that xyons have faster decay modes, depending on the spectrum 
of the DSB sector.  Suppose that DSB gauge interactions produce composite 
states $\phi_A$ and $\phi_B$, and that xyons can be converted into these 
states due to strong interactions of the DSB sector: $\varphi \rightarrow 
\phi_A + \phi_B$.  The (virtual) states $\phi_A$ and $\phi_B$ can then 
directly decay into the MSSM particles, depending on their 321 gauge 
quantum numbers.  For example, if $\phi_A$ and $\phi_B$ have the quantum 
numbers of $L$ and $D^\dagger$, respectively, they can decay into the 
MSSM states through operators of the form ${\cal L} \sim \int\!d^2\theta 
L \bar{\Phi}_A$ and $\int\!d^2\theta D \Phi_B$, where $\bar{\Phi}_A$ and 
$\Phi_B$ are chiral superfields containing appropriate components of the 
massive $\phi_A$ and $\phi_B$ supermultiplets.  Then, if $\bar{\Phi}_A$ 
and $\Phi_B$ are made out of $n_A$ and $n_B$ constituent fields 
($n_A, n_B \geq 2$), the decay rate of xyons will be roughly of order 
$\Lambda(\Lambda/M_*)^{n_A+n_B-4}$, which could be much faster than 
the direct decay discussed before.

The situation is similar in the holographic theories, except that in this 
case interactions of the DSB sector do not become weak in the UV so that 
the dimensions of the composite operators, such as $n_A$ and $n_B$ above, 
are now free parameters taking arbitrary values larger than $1$ (not 
necessarily integers due to large anomalous dimensions).  In the 5D 
viewpoint, the direct decay operators discussed before correspond to 
UV-brane operators of the form:
\begin{equation}
  S_{\varphi,1} \sim \int\!d^4x \int_0^{\pi R}\!\!dy\, 2 \delta(y) 
    \biggl[ \frac{1}{M_*}\!\!\int\!d^4\theta\, U^\dagger 
        {\cal P}'[{\cal B}] Q
      + \frac{1}{M_*}\!\!\int\!d^4\theta\, Q^\dagger 
        {\cal P}'[{\cal B}] E
      + \frac{1}{M_*}\!\!\int\!d^4\theta\, L^\dagger 
        {\cal P}'[{\cal B}] D
      + {\rm h.c.} \biggr],
\label{eq:direct-xyon}
\end{equation}
where ${\cal B} \equiv \partial_y e^{2V} - \sqrt{2}\, \Sigma^\dagger e^{2V} 
- \sqrt{2}\, e^{2V} \Sigma$, and ${\cal P}'[{\cal X}]$ is a projection 
operator extracting the $({\bf 3}, {\bf 2})_{-5/6}$ component of an $SU(5)$ 
adjoint ${\cal X}$ (here we have suppressed order one coefficients).  Since 
the wavefunction value of the canonically normalized xyon field is about 
$\sqrt{g_5^2 k}\, e^{-\pi kR} \approx \sqrt{\pi kR}\, (k'/k)$ at $y=0$, this 
leads to the xyon decay rate $\Gamma \simeq (m_\varphi^3/8\pi) (\pi kR) 
(k'/k M_*)^2 (m_q/m_\varphi)^2 \sim m_\varphi^3 (k'/k^2)^2 (m_q/m_\varphi)^2$, 
which implies that the dimension of the composite xyon field is given by 
$n=2$.  In fact, this statement is generally true in any theories where 
xyons are identified as the extra dimensional component of the broken 
unified gauge bosons.  We thus conclude that direct decays of xyons 
into the MSSM states are highly suppressed, leading to a lifetime much 
longer than the age of the universe if they are dominant decay modes.

Faster decay modes of xyons can, in principle, be available if some of 
the MSSM Higgs and/or matter fields propagate in the 5D bulk.  In the case 
that only the Higgs fields propagate in the bulk, xyon decay could occur 
as follows.  First, xyons can be converted into the MSSM Higgs-doublet 
and the (virtual) Higgs triplet states through the bulk gauge interaction, 
$S = \int\!d^4x \int\!dy\, \{ -e^{-3k|y|} \int\!d^2\theta \sqrt{2}\, 
H^c \Sigma H + {\rm h.c.} \}$, as $\varphi \rightarrow H_D + H_T^c$ (or 
$\varphi \rightarrow \bar{H}_D + \bar{H}_T^c$ through the corresponding 
interaction for $\{\bar{H}, \bar{H}^c\}$).  The Higgs triplet state can 
then decay into the MSSM particles, for example through the UV-brane 
operators $S = \int\!d^4x \int\!dy\, 2\delta(y) \{ (1/M_*)\int\!d^2\theta 
(\nabla_y H_T) QQ + (1/M_*)\int\!d^2\theta (\nabla_y H_T)UE + {\rm h.c.} 
\}$, where $\nabla_y H_T$ represents the triplet component of $\nabla_y H 
\equiv \partial_y H + \sqrt{2} \Sigma H$.%
\footnote{Possible operators such as $S = \int\!d^4x \int\!dy\, 2\delta(y) 
\{ \int\!d^2\theta H_T^c QL + \int\!d^2\theta H_T^c UD + {\rm h.c.} \}$ 
are forbidden if we impose a continuous $U(1)_R$ symmetry with the 
charges $Q_R(V,\Sigma,H,\bar{H})=0$, $Q_R(H^c,\bar{H}^c)=2$ and 
$Q_R(Q,U,D,L,E,N)=1$, which is well motivated to provide complete 
solutions to the doublet-triplet and dimension-five proton decay 
problems~\cite{Goldberger:2002pc}.  These operators are allowed if we do 
not impose $U(1)_R$.  Operators involving $H_T$ without the $y$-derivative, 
e.g. $S = \int\!d^4x \int\!dy\, 2\delta(y) \{ \int\!d^2\theta H_T QQ + 
\int\!d^2\theta H_T UE + {\rm h.c.} \}$, could also be allowed if the 
$SU(5)$ breaking on the UV brane is due to the Higgs mechanism.  These 
operators would lead to faster decays with rates $\Gamma \approx k' 
(k'/k)^{4c_H}$.  This, however, does not change our main conclusion here.}
This leads to a partial decay rate of xyons of order $\Gamma \approx 
k' (k'/k)^{4c_H+2}$, where $c_H$ is the bulk mass parameter for the Higgs 
multiplet which controls the wavefunction profile for the zero mode (for 
details see~\cite{Nomura:2004is}).  Given that the successful prediction 
for gauge coupling unification requires $c_H \geq 1/2$~\cite{Nomura:2004is}, 
we find that this decay mode cannot be much faster than the direct decay 
discussed before.  We thus conclude that {\it if matter fields are localized 
on the UV brane, which is practically equivalent to the condition that the 
entire DSB sector is even under $R$ parity, then xyons behave as stable 
particles at least for collider purposes}.

It is possible to consider the case where xyons decay much faster.  This 
occurs if some of the MSSM matter fields propagate in the bulk.  Suppose 
that $D$ and $L$ of the MSSM come from bulk hypermultiplets as follows. 
We introduce two hypermultiplets $\{ F, F^c \}$ and $\{ F', F'^c \}$ for 
each generation, transforming as ${\bf 5}^*$ under $SU(5)$.  By assigning 
suitable boundary conditions, it is possible to obtain four zero-mode 
states $D({\bf 3^*}, {\bf 1})_{1/3} \subset F$, $L({\bf 1}, {\bf 2})_{-1/2} 
\subset F'$, $L^c({\bf 1}, {\bf 2})_{1/2} \subset F^c$ and $D^c({\bf 3}, 
{\bf 1})_{-1/3} \subset F'^c$ from these two hypermultiplets.  We then 
introduce two chiral superfields $D'({\bf 3^*}, {\bf 1})_{1/3}$ and 
$L'({\bf 1}, {\bf 2})_{-1/2}$ on the IR brane (for each generation) 
together with mass terms of the form $S = \int\!d^4x \int\!dy\, 
2\delta(y-\pi R) \{ \int\!d^2\theta (D D^c + D' D^c + L L^c + L' L^c) 
+ {\rm h.c.} \}$.  This leaves three generations of $D$ and $L$ at low 
energies, giving a complete matter sector of the MSSM together with 
the $Q$, $U$ and $E$ fields located on the UV brane.  The bulk mass 
parameters of the $\{ F, F^c \}$ and $\{ F', F'^c \}$ fields are taken 
as $c_{F}, c_{F'} \geq 1/2$ to preserve the successful prediction for 
gauge coupling unification.%
\footnote{The successful prediction is not destroyed even for $c_{F}, 
c_{F'} < 1/2$, if for some reason $c_F = c_{F'}$.}
An interesting feature of this setup is that we can arrange the theory 
such that it does not lead to proton decay at a rate inconsistent 
with experiments even without imposing an ad hoc symmetry.  This 
occurs, for example, if the theory possesses a $U(1)_R$ symmetry 
of~\cite{Goldberger:2002pc} and its breaking is encoded only in the 
$F$-component VEV of the supersymmetry-breaking field $Z$, or simply 
if only one or two generations are significantly delocalized from 
the UV brane.  Potentially dangerous flavor changing neutral currents 
arising from the IR-brane mass mixings are suppressed sufficiently for 
$c_{F}, c_{F'} \geq 1/2$.  To avoid reintroduction of the supersymmetric 
flavor problem, however, we need $c_{F}, c_{F'} \simgt 0.8$, unless we 
introduce an additional ingredient into the theory, such as a flavor 
symmetry in the bulk and on the IR brane together with appreciable 
tree-level UV-brane kinetic terms for bulk matter.

Xyon decay in the bulk matter theory discussed above occurs through the 
IR-brane operator
\begin{equation}
  S_{\varphi,2} \sim \int\!d^4x \int_0^{\pi R}\!\!dy\, 2 \delta(y-\pi R) 
    \biggl[ e^{-2\pi kR} \frac{1}{M_*^2}\!\!\int\!d^4\theta\, 
      L^\dagger {\cal P}'[{\cal B}] D + {\rm h.c.} \biggr],
\label{eq:direct-xyon-2}
\end{equation}
where we have suppressed order one coefficients.  This leads to the xyon 
decay interactions of the form
\begin{equation}
  {\cal L}_{\rm 4D} \sim \xi\,
    \Bigl( (\partial_\mu \varphi)\, d^\alpha 
      \sigma^\mu_{\alpha\dot{\alpha}} l^{\dagger\dot{\alpha}} 
      -i (\partial_\mu \varphi) \{ \tilde{l}^* (\partial^\mu\tilde{q})
       - (\partial^\mu\tilde{l}^*) \tilde{q} \} \Bigr) + {\rm h.c.},
\label{eq:direct-xyon-3}
\end{equation}
where $\varphi$ is the xyon field transforming as $({\bf 3}, {\bf 2})_{-5/6}$ 
under 321, and $d$, $l$, $\tilde{d}$ and $\tilde{l}$ are the down-type 
quark, doublet lepton, right-handed down-type squark and left-handed 
slepton, respectively.  The coefficient $\xi$ is given by
\begin{equation}
  \xi \simeq \sqrt{\frac{(2c_F-1)(2c_{F'}-1)\, g_5^2 k}
    {(1-e^{(1-2c_F)\pi kR})(1-e^{(1-2c_{F'})\pi kR})}} 
    \left( \frac{k'}{M_*'^2} \right) e^{-(c_F+c_{F'}-1)\pi kR},
\label{eq:xi-1}
\end{equation}
where $M_*' \equiv M_* e^{-\pi kR}$, and this can be approximated 
further as
\begin{equation}
  \xi \simeq \left\{ \begin{array}{@{\,}ll}
    (1/\sqrt{\pi kR}) (k'/M_*'^2) 
    & \quad ({\rm for}\,\,\, c_F,c_{F'} \simeq 1/2) \\
  \sqrt{\pi kR}\, (k'/M_*'^2)\, e^{-(c_F+c_{F'}-1)\pi kR}
    & \quad ({\rm for}\,\,\, c_F,c_{F'} \gg 1/2).
  \end{array} \right. \quad
\label{eq:xi-2}
\end{equation}
The partial decay rates for xyons are given by
\begin{equation}
  \Gamma_{\varphi \rightarrow d+l} \simeq 
    \frac{\xi^2}{8\pi} (m_d^2+m_l^2)\, m_\varphi,
\end{equation}
for $\varphi \rightarrow d + l$ and
\begin{equation}
  \Gamma_{\varphi \rightarrow \tilde{d}+\tilde{l}} \simeq 
    \frac{\xi^2}{8\pi} \frac{(m_{\tilde{d}}^2-m_{\tilde{l}}^2)^2}{m_\varphi},
\end{equation}
for $\varphi \rightarrow \tilde{d} + \tilde{l}$, where we have set 
$m_d,m_l,m_{\tilde{d}},m_{\tilde{l}} \ll m_\varphi$.  For the superparticle 
spectrum given in Eqs.~(\ref{eq:gaugino}~--~\ref{eq:slepton}) (and thus 
the parameters of Eqs.~(\ref{eq:sample})) and $m_d=m_b$, we find that the 
dominant decay mode is that to a squark and a slepton.  For $m_\varphi 
\simeq (2\!\sim\!3)~{\rm TeV}$, the lifetime of xyons is given by 
\begin{equation}
  \tau_\varphi \simeq \left\{ \begin{array}{@{\,}ll}
    10^{-16}~{\rm sec}
    & \quad ({\rm for}\,\,\, c_F,c_{F'} \simeq 1/2) \\
    10^{-19} \times e^{2(c_F+c_{F'}-1)\pi kR}~{\rm sec}
    & \quad ({\rm for}\,\,\, c_F,c_{F'} \gg 1/2).
  \end{array} \right. \quad
\label{eq:xyon-lifetime}
\end{equation}
The shortest lifetime is obtained for $c_F = c_{F'} = 1/2$.  Note that 
even for $c_F = c_{F'} = 1/2$ the lifetime is long enough such that 
$\Gamma_\varphi/m_\varphi \ll 1$.  Therefore, we conclude that {\it the 
xyon always appears as a narrow particle rather than a broad resonance}. 
The lifetime of xyons becomes exponentially longer for larger values 
of $c_F$ and $c_{F'}$.  For example, the lifetime is already of order 
$\tau_\varphi \approx 10^{-4}~{\rm sec}$ for $c_F = c_{F'} = 0.8$. 

A final comment is in order.  We have implicitly assumed in our analysis 
that there are no light exotic particles, for example colored Higgs fields, 
into which xyons can decay.  This is a reasonable assumption because if 
the parameters of the theory obey naive dimensional analysis the masses of 
these exotic particles are of order $\pi k'$, which are much larger than 
the xyon mass.  If the masses of the exotic particles are somehow small, 
however, phenomenology of xyons (and the exotic particles) could change 
significantly.  We will not pursue such a possibility further in this paper.

\section{Experimental Signatures}
\label{sec:exp}

In this section we study experimental signatures of xyons.  We first study 
signals and reaches of xyons at hadron colliders.  We then discuss possible 
indirect effects of xyons in the case that the xyon mass lies in the range 
outside of the direct reach of the LHC.  We also discuss cosmological 
implications of our scenario.

\subsection{Collider search}
\label{subsec:collider}

Since xyons are generally colored, they hadronize after production, 
picking up the standard model quarks.  Here we mainly consider the case 
where xyons are long lived (the case where matter fields are localized 
on or towards the UV brane in the holographic 5D theories) and study 
experimental consequences of having these exotic hadrons. 

In the simplest case of $SU(5)$ xyons, the 321 quantum numbers of xyons 
are given by $({\bf 3}, {\bf 2})_{-5/6}$.  Let us denote the isospin 
up and down components of the $({\bf 3}, {\bf 2})_{-5/6}$ xyon as 
$\varphi_\uparrow$ and $\varphi_\downarrow$, respectively.  We first 
find that the contributions from the $SU(2)_L$ and $U(1)_Y$ $D$-terms 
to the xyon squared masses, $\delta m^2_{D,\varphi}$, are only of 
order $O(m_Z^4/(\pi k')^2)$: $\delta m^2_{D,\varphi_\uparrow} - 
\delta m^2_{D,\varphi_\downarrow} \simeq -(5/16)(m_Z^4/m_{\chi^{\rm 
XY}}^2) \cos^2\! 2\beta$, where $\tan\beta$ is the ratio of the VEVs 
of the two Higgs doublets.  This is because the auxiliary fields $D$ 
couple in the Lagrangian only in the form ${\cal L} \sim \chi^{\rm XY} 
D^{321} A_5^{\rm XY}$ (see Appendix~A).  This implies that the contributions 
to the mass splitting $m_{\varphi_\uparrow} - m_{\varphi_\downarrow}$ 
from the $D$-terms are negligible.

The dominant effect for the mass splitting between $\varphi_\uparrow$ 
and $\varphi_\downarrow$ then arises from loops of the gauge fields. 
The effect comes dominantly from one loop of the electroweak gauge 
bosons with the loop momenta of order $m_\varphi$, giving the mass 
splitting $m_{\varphi_\uparrow} - m_{\varphi_\downarrow} \simeq 2\pi 
(Q_\uparrow^2-Q_\downarrow^2) (e^2/16\pi^2) m_Z$.  Here, $Q_\uparrow=-1/3$ 
and $Q_\downarrow=-4/3$ are the electric charges of $\varphi_\uparrow$ 
and $\varphi_\downarrow$, respectively.  This gives the mass splitting 
$m_{\varphi_\uparrow} - m_{\varphi_\downarrow} \simeq -600~{\rm MeV}$. 
While a mass splitting of this size is not negligible, we find it useful 
to classify the hadronic states consisting of xyons and the light quarks, 
$u$ and $d$, using an approximate $SU(2)_\varphi \times SU(2)_q$ symmetry, 
where $\varphi_\uparrow$ and $\varphi_\downarrow$ form a doublet under 
$SU(2)_\varphi$ while $u$ and $d$ a doublet under $SU(2)_q$.  We then 
find that in a given $SU(2)_\varphi \times SU(2)_q$ multiplet the 
states containing $\varphi_\uparrow$ are lighter than those containing 
$\varphi_\downarrow$ by about $600~{\rm MeV}$. 

The lightest xyonic hadrons are expected to be one of the four 
fermionic mesons
\begin{equation}
  \tilde{T}^0 \equiv \varphi_\uparrow \bar{d}, \qquad 
  \tilde{T}^- \equiv \varphi_\uparrow \bar{u}, \qquad 
  \tilde{T}'^- \equiv \varphi_\downarrow \bar{d}, \qquad 
  \tilde{T}'^{--} \equiv \varphi_\downarrow \bar{u}, 
\end{equation}
which form a $({\bf 2}, {\bf 2})$ multiplet under $SU(2)_\varphi \times 
SU(2)_q$. Here, the superscripts represent the electric charges.  As we 
have seen, the masses of these xyonic mesons ({\it xymesons}) split due 
to the mass difference between $\varphi_\uparrow$ and $\varphi_\downarrow$ 
--- xymesons containing $\varphi_\downarrow$ are heavier than those 
containing $\varphi_\uparrow$ by about $600~{\rm MeV}$.  This implies 
that $\tilde{T}'^-$ ($\tilde{T}'^{--}$) decays into $\tilde{T}^0$ 
($\tilde{T}^-$) and a charged pion with the lifetime of about 
$10^{-12}~{\rm sec}$.  The mass splitting between $\tilde{T}^0$ and 
$\tilde{T}^-$ (and $\tilde{T}'^-$ and $\tilde{T}'^{--}$) is of order 
a few MeV, which comes from isospin breaking effects due to electromagnetic 
interactions and the $u$-$d$ mass difference.  Because the two effects 
work in the opposite direction, it is not clear which of $\tilde{T}^0$ 
and $\tilde{T}^-$ is lighter.  While the heavier one decays into the 
lighter one and leptons through weak interactions, its lifetime is 
of order $10^{-1}$ to $10^{2}$ seconds, so that both $\tilde{T}^0$ 
and $\tilde{T}^-$ are essentially stable for collider purposes.

There are also bosonic baryons formed by xyons and the standard model 
quarks. The lightest states of these xyonic baryons ({\it xybaryons}) 
will come either from a $({\bf 2}, {\bf 1})$ scalar multiplet 
\begin{equation}
  \tilde{U}_S^0 \equiv \varphi_\uparrow [u d], \qquad 
  \tilde{U}_S^- \equiv \varphi_\downarrow [u d], 
\end{equation}
or from a $({\bf 2}, {\bf 3})$ vector multiplet
\begin{eqnarray}
&& \tilde{U}_V^+ \equiv \varphi_\uparrow u u, \qquad 
   \tilde{U}_V^0 \equiv \varphi_\uparrow \{u d\}, \qquad
   \tilde{U}_V^- \equiv \varphi_\uparrow d d, 
\nonumber\\
&& \tilde{U}_V'^0 \equiv \varphi_\downarrow u u, \qquad
   \tilde{U}_V'^- \equiv \varphi_\downarrow \{u d\}, \qquad 
   \tilde{U}_V'^{--} \equiv \varphi_\downarrow d d.
\end{eqnarray}
Here, $\{\}$ and $[]$ denote symmetrization and antisymmetrization, 
respectively, and scalar and vector multiplets have spin-$0$ 
and spin-$1$, respectively.  Because of the $\varphi_\uparrow$-%
$\varphi_\downarrow$ mass difference, $\tilde{U}_S^-$ ($\{ \tilde{U}_V'^0, 
\tilde{U}_V'^-, \tilde{U}_V'^{--} \}$) is heavier than $\tilde{U}_S^0$ 
($\{ \tilde{U}_V^+, \tilde{U}_V^0, \tilde{U}_V^- \}$) by about 
$600~{\rm MeV}$.  The mass splittings among $\tilde{U}_V^+$, 
$\tilde{U}_V^0$ and $\tilde{U}_V^-$ (and $\tilde{U}_V'^0$, $\tilde{U}_V'^-$ 
and $\tilde{U}_V'^{--}$) are of order a few MeV, arising from isospin 
breaking effects.  Finally, the mass splitting between particles in 
the $({\bf 2}, {\bf 1})$ multiplet and those in the $({\bf 2}, {\bf 3})$ 
multiplet arises from spin-dependent interactions, which is expected to 
be a few hundred MeV.  Thus, the heavier will decay into the lighter 
and pions by strong interactions.  This implies that only the lighter of 
$\tilde{U}_S^0$ and $\{ \tilde{U}_V^+, \tilde{U}_V^0, \tilde{U}_V^- \}$ 
behaves as a stable particle(s) at colliders.  The stability of the lighter 
is ensured by baryon number conservation (unless the mass of the lightest 
xybaryon is larger than the sum of the nucleon and the lightest xymeson 
masses, which we think is unlikely).  The naive nonrelativistic quark 
model suggests that $\tilde{U}_S^0$ is lighter.  The mass splitting 
between the lightest xybaryons and xymesons are expected to be of order 
a few hundred MeV. 

In fact, we can work out the spectrum of the lightest xyhadrons in more 
detail, if we apply the empirical mass formula for the lowest-lying 
hadronic states with no radial excitation or orbital angular 
momentum~\cite{Close:1979bt}.  We expect that the masses of the 
$\varphi_\uparrow$-xyhadron states $\{ \tilde{T}^0, \tilde{T}^- \}$, 
$\tilde{U}_S^0$ and $\{ \tilde{U}_V^+, \tilde{U}_V^0, \tilde{U}_V^- \}$ 
are roughly given by $m_{\varphi_\uparrow} + (300~{\rm MeV})$, 
$m_{\varphi_\uparrow} + (460~{\rm MeV})$ and $m_{\varphi_\uparrow} 
+ (650~{\rm MeV})$, respectively, with the mass splittings among the 
states inside a curly bracket of order a few MeV.  This spectrum ensures 
the stability of the lightest xybaryon and allows a vector xybaryon to 
decay into a scalar xybaryon and a pion through strong interactions. 
A similar pattern is repeated for the xyhadrons containing 
$\varphi_\downarrow$: the masses of $\{ \tilde{T}'^-, \tilde{T}'^{--} \}$, 
$\tilde{U}_S^-$ and $\{ \tilde{U}_V'^0, \tilde{U}_V'^-, \tilde{U}_V'^{--} 
\}$ are roughly given by $m_{\varphi_\downarrow} + (300~{\rm MeV})$, 
$m_{\varphi_\downarrow} + (460~{\rm MeV})$, and $m_{\varphi_\downarrow} 
+ (650~{\rm MeV})$, with the mass splittings among the states inside 
a curly bracket again of order a few MeV.  Here, $m_{\varphi_\downarrow} 
\simeq m_{\varphi_\uparrow} + (600~{\rm MeV})$. 

Now, let us consider signals of these xyhadronic states at hadron colliders. 
Since both of the two lightest xymesons, $\tilde{T}^0$ and $\tilde{T}^-$, 
as well as the lightest xybaryon (presumably $\tilde{U}_S^0$) are 
effectively stable, we have both neutral and charged stable heavy particles 
at colliders.  While all these particles undergo hadronic interactions 
in the detectors, neutral ones are difficult to see in practice (although 
they may be seen as intermittent highly ionizing tracks through isospin 
and charge exchange with background material).  On the other hand, 
charged particles leave highly ionizing tracks both in the inner tracking 
region and the outer muon system, so that they can be detected relatively 
easily.  In fact, these signals are quite common in warped unified 
theories~\cite{Goldberger:2002pc}, which often lead to stable colored 
particles at colliders.  The production of these particles occurs through 
strong interaction processes, and the reach in their masses is roughly 
$2~{\rm TeV}$ at the LHC. 

A more precise estimate for the reach in xyon masses can be obtained 
from a detailed study of the hadronic production of colored particles 
given in Ref.~\cite{Cheung:2002uz}.  Here we require that charged 
stable xyhadrons reach the outer muon system to be observed, which gives 
conservative estimates for the reach in xyon mass.  We also assume, for 
simplicity, that all the produced xyons hadronize into xymesons, although 
we expect that the estimates are not much affected if some (presumably 
small) fraction of xyons hadronizes into xybaryons.  In the case of 
$SU(5)$ xyons, this implies that roughly $1/2$ of the produced xyons 
hadronize into charged particles that can be detected in both the inner 
and outer systems.  Here, we have assumed that xyhadrons in a single 
$SU(2)_q$ multiplet are produced with an almost equal probability in 
hadronization processes.  Since the signals of highly ionizing tracks 
caused by massive charged particles are almost background free, the 
reach is essentially determined by the production rate.  If we require 
the observation of 4 (10) events for the ``discovery'' of xyons, we find 
that the reach in xyon masses is about $1.8$, $2.0$ and $2.2~{\rm TeV}$ 
($1.6$, $1.8$ and $2.0~{\rm TeV}$) at the LHC with integrated luminosity 
of $100$, $300$ and $1000~{\rm fb}^{-1}$, respectively.  Comparing 
with the theoretically expected range for the xyon mass given in 
Fig.~\ref{fig:xyon}, we find that xyons may be observed at the LHC 
if the parameter $r_{\lambda,i}$ takes values somewhat smaller than 
unity, e.g. $r_{\lambda,i} \simlt 0.5$. 

The discovery of xyons at the VLHC is much more promising.  Requiring 
10~events for the discovery, we find that the VLHC with a center-of-mass 
energy of $50~{\rm TeV}$ ($200~{\rm TeV}$) has a reach in xyon masses 
of about $3.7$, $4.3$ and $5.0~{\rm TeV}$ ($9.0$, $11$ and $13~{\rm TeV}$) 
for integrated luminosity of $100$, $300$ and $1000~{\rm fb}^{-1}$, 
respectively.  Given the fact that the xyon mass saturates for large 
$r_{\lambda,i}$ for a fixed value of $k'$ (see Fig.~\ref{fig:xyon}), 
we find that the VLHC with a center-of-mass energy of $50~{\rm TeV}$ 
($200~{\rm TeV}$) can cover the entire parameter region of $r_{\lambda,i}$ 
for $k' \simlt 13~{\rm TeV}$ ($33~{\rm TeV}$).  Since we naturally 
expect $k' \approx 10~{\rm TeV}$ (to obtain weak-scale superparticle 
masses with all the parameters obeying naive dimensional analysis; 
see subsection~\ref{subsec:xyon-mass-2}), the discovery of xyons 
at the VLHC is quite promising.%
\footnote{If the 321 gauge couplings are weak at the unification scale, 
as can be the case in the holographic theories with $SU(5)$ broken by 
the Higgs mechanism on the UV brane, $k'$ can be larger but only by 
a factor of two or so.}

There may also be other signals that can be used to detect xyons.  When 
antixymesons or xybaryons traverse a detector, they may exchange isospin 
and charge with the background material through hadronic interactions, 
and so may make transitions between neutral and charged states.  This 
leaves a distinct signature of intermittent highly ionizing tracks. 
Neutral xyhadrons also give a signature of jets plus transverse missing 
energy, which may be observed at the VLHC.  If xyons pick up strange 
quarks after their production, they form strange xyhadrons.  These 
xyhadrons are expected to be heavier than non-strange xyhadrons by about 
$100~{\rm MeV}$, so that the lightest strange xymeson (or xybaryon) 
decays into a non-strange one through weak interactions.  Assuming 
that the mass difference is smaller than the pion mass, the lifetime 
of the decay will be of order $10^{-8}\!\sim\!10^{-7}~{\rm sec}$ with 
the final state containing leptons (otherwise it will be of order 
$10^{-12}\!\sim\!10^{-10}~{\rm sec}$ with a charged pion in the final 
state).  Since the decay could change the charges of the xyhadronic 
states, these xyhadrons may leave tracks in the inner tracking region 
but not in the outer muon system, or vice versa.

Testing the $SU(2)_L$-doublet nature of xyons will not be easy. 
One possibility might be to use the decay of $\varphi_\downarrow$ 
into $\varphi_\uparrow$.  As we have seen, a xyhadron containing 
$\varphi_\downarrow$ is heaver than the one with $\varphi_\downarrow$ 
replaced by $\varphi_\uparrow$, by about $600~{\rm MeV}$.  This implies 
that some xyhadrons, for example $\tilde{T}'^-$, $\tilde{T}'^{--}$ and 
$\tilde{U}_S^-$, decay through weak interactions with the lifetime of 
about $10^{-12}~{\rm sec}$, corresponding to the decay length of about 
sub-millimeters.  The final state contains either a charged or neutral 
xyhadron and a charged pion.  Therefore, if one could somehow see the 
final-state xyhadrons and the soft charged pions, arising from decays 
of pair-produced xyons, and determine the decay points precisely, one 
might be able to see the $SU(2)_L$-doublet nature of xyons.

We finally consider the case in which the lightest xyon, $\varphi_\uparrow$, 
is unstable at a collider timescale, as in the case of the holographic 
theories with matter fields propagating in the bulk.  In this case, xyhadrons 
may decay into the MSSM particles inside the detector.  In particular, 
the lightest xymeson and xybaryon can also decay inside the detector. 
The lifetime is highly sensitive to the parameters of the theory.  While 
we naturally expect that it is still longer than about $10^{-4}~{\rm sec}$ 
to avoid reintroduction of the supersymmetric flavor problem, it can in 
principle be any number larger than about $10^{-16}~{\rm sec}$.  The decay 
products will consist of an energetic (s)quark and an energetic (s)lepton. 
This, therefore, potentially provides a distinct signature at hadron 
colliders.

\subsection{Super-oblique corrections}
\label{subsec:super-oblique}

Since the xyon supermultiplet has large mass splittings among its 
components, it can potentially leave non-negligible effects on some 
parameters at energies lower than the xyon mass.  These effects 
could become important if xyons are so heavy that they cannot be 
directly produced at the LHC.  Here we briefly discuss these effects 
and estimate their sizes.  We find, unfortunately, that they are not 
large and thus cannot be used to probe xyons indirectly.

Consider a generic supermultiplet charged under some gauge group. 
If this multiplet has large mass splittings among its components, 
then it induces a difference between the couplings of the gauge boson 
and the gaugino at loop level.  This type of correction is called 
a super-oblique correction, which does not decouple as the multiplet gets 
heavy as long as the fractional mass differences among the components 
stay the same~\cite{Cheng:1997sq}.  At the leading-log level, the 
corrections are estimated using the renormalization group equations for 
the gauge and gaugino couplings, which are equal in the supersymmetric 
limit but not if supersymmetry is broken.  In the case of $SU(5)$ xyons, 
we find that the differences between the gauge coupling $g_i$ and the 
gaugino coupling $\tilde{g}_i$ at a scale below the xyon mass are 
given by
\begin{equation}
  \tilde{U}_i \equiv \frac{\tilde{g}_i}{g_i}-1 
    \simeq \frac{g_i^2 \hat{b}_i}{16\pi^2} 
    \ln\frac{M_{\tilde{\varphi}}}{m_\varphi},
\end{equation}
where $(\hat{b}_1, \hat{b}_2, \hat{b}_3) = (5/6, 1/2, 1/3)$, $i=1,2,3$ 
represents $U(1)_Y$, $SU(2)_L$ and $SU(3)_C$, respectively, and 
$M_{\tilde{\varphi}}$ is the mass scale for the superpartners 
of xyons, $\tilde{\varphi} = \lambda'^{\rm XY}, \chi^{\rm XY}$. 
Since $\ln(M_{\tilde{\varphi}}/m_\varphi) \approx 2$, we find that 
the corrections are not large --- for example, $(\tilde{U}_1, 
\tilde{U}_2, \tilde{U}_3) = (0.23\%, 0.27\%, 0.48\%)$ for 
$\ln(M_{\tilde{\varphi}}/m_\varphi) = 2$.  We obtain somewhat larger 
corrections for xyons associated with the IR breaking of $SO(10) 
\rightarrow SU(4)_C \times SU(2)_L \times SU(2)_R$: $(\hat{b}_1, 
\hat{b}_2, \hat{b}_3) = (13/15, 1, 2/3)$ and $(\tilde{U}_1, 
\tilde{U}_2, \tilde{U}_3) = (0.24\%, 0.54\%, 0.97\%)$ for 
$\ln(M_{\tilde{\varphi}}/m_\varphi) = 2$, but these are still small. 
In fact, corrections of these sizes are expected to arise generically 
as threshold corrections from the DSB sector.  For example, 
IR-brane operators of the form ${\cal L} \approx 2\delta(y-\pi R) 
\int\!d^4\theta\, Z^\dagger Z ({\cal W}_i^\alpha {\cal D}^2 
{\cal W}_{i \alpha} + {\rm h.c.})$ naturally give contributions 
to $\tilde{U}_i$ comparable to those from xyons.  Therefore, we 
find it is difficult to use super-oblique corrections to probe xyons 
indirectly when they are not produced at colliders.  

The smallness of the xyon contributions to $\tilde{U}_i$ is generic, 
because it mainly comes from the fact that xyons are scalars and 
thus do not contribute significantly to the group-theoretical 
factors $\hat{b}_i$.  It is also generically true that the mass 
splittings in the xyon supermultiplet are not very large, i.e. 
$\ln(M_{\tilde{\varphi}}/m_\varphi)$ is not very large, and that 
the multiplicity of xyon supermultiplets is not very large either.

\subsection{Cosmology}

Cosmological implications of xyons depend very much on their lifetime. 
We first consider the case in which the lifetime of xyons is much 
longer than the age of the universe.  Assuming that the thermal history 
of the universe is standard below the temperature of about $10~{\rm TeV}$, 
the relic abundance of xyonic particles (xyhadrons) today are estimated 
as follows.  Below the temperature of about $T \simeq m_\varphi/28 
\approx 100~{\rm GeV}$, annihilation of xyons into gluons becomes 
ineffective and the xyon abundance freezes out.  If there is no subsequent 
annihilation of xyons, this will lead to the present xyon energy density 
of about $\Omega_\varphi \simeq (0.01\!\sim\!1)$ for $m_\varphi \simeq 
(1\!\sim\!10)~{\rm TeV}$.  However, there may be periods of further 
annihilations at later stages in the evolution of the universe, which 
could significantly reduce $\Omega_\varphi$.  The most important among 
these would come from nonperturbative QCD effects at a temperature 
of about $T \simeq \Lambda_{\rm QCD} \approx 300~{\rm MeV}$.

At $T \simeq \Lambda_{\rm QCD}$, xyons will hadronize into xyhadrons. 
The annihilation cross sections of these xyhadrons are difficult to 
estimate because they are determined by nonperturbative QCD effects. The 
largest possible cross section, which leads to the smallest relic abundance, 
will result if it is the same order as the nucleon-nucleon cross sections 
at low energies, $\sigma_{\rm ann.} \sim (m_\pi^2 \beta)^{-1}$, where 
$m_\pi$ is the pion mass and $\beta$ the relative velocity between the 
two xyhadrons.  Since nucleon-nucleon scatterings at low energies are 
mostly quark rearrangement processes, it is not clear if the xyon 
annihilation cross section, which requires annihilation of xyon cores, 
takes a value similar to that of the low-energy nucleon-nucleon cross 
sections.  However, it may be possible that, when two xyhadrons approach, 
they form a bound state (xynuclei) with a cross section of order 
$(m_\pi^2 \beta)^{-1}$, and then this bound state reorganizes itself 
into $\varphi \bar{\varphi}$ and the usual hadrons for energetic 
reasons, leading to rapid xyon annihilation.  If this happens, the xyon 
annihilation cross section is, in fact, given by $\sigma_{\rm ann.} 
\sim (m_\pi^2 \beta)^{-1}$, leading to a very small abundance of 
$\Omega_\varphi \simeq 10^{-10}$~\cite{Baer:1998pg}.  Since we do not 
really know the low-energy dynamics of xyhadrons, the estimate of the 
xyon relic abundance is subject to a rather large uncertainty:
\begin{equation}
  10^{-10} \simlt \Omega_{\varphi} \simlt 1.
\label{eq:relic}
\end{equation}
Note that, if nonperturbative annihilation at the QCD era is effective, 
we generically expect the value of $\Omega_\varphi$ in the lower part 
of this range.  For example, the relic abundance can be as small as 
$\Omega_\varphi \simeq 10^{-8}$ even for $\sigma_{\rm ann.} \sim 
m_\pi^{-2}$ without the $1/\beta$ enhancement. 

While there is a recent argument that the nonperturbative enhancement of 
annihilation at the QCD era is unlikely to occur~\cite{Arvanitaki:2005fa}, 
we first discuss observational constraints on xyons in the case 
that the relic abundance of xyons is given by the lower range of 
Eq.~(\ref{eq:relic}).  For these small values for the abundance and 
the xyon mass range of our interest, most of the constraints from 
direct search experiments are satisfied~\cite{Perl:2001xi}.  Strong 
constraints, however, come from heavy isotope searches.  For $1~{\rm TeV} 
\simlt m_\varphi \simlt 10~{\rm TeV}$, the relevant bounds are those 
in Ref.~\cite{Hemmick:1989ns}.%
\footnote{For $m_\varphi \simlt 1~{\rm TeV}$, the strongest bound 
would come from that in Ref.~\cite{Smith:1982qu}.}
Some of the dark matter search experiments also give strong constraints. 
Implications of these bounds, however, differ depending on which of 
$\tilde{T}^0$ and $\tilde{T}^-$ is the lightest xymeson.

In the case that $\tilde{T}^-$ is the lightest xymeson, most of the 
relic xyons will be in the form of $\tilde{T}^-$ with some of them 
in the form of the lightest xybaryon, $\tilde{U}_S^0$.  The charged 
xymeson, $\tilde{T}^-$, is then bound with a proton at the time of 
nucleosynthesis, while its antiparticle is bound with an electron 
at the recombination era~\cite{DeRujula:1989fe}.  At the time of galaxy 
formation, the antixymeson-electron bound state dissipates energy through 
radiation and collapses into the galactic disk.  This implies that its 
local abundance around the earth will be enhanced compared with the cosmic 
abundance of Eq.~(\ref{eq:relic}) by about a factor of $10^7$ or more. 
On the other hand, the local abundance of the xymeson-proton bound state 
is about $10^5$ times the cosmic abundance, as it is determined purely 
by gravitational clusterings.  With these local abundances, the bounds 
in~\cite{Hemmick:1989ns}, as well as data from plastic track detectors, 
exclude the existence of the stable charged xymeson.  This case, therefore, 
will be viable only if there is a significant entropy production below 
the temperature of order TeV in the history of the universe.

What if the lightest xymeson is $\tilde{T}^0$?  In this case, the 
relic xyons are mostly in the form of $\tilde{T}^0$ with a small 
fraction in the form of $\tilde{U}_S^0$.  The local xyon density is 
about $10^5$ times the cosmic abundance of Eq.~(\ref{eq:relic}).  With 
$\Omega_\varphi$ in the lower range of Eq.~(\ref{eq:relic}), most of 
the constraints from direct searches at balloon, satellite, ground- 
and underground-based experiments are satisfied (a roughly upper-half 
portion of the range of Eq.~(\ref{eq:relic}) could be excluded by the 
bounds from xyhadron annihilation in the halo~\cite{Mohapatra:1997sc}.) 
The significant constraints, however, may come from heavy isotope searches 
and dark matter search experiments located at shallow sites. 

Let us first consider the constraints from heavy isotope search 
experiments.  Suppose now that the relic xyhadrons are bound in nuclei, 
forming anomalous heavy isotopes.  Then the number density of these 
anomalous heavy xynuclei compared to that of the ordinary nuclei in the 
earth can be estimated from the expected xyhadron flux on the earth's 
surface as 
\begin{equation}
  r \sim 10^{-18}\, \Omega_\varphi \left( \frac{\rm TeV}{m_\varphi} \right)
    \left( \frac{t_{\rm acc}}{\rm yr} \right),
\end{equation}
where $t_{\rm acc}$ is the time period over which xynuclei accumulate in 
a sample of matter without being removed~\cite{Starkman:1990nj}.  For the 
lowest possible value of $\Omega_\varphi \simeq 10^{-10}$ and a reasonable 
time period $t_{\rm acc} \simeq (10^{8}\!\sim\!10^{10})~{\rm yr}$, 
we obtain $r \sim (10^{-20}\!\sim\!10^{-18})({\rm TeV}/m_\varphi)$. These 
values are marginally consistent with the bounds in~\cite{Hemmick:1989ns}, 
given the possibility of $r$ being further reduced by a factor of 100 
for hydrogen due to geochemical processes.  There are also constraints 
from the amount of primordial heavy isotopes containing xyhadrons, 
generated during nucleosynthesis, which give $\Omega_\varphi \simlt 
10^{-7}$~\cite{Mohapatra:1998nd}.%
\footnote{The constraints on the lightest xymeson, which has isospin 
$1/2$, may be different from those on a heavy colored particle with 
isospin $0$, especially because the xymeson could bind with the proton 
due to one-pion exchange.  We estimate, however, that the difference 
is not very large.}
Overall, the thermal xyon relics seem to be consistent with heavy 
isotope searches, but only if $\Omega_\varphi$ is in the lower edge 
of the region in Eq.~(\ref{eq:relic}).  Note, however, that, in contrast 
with $\tilde{T}^-$, neutral xyhadrons $\tilde{T}^0$ and $\tilde{U}_S^0$ 
may not bind with ordinary matter to form heavy isotopes.  While 
these particles feel the ordinary nuclear force, it is possible that 
they are not bound into nuclei, especially into lighter ones, for 
energetic reasons.  If this is the case, the bounds from heavy isotope 
searches disappear.

The constraints on the xyon relic density also come from dark matter 
search experiments.  Since xyhadrons are strongly interacting, their 
energies are significantly degraded during propagations in matter due 
to collisions, so they cannot reach deep underground.  In order for 
an experiment to be able to constrain the xyon abundance, therefore, it 
must be located at a relatively shallow site, a depth smaller than about 
100 meters of water equivalent.  This almost singles out the relevant 
experiment to be the CDMS experiment at Stanford~\cite{Abrams:2002nb}. 
Assuming that the xyhadron-nuclei cross sections are given by the 
geometric cross sections of $\sigma \sim \pi A^{2/3} m_\pi^{-2}$, 
where $A$ is the nucleon number of the target nucleus, we estimate that 
xyhadrons typically undergo $(30\!\sim\!40)$ collisions before reaching 
the detector.  A significant energy degradation, on the other hand, 
occurs only after about 100 or more collisions, implying that the 
CDMS detector can constrain the xyon relic abundance.  By estimating the 
number of events expected from cosmic neutral xyhadrons and comparing 
it with the nuclear-recoil data, we find that the lower edge of the 
region in Eq.~(\ref{eq:relic}) may be marginally consistent with the 
data, but not for larger values.  Given the crudeness of our estimate, 
we can consider that the thermal xyon relics are consistent with 
the data if $\Omega_\varphi \sim 10^{-10}$, although a more careful 
analysis will be need to be really conclusive.%
\footnote{If the cross sections between xyhadrons and nuclei are much 
larger than the geometric cross sections, e.g. by an order of magnitude 
or more, the constraint from CDMS will disappear as cosmic xyhadrons 
will not be able to reach the detector without significant energy 
degradations.  The constraint also becomes weaker, if the cross sections 
are much smaller than the geometric ones, because the expected event 
rates then decreases.}

We have seen that the existence of stable xyons may be consistent 
with all the observations if the xyon relic abundance is as small as 
$\Omega_\varphi \sim 10^{-10}$ due to a large nonperturbative enhancement 
of the annihilation cross section at the QCD era.  This enhancement, 
however, may not occur.  In this case, there are essentially two ways 
to make cosmology consistent.  One is to consider ``non-standard'' 
cosmology, such as a significant late-time entropy production below the 
temperature of about a TeV.  The other is to consider the case where 
xyons are unstable.  We here discuss the latter case.  If the xyon 
lifetime is shorter than $\simeq 10^{-1}~{\rm sec}$, xyons decay before 
nucleosynthesis and there is no cosmological constraint.  In the case 
that the lifetime is longer than $\simeq 10^{-1}~{\rm sec}$ (and still 
shorter than the age of the universe), there are potential constraints 
coming, for example, from the destruction of the light nuclei synthesized 
during nucleosynthesis and the distortions of the diffuse gamma-ray 
background and the cosmic microwave background.  In the case that 
the xyon relic abundance is determined by perturbative annihilation, 
$\Omega_\varphi \simeq (0.01\!\sim\!1)$ for $m_\varphi \simeq 
(1\!\sim\!10)~{\rm TeV}$, these constraints require the lifetime of 
xyons to be shorter than about $100~{\rm sec}$~\cite{Arvanitaki:2005fa}. 
Xyons with such lifetimes can naturally arise if matter fields propagate 
in the bulk in the holographic 5D theories ($c \simlt 0.9$ if the 
relevant matter fields have the same bulk mass parameter).  If the 
xyon lifetime is longer than $100~{\rm sec}$, the xyon relic density 
is constrained to be in the lower range of Eq.~(\ref{eq:relic}), although 
not necessarily at the lowest edge of $\Omega_{\varphi} \simeq 10^{-10}$.

In the case that the xyon lifetime is shorter than $\simeq 100~{\rm sec}$ 
or $\Omega_{\varphi} \simeq 10^{-10}$ with stable neutral xymesons, 
the thermal history of the universe can be standard.  This opens up 
the possibility of having dark matter whose abundance is determined by 
conventional analyses.  For example, dark matter may be the QCD axion 
with a decay constant of order $f_{\rm PQ} \simeq 10^{12}~{\rm GeV}$. 
A more attractive possibility will be a dark matter candidate whose 
interaction strengths are set by the TeV scale, because then the observed 
amount of dark matter is naturally reproduced.  In fact, such a particle 
can arise naturally in our scenario from fields localized on the IR 
brane~\cite{Nomura:2004zs}.%
\footnote{An alternative possibility is the pedestrian dark matter 
discussed in~\cite{Chacko:2005ra}, which can arise from the extended 
Higgs sector of the model.}
Suppose we introduce a pair of singlet chiral superfields $S$ and 
$\bar{S}$ on the IR brane and introduce a vector-like symmetry acting 
(only) on them: either a $U(1)$ symmetry, $S(+1)$ and $\bar{S}(-1)$, 
or a $Z_2$ symmetry under which both $S$ and $\bar{S}$ are odd. 
The mass and interactions for these fields are then given by 
\begin{equation}
  S_{\rm DM} = \int\!d^4x \int_0^{\pi R}\!\!dy\,
    2 \delta(y-\pi R) \biggl[ e^{-2\pi kR} \int\!d^2\theta\, M_S S \bar{S} 
    + \sum_{i=1,2,3} \int\!d^2\theta\, \frac{1}{M_i^2} S \bar{S} \, 
      {\rm Tr}[ {\cal W}_i^\alpha {\cal W}_{i \alpha} ] 
    + {\rm h.c.} \biggr],
\label{eq:DM}
\end{equation}
where we have assumed, for simplicity, that $S$ and $\bar{S}$ carry $U(1)$ 
charges.  Using naive dimensional analysis, we find $M_S \approx M_*$ 
and $M_i^2 \approx (M_*/\pi k)^3 k^2$.  The mass splitting between the 
fermionic and bosonic components of $S$ and $\bar{S}$ comes from couplings 
of $S$ and $\bar{S}$ to the supersymmetry breaking field $Z$.  Assuming 
that the fermionic component is the lightest, dark matter is a single 
Dirac fermion $\Psi = \{ \psi_S, \psi_{\bar{S}} \}$, which annihilates 
into the MSSM gauginos through the four-Fermi operators suppressed by 
$M_i'^2 \equiv M_i^2 e^{-2\pi kR} \approx (M_*/\pi k)^3 k'^2$.  The mass 
of this particle is close to the cutoff scale, $M_\Psi = M_S e^{-\pi kR} 
\approx M_*'$.  The annihilation of $\Psi$ occurs through the $s$-wave with 
the thermally averaged cross section given by $\langle \sigma v \rangle 
\approx (n/8\pi)(M_\Psi^2/M_i'^4)$, where $n$ is the multiplicity of 
the final-state gauginos.  Using the values $M_*/\pi k \approx 3$, 
$k' \approx 10~{\rm TeV}$ and $n \approx 10$, we obtain the relic $\Psi$ 
abundance of $\Omega_\Psi \approx (0.1\!\sim\!1)$.  In fact, the correct 
abundance is reproduced in a wide range of parameters with the dark matter 
mass of $M_\Psi \approx (10\!\sim\!100)~{\rm TeV}$.  From the 4D point 
of view, our dark matter $\Psi$ is a stable composite state of the DSB 
sector~\cite{Dimopoulos:1996gy}, but with the important new ingredient 
that it ``directly'' couples to the MSSM states through mixings between 
the elementary states and the composite DSB states.  Annihilation could 
also occur into the Higgs fields, if the Higgs doublets have nearly 
conformally-flat wavefunctions.  The IR-brane dark matter may also be 
a Majorana fermion (no $\bar{S}$ field).  It is interesting that we 
are led to a picture of dark matter with a characteristic scale of 
$\approx (10\!\sim\!100)~{\rm TeV}$, rather than the conventional one 
of $\approx (0.1\!\sim\!1)~{\rm TeV}$.%
\footnote{The annihilation of such dark matter in the galactic center 
may provide the origin of the recently observed high-energy $\gamma$-ray 
flux from that region~\cite{Aharonian:2004wa}.  The high-energy 
$\gamma$-ray flux as a signal of dark matter annihilation has 
recently been considered in the context of the lightest messenger 
dark matter~\cite{Hooper:2004fh}.}
Unfortunately, the direct detection of our IR-brane dark matter will 
be difficult because the cross sections with nuclei are suppressed by 
large mass parameters $M_i' = O(10\!\sim\!100~{\rm TeV})$.

It is interesting to note that the class of theories discussed here 
is free from dangerous relics such as the gravitino and moduli. 
Since the supersymmetry-breaking scale is very low, $\Lambda \approx 
(10\!\sim\!100)~{\rm TeV}$, we expect that the gravitino (and moduli, 
if any) is very light $m_{3/2} \simeq \Lambda^2/M_{\rm Pl} \approx 
(0.1\!\sim\!10)~{\rm eV}$.  In the particular context of the 5D 
theory discussed in subsection~\ref{subsec:holographic}, we find 
\begin{equation}
  m_{3/2} \simeq \frac{\pi}{4\sqrt{3}} 
    \biggl( \frac{M_*}{\pi k} \biggr)^2 \frac{k'^2}{M_{\rm Pl}}
  \simeq 0.2\, \biggl( \frac{M_*/\pi k}{3} \biggr)^2 
    \biggl( \frac{k'}{10~{\rm TeV}} \biggr)^2~{\rm eV},
\end{equation}
where we have used naive dimensional analysis to obtain $F_Z \simeq 
M_*^2/4\pi$.  Such a light gravitino does not produce the ``gravitino 
problem'', as its thermal relic abundance is small.  It also evades the 
bound, $m_{3/2} \simlt 16~{\rm eV}$, recently derived from the analysis 
of the matter power spectrum and the cosmic microwave background 
data~\cite{Viel:2005qj}.  We note that in warped theories the radion 
does not cause any cosmological problem either, because its mass and 
interaction strengths are both dictated by the TeV scale, so that it 
decays before nucleosynthesis.  These features allow the theory to 
have a high reheating temperature after inflation without contradicting 
with the observations.  In particular, this implies that the theory 
may accommodate baryogenesis at high temperatures without any conflict. 
For example, assuming that there is no significant entropy production 
associated with the phase transition in the DSB sector (the viability 
of this assumption depends on the dynamics of the phase transition, 
especially the mechanism of radius stabilization~\cite{Creminelli:2001th}), 
the theory accommodates conventional thermal leptogenesis at the 
temperature $T \simeq 10^{10}~{\rm GeV}$~\cite{Fukugita:1986hr}.%
\footnote{The mechanism of soft leptogenesis has been discussed very 
recently in~\cite{Grossman:2005yi}, which may work regardless of the 
entropy production if the right-handed neutrinos have Majorana masses 
of order $(1\!\sim\!10)~{\rm TeV}$ and the temperature after the 
phase transition is as high as the right-handed neutrino masses.}

\section{Discussion and Conclusions}
\label{sec:concl}

In this paper we have argued that the requirement for the absence 
of fine-tuning in supersymmetric theories naturally leads to a class 
of theories that predicts exotic scalar particles with mass in the 
multi-TeV region.  The key ingredients to evade fine-tuning in 
supersymmetric theories are the following:
\begin{itemize}
\item[(1)]
There is an additional contribution(s) to the mass of the physical 
Higgs boson (to the Higgs quartic couplings) other than that from 
the $SU(2)_L \times U(1)_Y$ $D$-terms of the MSSM.
\item[(2)]
The mediation scale of supersymmetry breaking is low: $M_{\rm mess} 
= O(10\!\sim\!100~{\rm TeV})$.
\item[(3)]
The masses of the squarks and sleptons do not respect the unified 
mass relations arising from the simple ``$SU(5)$-symmetric'' 
supersymmetry breaking sector.
\end{itemize}

To evade the fine-tuning, (i) is absolutely necessary.  However, if 
we want to preserve one of the major successes of supersymmetry --- the 
constraints from the precision electroweak data are satisfied relatively 
straightforwardly --- then this is not enough.  We need to satisfy 
(ii) and/or (iii), most likely both.  In fact, for $M_{\rm Higgs} 
\simlt 250~{\rm GeV}$ as suggested by the data, the direct effect of 
raising the Higgs-boson mass to reduce fine-tuning, manifested as the 
increase of the left-hand-side of Eq.~(\ref{eq:ewsb-cond}), is not so 
significant.  Rather, the virtue of having the additional source of 
the Higgs quartic coupling lies in the fact that we do not need large 
top squark masses to have large radiative corrections to the physical 
Higgs-boson mass to evade the experimental bound of $M_{\rm Higgs} 
\simgt 114~{\rm GeV}$.  Since the amount of fine-tuning is determined 
almost completely by the masses of the top squarks, this could help 
a lot unless there is some other requirement that bounds the top squark 
masses from below.  This is the place where the condition (iii) comes in 
--- given the direct search bound of $m_{\tilde{e}} \simgt 100~{\rm GeV}$, 
naturalness requires that either the top quark and Higgs boson are both 
rather heavy, $m_t \simeq (180\!\sim\!182)~{\rm GeV}$ and $M_{\rm Higgs} 
\simeq (200\!\sim\!250)~{\rm GeV}$, or that we have to break the unified 
mass relation to make the top squarks sufficiently light to evade 
fine-tuning.  In this paper we have focused on the latter.

The low mediation scale, together with the absence of supersymmetric 
flavor changing neutral currents, suggests that supersymmetry breaking 
is mediated from the dynamical supersymmetry breaking (DSB) sector to 
the visible sector through standard model gauge interactions.  The 
simplest theories of this kind arise if the DSB sector is charged under 
the standard model gauge group so that the superparticle masses are 
automatically generated once supersymmetry is broken in this sector. 
Since the DSB sector is charged under the standard model gauge group, 
it contributes to the evolution of the $SU(3)_C \times SU(2)_L \times 
U(1)_Y$ (321) gauge couplings.  The successful supersymmetric prediction 
for gauge coupling unification is then preserved if this sector possesses 
an approximate global $SU(5)$ symmetry above $M_{\rm mess}$, which 
contains the 321 gauge group as a subgroup.   In fact, this becomes 
almost the necessary requirement if we want to formulate this class 
of theories using the ``dual'' description in higher dimensional 
warped spacetime, because the DSB sector is then strongly coupled 
over a wide energy interval above $M_{\rm mess}$ and the contribution 
to the evolution of the 321 gauge couplings is controlled only by 
imposing the global $SU(5)$ symmetry to this sector. 

The presence of the global $SU(5)$ symmetry above $M_{\rm mess}$ and 
the absence of the unified mass relations for the squarks and sleptons 
implies that $SU(5)$ is broken dynamically to the 321 subgroup at the 
scale $M_{\rm mess}$.  This leads to light scalar states in the DSB 
sector, {\it xyons}, which are the pseudo-Goldstone bosons of this 
symmetry breaking.  In the minimal case of the $SU(5)$ DSB sector, the 
321 quantum numbers of xyons are given by $({\bf 3}, {\bf 2})_{-5/6}$. 
We have estimated the mass of xyons in generic cases and calculated it 
in a class of calculable theories formulated in higher dimensions.  We 
have found that the mass squared of xyons is positive in most of the 
parameter region, implying that the dynamical breaking of $SU(5)$ tends to 
align with the explicit breaking given by the gauging of the 321 subgroup. 
We have also found that the xyon mass is naturally in the multi-TeV region, 
$\approx (1\!\sim\!5)~{\rm TeV}$, assuming that the DSB sector obeys 
naive dimensional analysis.  In fact, it is natural to expect that 
the sizes of the operators in the DSB sector are determined by naive 
dimensional analysis; otherwise, incalculable threshold corrections 
to the 321 gauge couplings at $M_{\rm mess}$ would likely destroy 
the successful prediction associated with gauge coupling unification.

\begin{figure}[t]
\begin{center} 
\begin{picture}(120,45)(0,170)
  \Photon(5,190)(60,190){3.5}{6}   \Line(5,190)(60,190)
  \Text(0,198)[b]{$\lambda$} \Text(120,202)[b]{$\lambda'$} 
  \GBox(60,184)(115,196){0.85}
  \Text(90,191)[]{\tiny SUSY}  \Line(80,188)(100,193)
  \GCirc(60,190){6}{0}
\end{picture}
\caption{The diagram giving Dirac masses for the gauginos, where $\lambda'$ 
 represents the Dirac partners of the MSSM gaugino states, $\lambda$.}
\label{fig:Dirac}
\end{center}
\end{figure}
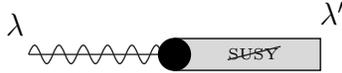

We emphasize that our argument is quite general and relies only on 
the assumption that the squark and slepton masses are generated through 
standard model gauge interactions.  In particular, it is independent 
of the physics providing the extra Higgs quartic couplings, the sector 
that has a large model dependence.  Our framework is also independent 
of the way the gauginos obtain masses.  For instance, the gauginos can 
obtain Dirac masses through the diagrams of Fig.~\ref{fig:Dirac} if the 
DSB sector provides the Dirac partners of the MSSM gaugino states. 
The masses of the gauginos are then given by 
\begin{equation}
  M_i \simeq g_i \frac{\hat{b}^{1/2}}{4\pi}\, 
    (\hat{\zeta}_i M_\rho),
\label{eq:Dirac-gaugino}
\end{equation}
in which case the ratios of the gaugino masses to the squark and 
slepton masses are significantly larger than in the Majorana case of 
Eq.~(\ref{eq:gaugino-masses}).  It is also possible that some of the 321 
gauginos are Majorana fermions while the others are Dirac or pseudo-Dirac 
fermions.  The point is that, even in these cases, the diagrams giving 
the squark and slepton masses are still those of Fig.~\ref{fig:gauge-med}b, 
so that the squark and slepton masses are given by the expressions in 
Eq.~(\ref{eq:scalar-masses}).  Therefore, to avoid the unwanted unified 
mass relations of Eq.~(\ref{eq:te-ratio}) we still need to break $SU(5)$ 
at the scale $M_{\rm mess}$ (we need to break $\hat{\zeta}_1 = \hat{\zeta}_2 
= \hat{\zeta}_3$).  The values for the xyon mass are not much affected 
either, because in the 4D picture the xyon mass arises from the diagrams 
with one loop of the 321 gauge multiplets, whose sizes do not depend much 
on the structure of the gaugino masses.

In fact, the class of theories discussed above can be constructed 
in a simple way as follows.  In our warped spacetime with the bulk 
$SU(5)$ gauge group reduced to 321 both on the UV and the IR branes, 
we can introduce an extra $U(1)$ gauge field together with three chiral 
superfields $A_3({\bf 8}, {\bf 1})_0$, $A_2({\bf 1}, {\bf 3})_0$ and 
$A_1({\bf 1}, {\bf 1})_0$ on the IR brane.  We then break the extra 
$U(1)$ via the Higgs mechanism.  Specifically, we introduce a pair 
of chiral superfields $\Phi$ and $\bar{\Phi}$ on the IR brane, which 
have $U(1)$ charges of $+1$ and $-1$, respectively, and have different 
supersymmetry-breaking masses, $m_\Phi^2 \neq m_{\bar{\Phi}}^2$, through 
different couplings to the supersymmetry-breaking field, ${\cal L} \sim 
2\delta(y-\pi R) \int\!d^4\theta\, (\kappa Z^\dagger Z \Phi^\dagger \Phi + 
\bar{\kappa} Z^\dagger Z \bar{\Phi}^\dagger \bar{\Phi})$ with $\kappa \neq 
\bar{\kappa}$.  Assuming that $m_\Phi^2$ and $m_{\bar{\Phi}}^2$ satisfy 
$m_\Phi^2 m_{\bar{\Phi}}^2 < 0$ and $m_\Phi^2 + m_{\bar{\Phi}}^2 > 0$, 
we have a stable vacuum in which a non-vanishing $D$-term for the $U(1)$ 
is generated.  This $D$-term then gives Dirac masses for the gauginos 
through the operators ${\cal L} \sim 2\delta(y-\pi R) \{ \int\!d^2\theta\, 
\sum_{i=1,2,3} {\cal W}'^\alpha {\cal W}_{i \alpha} A_i + {\rm h.c.} \}$, 
where ${\cal W}'_\alpha$ is the field-strength superfield for the 
extra $U(1)$~\cite{Fox:2002bu}.%
\footnote{We assume that the tree-level kinetic mixing between the 
extra $U(1)$ and hypercharge is suppressed.  Potentially dangerous 
effects of introducing the singlet field $A_1$ can also be suppressed, 
for example, if $Z$ is charged under an $R$ symmetry and the theory 
possesses an approximate symmetry $A_1 \rightarrow -A_1$ (i.e. the 
coefficients of the operators containing the odd number of $A_1$ are 
somewhat suppressed, say by an order of magnitude).  This leads to 
a scenario in which the Dirac bino is significantly lighter than the 
other gauginos and the right-handed selectron masses are generated 
mainly by the $D$-term VEV of $U(1)_Y$.}
This model is similar to that discussed in~\cite{Chacko:2004mi}. 
The main difference, however, is that here we work in the context 
of~\cite{Nomura:2004is} so that $SU(5)$ is broken on the IR brane and 
the coefficients of the above gaugino mass operators do not respect 
$SU(5)$.  This is required to break the unwanted unified mass 
relations for the squarks and sleptons, and thus to eliminate the 
fine-tuning. Another interesting feature of the present model is that 
the gauginos can be purely Dirac fermions, which contrasts with the 
case in~\cite{Chacko:2004mi} where the gauginos can only be pseudo-Dirac 
fermions due to an unbroken $SU(5)$ on the IR brane.  This provides 
a rather simple solution to the supersymmetric $CP$ problem; with the 
minimal two Higgs-doublet structure for the Higgs sector, all the complex 
phases can be absorbed into the phases of the fields by the usual $R$ 
and Peccei-Quinn rotations and the chiral rotations for the Dirac gaugino 
fields.

We can also go further in the Higgs sector.  The point here is that 
physics on the IR brane does not affect physics above $M_{\rm mess}$, so 
that we can do whatever we want on the IR brane without destroying the 
successes of supersymmetric theories --- we can introduce arbitrary 321 
multiplets with arbitrary couplings on the IR brane.  For example, we 
can introduce an additional pair of Higgs doublets together with a singlet 
field on the IR brane to push the mass of the physical Higgs boson up 
to about $250~{\rm GeV}$~\cite{Birkedal:2004zx} and/or modify the Higgs 
potential by introducing an IR-brane superpotential coupling(s) of the 
form $T_{-1} H_D H_D$, $T_{+1} \bar{H}_D \bar{H}_D$ or $T_0 H_D \bar{H}_D$, 
where $T_Y$ represents an $SU(2)_L$-triplet superfield with the hypercharge 
$Y$.  In fact, this large freedom for physics on the IR brane, which is 
equivalent to the freedom for the IR dynamics of the DSB sector, opens 
up a very large class of model building possibilities for supersymmetric 
theories without fine-tuning, in the framework discussed in this paper. 

In any of the theories described above, xyons are naturally expected 
to be in the multi-TeV region.  We have studied in detail the experimental 
signatures of these particles at the LHC and at the VLHC.  We have found 
that a generic signature is highly ionizing tracks caused by stable 
charged bound states of xyons: the lightest charged xymeson $\tilde{T}^-$. 
We have found that the reach of the LHC in the xyon mass is about 
$(2.0\!\sim\!2.2)~{\rm TeV}$, so that xyons may be discovered at the 
LHC.  At the VLHC, the reach in the xyon mass is about $5.0~{\rm TeV}$ 
($13~{\rm TeV}$) for a center-of-mass energy of $50~{\rm TeV}$ 
($200~{\rm TeV}$).  In the holographic 5D theories, this covers the 
entire parameter region for the IR-brane scale of $k' \simlt 13~{\rm TeV}$ 
($33~{\rm TeV}$).  Since we naturally expect $k' \approx 10~{\rm TeV}$ 
to obtain weak-scale superparticle masses, the discovery of xyons at the 
VLHC is quite promising. 

We have also discussed cosmology in our scenario.  We have found 
that if the stable lightest xymeson is neutral and there is strong 
nonperturbative xyon annihilation at the QCD era, the thermal history 
of the universe can be standard.  Alternatively, the xyon lifetime 
may be shorter than about $100~{\rm sec}$, in which case the thermal 
history may also be standard even without nonperturbative annihilation. 
In these cases, dark matter can naturally arise as thermal relic 
particles.  In particular, we have considered a class of dark matter, 
which arises as a composite state of the DSB sector but interacts 
with the MSSM states through mixings between the elementary states 
and the composite DSB states.  The correct dark matter abundance is 
naturally reproduced, but direct detection will be difficult because 
the characteristic mass scale for this dark matter is of order 
$(10\!\sim\!100)~{\rm TeV}$, rather than the conventional one of 
order $(0.1\!\sim\!1)~{\rm TeV}$.  A consistent cosmological scenario 
can be obtained, since the theories do not have any dangerous relics 
and the reheating temperature after inflation can be high.  In particular, 
the gravitino does not cause any cosmological problem as it is very 
light with $m_{3/2} \simeq (0.1\!\sim\!10)~{\rm eV}$.

We find it is quite encouraging that we have obtained a consistent 
picture of supersymmetric theories that do not suffer from fine-tuning 
in electroweak symmetry breaking.  A thorough study of the parameter 
space of the class of theories discussed here is warranted.  Much model 
building work will also be possible in the present framework.  In 
particular, detailed exploration of the Higgs sector will be one 
of the important issues.

\section*{Acknowledgments}

This work was supported in part by the Director, Office of Science, Office 
of High Energy and Nuclear Physics, of the US Department of Energy under 
Contract DE-AC03-76SF00098 and DE-FG03-91ER-40676.  The work of Y.N. was 
also supported by the National Science Foundation under grant PHY-0403380 
and by a DOE Outstanding Junior Investigator award.

\newpage

\section*{Appendix A}

In this Appendix we present the calculations of the xyon mass 
in the holographic higher dimensional theories discussed in 
section~\ref{subsec:holographic}.  We also present, for completeness, 
the expressions for the masses of the MSSM superparticles and the 
superpartners of xyons.

\subsection*{A.1~~~Lagrangian of the gauge sector}

The Lagrangian of the gauge sector of the theory is given 
by~\cite{Nomura:2004is}
\begin{equation}
  S = \int\!d^4x \int_0^{\pi R}\!\!dy\,\, \Bigl\{ {\cal L}_{\rm bulk} 
    + 2\delta(y){\cal L}_{\rm UV} + 2\delta(y-\pi R){\cal L}_{\rm IR} \Bigr\},
\label{eq:gauge-Lag}
\end{equation}
where
\begin{eqnarray}
  && {\cal L}_{\rm bulk} 
    = \biggl\{ \frac{1}{4\kappa g_5^2} \int\!d^2\theta\, 
      {\rm Tr}[ {\cal W}^\alpha {\cal W}_\alpha ] + {\rm h.c.} \biggr\}
    + \frac{e^{-2k|y|}}{4\kappa g_5^2} \int\!d^4\theta\, {\rm Tr}[ {\cal A}^2 ],
\label{eq:gauge-Lag-1}\\
  && {\cal L}_{\rm UV} 
    = \sum_{i=1,2,3} \biggl\{ 
      \frac{1}{4\kappa \tilde{g}_{0,i}^2} \int\!d^2\theta\, 
      {\rm Tr}[ {\cal W}_i^\alpha {\cal W}_{i \alpha} ] + {\rm h.c.} \biggr\},
\label{eq:gauge-Lag-2}\\
  && {\cal L}_{\rm IR} 
    = \sum_{i=1,2,3} \biggl\{ 
      -\int\!d^2\theta\, \frac{\zeta_i}{2\kappa M_*} Z \, 
      {\rm Tr}[ {\cal W}_i^\alpha {\cal W}_{i \alpha} ] + {\rm h.c.} \biggr\}
\nonumber\\
  &&  \qquad\quad {} + e^{-2\pi kR} \int\!d^4\theta\, 
    \biggl( \frac{\eta}{4\kappa M_*} Z^\dagger 
      + \frac{\eta^*}{4\kappa M_*} Z 
      - \frac{\rho}{8\kappa M_*^2} Z^\dagger Z \biggr)
      {\rm Tr}[ {\cal P}[{\cal A}] {\cal P}[{\cal A}]],
\label{eq:gauge-Lag-3}
\end{eqnarray}
and
\begin{equation}
  {\cal A} \equiv e^{-2V}\!(\partial_y e^{2V}) 
    - \sqrt{2}\, e^{-2V} \Sigma^\dagger e^{2V} - \sqrt{2}\, \Sigma,
\label{eq:def-A}
\end{equation}
(here we have neglected unimportant IR-brane localized gauge kinetic terms). 
The trace is over the $SU(5)$ space and ${\cal P}[{\cal X}]$ is a projection 
operator: with ${\cal X}$ an adjoint of $SU(5)$, ${\cal P}[{\cal X}]$ 
extracts the $({\bf 3}, {\bf 2})_{-5/6} + ({\bf 3}^*, {\bf 2})_{5/6}$ 
component of ${\cal X}$ under the decomposition to 321.  The $SU(5)$ 
field-strength superfield is defined by ${\cal W}_\alpha \equiv 
-(1/8)\bar{\cal D}^2(e^{-2V} {\cal D}_\alpha e^{2V})$, and 
${\cal W}_{i \alpha}$ with $i=1,2,3$ are the field-strength superfields 
for the $U(1)_Y$, $SU(2)_L$ and $SU(3)_C$ subgroups (${\cal W}_{i \alpha} 
\subset {\cal W}_\alpha$).  The normalization for the $SU(5)$ generators 
is given by ${\rm Tr}[ T^A T^B ] = \kappa\, \delta^{AB}$ where $A,B = 
1,\cdots,24$.

The Lagrangian in components is obtained by expanding 
Eqs.~(\ref{eq:gauge-Lag-1}~--~\ref{eq:gauge-Lag-3}) in component fields: 
\begin{eqnarray}
  V &=& - \theta^\alpha \sigma^\mu_{\alpha\dot{\alpha}} 
      \bar{\theta}^{\dot{\alpha}} A_\mu 
    - i e^{-\frac{3}{2}k|y|} \bar{\theta}^2 \theta^\alpha \lambda_\alpha 
    + i e^{-\frac{3}{2}k|y|} \theta^2 \bar{\theta}_{\dot{\alpha}} 
      \lambda^{\dagger\dot{\alpha}} 
    + \frac{1}{2} e^{-2k|y|} \theta^2 \bar{\theta}^2 D,
\\
  \Sigma &=& \frac{1}{\sqrt{2}}(\chi + i A_5)
    - \sqrt{2}i e^{-\frac{1}{2}k|y|} \theta^\alpha \lambda'_\alpha 
    + e^{-k|y|} \theta^2 F_\Sigma,
\end{eqnarray}
where $A_\mu = A_\mu^A T^A$, $A_5 = A_5^A T^A$, $\lambda_\alpha 
= \lambda^{A}_\alpha T^A$, $\lambda'_\alpha = \lambda'^A_\alpha T^A$, 
$\chi = \chi^A T^A$, $D = D^A T^A$, and $F_\Sigma = F_\Sigma^A T^A$. 
The result is again written in the form of Eq.~(\ref{eq:gauge-Lag}) with
\begin{eqnarray}
  && {\cal L}_{\rm bulk} 
  = \frac{1}{g_5^2} \biggl[ -\frac{1}{4} F^{\mu\nu A}F_{\mu\nu}^A 
    -\frac{1}{2} e^{-2k|y|} F^{\mu 5 A}F_{\mu 5}^A 
    -\frac{1}{2\xi} \Bigl\{ (\partial^\mu A_\mu^A)
      +\xi\, \partial^y (e^{-2k|y|} A_5^A) \Bigr\}^2 
  \nonumber\\
  && \qquad\qquad
  {} -\frac{1}{2} e^{-2k|y|} ({\cal D}^\mu \chi^A) ({\cal D}_\mu \chi^A)
    -i e^{-3k|y|} (\lambda^{\dagger A}_{\dot{\alpha}}\bar{\sigma}^{\mu 
      \dot{\alpha}\alpha} \!\!\lrD_\mu\!\! \lambda^{A}_\alpha)
    -i e^{-3k|y|} (\lambda'^{\dagger A}_{\dot{\alpha}}\bar{\sigma}^{\mu 
      \dot{\alpha}\alpha} \!\!\lrD_\mu\!\! \lambda'^{A}_\alpha)
  \nonumber\\
  && \qquad\qquad
  {} -e^{-4k|y|} (\lambda'^{\alpha A} \!\!\lrD_y\!\! \lambda^{A}_{\alpha}) 
    -e^{-4k|y|} (\lambda'^{\dagger A}_{\dot{\alpha}} \!\!\lrD_y\!\! 
      \lambda^{\dagger\dot{\alpha}A}) 
    -\frac{k}{2}\epsilon(y) e^{-4k|y|} 
      (\lambda^{\alpha A} \lambda'^{A}_{\alpha} 
      +\lambda^{\dagger A}_{\dot{\alpha}} \lambda'^{\dagger\dot{\alpha}A})
  \nonumber\\
  && \qquad\qquad
  {} +\frac{1}{2} e^{-4k|y|} D^A D^A 
    +e^{-4k|y|} F_\Sigma^{\dagger A} F_\Sigma^A
    +e^{-4k|y|} (D^A \!\!\lrD_y\!\!  \chi^A)
  \nonumber\\
  && \qquad\qquad
  {} +i e^{-4k|y|} f^{ABC} \lambda'^{\alpha A} \chi^B \lambda^{C}_{\alpha}
    -i e^{-4k|y|} f^{ABC} \lambda'^{\dagger A}_{\dot{\alpha}} \chi^B 
     \lambda^{\dagger\dot{\alpha}C} \biggr],
\label{eq:comp-Lag-1}\\
  && {\cal L}_{\rm UV} 
    = \sum_{i=1,2,3} \frac{1}{\tilde{g}_{0,i}^2} 
    \biggl\{ -\frac{1}{4} F_i^{\mu\nu a}F_{i\mu\nu}^a 
    -i (\lambda^{\dagger a}_{i\dot{\alpha}}\bar{\sigma}^{\mu 
      \dot{\alpha}\alpha} \!\!\lrD_\mu\!\! \lambda^{a}_{i\alpha})
    +\frac{1}{2} D_i^a D_i^a \biggr\},
\label{eq:comp-Lag-2}\\
  && {\cal L}_{\rm IR} 
    = \sum_{i=1,2,3} \biggl\{ 
    -\frac{1}{2}\frac{\zeta_i F_Z}{M_*} e^{-4\pi kR} 
      \lambda^{\alpha a}_i \lambda^{a}_{i \alpha} 
    -\frac{1}{2}\frac{\zeta_i^* F_Z^*}{M_*} e^{-4\pi kR} 
      \lambda^{\dagger a}_{i\dot{\alpha}} \lambda^{\dagger\dot{\alpha}a}_{i} 
    \biggr\}
  \nonumber\\
  && \qquad\qquad
  {} -\frac{1}{2}\frac{\eta F_Z^*}{M_*} e^{-4\pi kR} 
      \lambda'^{\alpha\hat{a}} \lambda'^{\hat{a}}_{\alpha} 
    -\frac{1}{2}\frac{\eta^* F_Z}{M_*} e^{-4\pi kR} 
      \lambda'^{\dagger\hat{a}}_{\dot{\alpha}} 
      \lambda'^{\dagger\dot{\alpha}\hat{a}} 
  \nonumber\\
  && \qquad\qquad
  {} -\sqrt{2}\frac{\eta F_Z^*}{M_*} e^{-4\pi kR} 
      \chi^{\hat{a}} F_\Sigma^{\hat{a}}
    -\sqrt{2}\frac{\eta^* F_Z}{M_*} e^{-4\pi kR} 
      \chi^{\hat{a}} F_\Sigma^{\dagger \hat{a}}
    -\frac{1}{2} \frac{\rho |F_Z|^2}{M_*^2} e^{-4\pi kR} 
      \chi^{\hat{a}} \chi^{\hat{a}},
\label{eq:comp-Lag-3}
\end{eqnarray}
where $F_Z$ is defined by $\langle Z \rangle = - e^{-\pi kR} F_Z \theta^2$, 
and the indices $a$ and $\hat{a}$ run over the 321 and XY components, 
respectively: $A = \{ a, \hat{a} \}$ (summation convention implied for 
$A$, $a$ and $\hat{a}$).  The function $\epsilon(y)$ takes $+1$ for 
$y>0$ and $-1$ for $y<0$, ${\cal D}_\mu$ and ${\cal D}_y$ are the gauge 
covariant derivatives, and $\lrD_\mu$ and $\lrD_y$ are defined by 
$(\varphi_1 \!\!\lrD_M\!\! \varphi_2) \equiv (1/2)\{\varphi_1 
({\cal D}_M \varphi_2) - ({\cal D}_M \varphi_1) \varphi_2\}$ for 
arbitrary fields $\varphi_1$ and $\varphi_2$, where $M=\mu,y$.  Here, 
we have added the gauge-fixing term
\begin{equation}
  S_{\rm GF} = \int\!d^4x \int_0^{\pi R}\!\!dy\,\, \frac{1}{g_5^2}
    \Biggl[ -\frac{e^{-4k|y|}}{2\xi} \biggl\{ e^{2k|y|}(\partial^\mu A_\mu^A) 
    +\xi e^{2k|y|} \partial^y (e^{-2k|y|} A_5^A) \biggr\}^2 \Biggr],
\end{equation}
in the original Lagrangian to obtain Eqs.~(\ref{eq:comp-Lag-1}~--%
~\ref{eq:comp-Lag-3}).

\subsection*{A.2~~~Superparticle masses}

The masses of the gauginos and their KK towers are obtained by solving 
the equations of motion derived from Eqs.~(\ref{eq:comp-Lag-1}~--~%
\ref{eq:comp-Lag-3}).  The equation determining these masses, $m_n$, 
are given by~\cite{Nomura:2003qb}
\begin{equation}
  \frac{J_0\left(\frac{m_n}{k}\right)
    + \frac{g_5^2}{\tilde{g}_{0,i}^2}m_n J_1\left(\frac{m_n}{k}\right)}
  {Y_0\left(\frac{m_n}{k}\right)
    + \frac{g_5^2}{\tilde{g}_{0,i}^2}m_n Y_1\left(\frac{m_n}{k}\right)}
  = \frac{J_0\left(\frac{m_n}{k'}\right)
    + g_5^2 M_{\lambda,i} J_1\left(\frac{m_n}{k'}\right)}
    {Y_0\left(\frac{m_n}{k'}\right)
    + g_5^2 M_{\lambda,i} Y_1\left(\frac{m_n}{k'}\right)},
\label{eq:KKmass-321gauginos}
\end{equation}
where $M_{\lambda,i} \equiv \zeta_i F_Z/M_*$ (see 
Eq.~(\ref{eq:susy-br-para-1})).  For $m_n/k' \ll 1$ and $g_5^2 
m_n M_{\lambda,i}/k' \ll 1$, which are satisfied for the lowest 
modes if the sizes of various parameters are determined by 
naive dimensional analysis ($r_{\lambda,i} \sim 1$; see 
Eq.~(\ref{eq:susy-br-para-2})), the 321 gaugino masses are given 
by $m_{\lambda_i^{321}} \simeq g_i^2 M_{\lambda,i} e^{-\pi kR}$, 
in agreement with Eq.~(\ref{eq:gaugino-masses-5D}) obtained by 4D 
considerations.  Here, $g_i = (\pi R/g_5^2+1/\tilde{g}_{0,i}^2)^{-1/2}$ 
are the 4D gauge couplings.  The squark and slepton masses are 
generated at one loop and given by
\begin{equation}
  m_{\tilde{f}}^2 = 
    \frac{1}{2\pi^2}\! \sum_{i=1,2,3}\! C_i^{\tilde{f}}
    \int_0^\infty\! dq\, q^3 \biggl\{ f^{321,i}(1/k,1/k;q) 
    - f^{321,i}(1/k,1/k;q) \Bigr|_{M_{\lambda,i}=0} \biggr\},
\label{eq:MSSMscalar}
\end{equation}
where $\tilde{f} = \tilde{q}, \tilde{u}, \tilde{d}, \tilde{l}, \tilde{e}$ 
represents the MSSM squarks and sleptons, and the $C_i^{\tilde{f}}$ are 
the group theoretical factors given by $(C_1^{\tilde{f}}, C_2^{\tilde{f}}, 
C_3^{\tilde{f}}) = (1/60,3/4,4/3)$, $(4/15,0,4/3)$, $(1/15,0,4/3)$, 
$(3/20,3/4,0)$ and $(3/5,0,0)$ for $\tilde{f} = \tilde{q}, \tilde{u}, 
\tilde{d}, \tilde{l}$ and $\tilde{e}$, respectively.  The functions 
$f^{321,i}(z,z';q)$ arise from the 321 gaugino propagators whose 
explicit form are given in Appendix~B.1 (see Eqs.~(\ref{eq:fz},%
~\ref{eq:ABC},~\ref{eq:XYZ321})). 

The masses for the superpartners of xyons, $\lambda'^{\rm XY}$ and 
$\chi^{\rm XY}$, are given as the lowest solutions of 
\begin{equation}
  \frac{J_1\left(\frac{m_n}{k}\right)}
  {Y_1\left(\frac{m_n}{k}\right)}
  = \frac{J_1\left(\frac{m_n}{k'}\right)
    - g_5^2 M_{\lambda,X} J_0\left(\frac{m_n}{k'}\right)}
    {Y_1\left(\frac{m_n}{k'}\right)
    - g_5^2 M_{\lambda,X} Y_0\left(\frac{m_n}{k'}\right)},
\label{eq:KKmass-XYfermion}
\end{equation}
and
\begin{equation}
  \frac{J_1\left(\frac{m_n}{k}\right)}
  {Y_1\left(\frac{m_n}{k}\right)}
  = \frac{J_1\left(\frac{m_n}{k'}\right)
    - \frac{g_5^2 M_{\chi,X}^2 k'}{m_n k} J_0\left(\frac{m_n}{k'}\right)}
    {Y_1\left(\frac{m_n}{k'}\right)
    - \frac{g_5^2 M_{\chi,X}^2 k'}{m_n k} Y_0\left(\frac{m_n}{k'}\right)},
\label{eq:KKmass-XYscalar}
\end{equation}
respectively~\cite{Nomura:2004is}.  Here, $M_{\lambda,X} \equiv 
\eta F_Z^*/M_*$, $M_{\chi,X}^2 \equiv \rho' |F_Z|^2/M_*^2$ (see 
Eq.~(\ref{eq:susy-br-para-1})), and $\rho' \equiv \rho + 8 g_5^2 
|\eta|^2 \delta(0)$ (see Eq.~(\ref{eq:rho-prime})).  For natural sizes 
of the couplings, suggested by naive dimensional analysis, these masses 
are of order $\pi k'$.

\subsection*{A.3~~~Xyon mass}

To calculate the mass of xyons, we need to know their wavefunctions. 
In our 5D theory, xyons correspond to the lowest modes for 
$A_5^{\rm XY}$.  The wavefunctions of these modes are obtained by 
solving the equations of motion for $A_5^{\rm XY}$, given by the 
Lagrangian of Eq.~(\ref{eq:comp-Lag-1}).  Writing xyon fields as 
$\varphi^{\hat{a}}(x)$, we find in the Feynman-'t~Hooft gauge ($\xi=1$)
\begin{equation}
  A_5^{\hat{a}}(x,y) = \sqrt{\frac{2 g_5^2 k}{e^{2\pi kR}-1}}
    e^{2k|y|} \varphi^{\hat{a}}(x),
\label{eq:xyon-wavef-y}
\end{equation}
where $\varphi^{\hat{a}}(x)$ is canonically normalized in 4D.  

\begin{figure}[t]
\begin{center} 
\begin{picture}(440,105)(0,-20)
  \Text(40,-5)[t]{\large (a)}
  \DashLine(0,50)(20,50){3} \Text(0,57)[b]{$A_5^{\rm XY}$}
  \DashLine(60,50)(80,50){3} \Text(80,57)[b]{$A_5^{\rm XY}$}
  \CArc(40,50)(20,0,360)
  \Text(40,75)[b]{$\lambda^{321}$} \Text(40,25)[t]{$\lambda'^{\rm XY}$}
  \Text(160,-5)[t]{\large (b)}
  \DashLine(120,50)(140,50){3} \Text(120,57)[b]{$A_5^{\rm XY}$}
  \DashLine(180,50)(200,50){3} \Text(200,57)[b]{$A_5^{\rm XY}$}
  \CArc(160,50)(20,0,360)
  \Text(160,75)[b]{$\lambda^{\rm XY}$} \Text(160,25)[t]{$\lambda'^{321}$}
  \Text(280,-5)[t]{\large (c)}
  \DashLine(240,50)(260,50){3} \Text(240,57)[b]{$A_5^{\rm XY}$}
  \DashLine(300,50)(320,50){3} \Text(320,57)[b]{$A_5^{\rm XY}$}
  \DashCArc(280,50)(20,0,360){3}
  \Text(280,75)[b]{$\chi^{321}$} \Text(280,25)[t]{$\chi^{\rm XY}$}
  \Text(400,-5)[t]{\large (d)}
  \DashLine(360,30)(440,30){3}
  \Text(360,25)[t]{$A_5^{\rm XY}$} \Text(440,25)[t]{$A_5^{\rm XY}$}
  \DashCArc(400,50)(20,0,360){3} \Text(400,75)[b]{$\chi^{\rm XY}$}
\end{picture}
\caption{One-loop diagrams relevant for the calculation of the xyon mass.}
\label{fig:xyon-mass}
\end{center}
\end{figure}
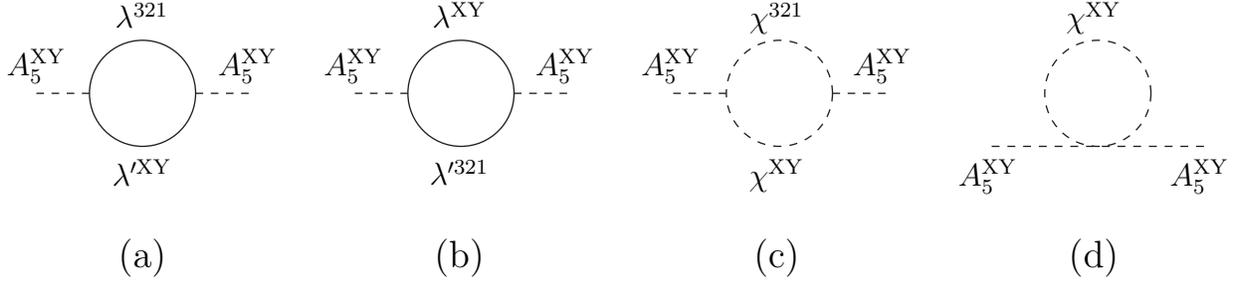

The mass of xyons are calculated most easily using the on-shell 
Lagrangian, which is obtained from Eqs.~(\ref{eq:comp-Lag-1}~--%
~\ref{eq:comp-Lag-3}) by integrating out the auxiliary fields:
\begin{equation}
  S = \int\!d^4x \int_0^{\pi R}\!\!dy\,\, 
    \Bigl\{ {\cal L}_{\rm gauge} + {\cal L}_{\rm gaugino} 
    + {\cal L}_{\rm scalar} + {\cal L}_{\rm Yukawa} \Bigr\}.
\label{eq:onshell}
\end{equation}
Here, 
\begin{eqnarray}
  && {\cal L}_{\rm gauge} 
  = \frac{1}{g_5^2} \biggl[ -\frac{1}{4} F^{\mu\nu A}F_{\mu\nu}^A 
    -\frac{1}{2} e^{-2k|y|} F^{\mu 5 A}F_{\mu 5}^A 
    -\frac{1}{2} \Bigl\{ (\partial^\mu A_\mu^A)
      +\partial^y (e^{-2k|y|} A_5^A) \Bigr\}^2 \biggr],
\label{eq:onshell-1}\\
  && {\cal L}_{\rm gaugino} 
  = \frac{1}{g_5^2} \biggl[ -i e^{-3k|y|} 
    (\lambda^{\dagger A}_{\dot{\alpha}}\bar{\sigma}^{\mu 
      \dot{\alpha}\alpha} \!\!\lrD_\mu\!\! \lambda^{A}_\alpha)
    -i e^{-3k|y|} (\lambda'^{\dagger A}_{\dot{\alpha}}\bar{\sigma}^{\mu 
      \dot{\alpha}\alpha} \!\!\lrD_\mu\!\! \lambda'^{A}_\alpha)
  \nonumber\\
  && \qquad\qquad\quad 
  {} -e^{-4k|y|} (\lambda'^{\alpha A} \!\!\lrD_y\!\! \lambda^{A}_{\alpha}) 
    -e^{-4k|y|} (\lambda'^{\dagger A}_{\dot{\alpha}} \!\!\lrD_y\!\! 
      \lambda^{\dagger\dot{\alpha}A}) 
    -\frac{k}{2}\epsilon(y) e^{-4k|y|} 
      (\lambda^{\alpha A} \lambda'^{A}_{\alpha} 
      +\lambda^{\dagger A}_{\dot{\alpha}} \lambda'^{\dagger\dot{\alpha}A}) 
    \biggr]
  \nonumber\\
  && \qquad\qquad 
  {} +2\delta(y-\pi R) \biggl[ \sum_{i=1,2,3} \biggl\{ 
    -\frac{1}{2} e^{-4\pi kR} M_{\lambda,i} 
      \lambda^{\alpha a}_i \lambda^{a}_{i \alpha} 
    -\frac{1}{2} e^{-4\pi kR} M_{\lambda,i}^* 
      \lambda^{\dagger a}_{i\dot{\alpha}} \lambda^{\dagger\dot{\alpha}a}_{i} 
    \biggr\}
  \nonumber\\
  && \qquad\qquad\quad
  {} -\frac{1}{2} e^{-4\pi kR} M_{\lambda,X} 
      \lambda'^{\alpha\hat{a}} \lambda'^{\hat{a}}_{\alpha} 
    -\frac{1}{2} e^{-4\pi kR} M_{\lambda,X}^* 
      \lambda'^{\dagger\hat{a}}_{\dot{\alpha}} 
      \lambda'^{\dagger\dot{\alpha}\hat{a}}  \biggr],
\label{eq:onshell-2}\\
  && {\cal L}_{\rm scalar} 
  = \frac{1}{g_5^2} \biggl[ 
    -\frac{1}{2} e^{-2k|y|} ({\cal D}^\mu \chi^A) ({\cal D}_\mu \chi^A)
    -\frac{1}{2} e^{-4k|y|} ({\cal D}^y \chi^A) ({\cal D}_y \chi^A)
  \nonumber\\
  && \qquad\qquad\quad
  {} +2 k^2 e^{-4k|y|} \chi^A \chi^A 
    -2 k e^{-4k|y|} \{ \delta(y)-\delta(y-\pi R) \} \chi^A \chi^A \biggr]
  \nonumber\\
  && \qquad\qquad 
  {} +2\delta(y-\pi R) \biggl[ -\frac{1}{2} e^{-4\pi kR} 
      M_{\chi,X}^2 \chi^{\hat{a}} \chi^{\hat{a}} \biggr],
\label{eq:onshell-3}\\
  && {\cal L}_{\rm Yukawa} 
  = \frac{1}{g_5^2} \biggl[ 
     i e^{-4k|y|} f^{ABC} \lambda'^{\alpha A} \chi^B \lambda^{C}_{\alpha}
    -i e^{-4k|y|} f^{ABC} \lambda'^{\dagger A}_{\dot{\alpha}} \chi^B 
     \lambda^{\dagger\dot{\alpha}C} \biggr],
\label{eq:onshell-4}
\end{eqnarray}
where $M_{\lambda,i} = \zeta_i F_Z/M_*$, $M_{\lambda,X} = \eta F_Z^*/M_*$, 
$M_{\chi,X}^2 = \rho' |F_Z|^2/M_*^2$, and we have taken the Feynman-'t~Hooft 
gauge, $\xi=1$.  We have also suppressed the contributions from the UV-brane 
localized gauge kinetic terms, for the simplicity of the presentation.  

\ From Eqs.~(\ref{eq:onshell}~--~\ref{eq:onshell-4}), we find that 
at one loop the xyon mass receives contributions from the diagrams 
with $A_\mu^{321}$-$A_5^{\rm XY}$, $A_\mu^{\rm XY}$-$A_5^{321}$, 
$A_\mu^{321}$-$A_\mu^{\rm XY}$, $\lambda^{321}$-$\lambda'^{\rm XY}$, 
$\lambda^{\rm XY}$-$\lambda'^{321}$ and $\chi^{321}$-$\chi^{\rm XY}$ 
loops using two 3-point vertices, and from the diagrams with 
$A_\mu^{321}$, $A_\mu^{\rm XY}$, $\chi^{321}$ and $\chi^{\rm XY}$ 
loops using one 4-point vertex (and the ghost loops).  Instead of 
calculating all these loops, here we use the following trick to 
simplify the calculation of the xyon mass.  We use the fact that the 
xyon mass is not generated in the supersymmetric limit.  The xyon 
mass, then, can be obtained by calculating the diagrams that contain 
the effects of supersymmetry breaking and subtracting the values of 
the same diagrams in the supersymmetric limit.  In this procedure, we 
have to calculate only the diagrams given in Fig.~\ref{fig:xyon-mass}. 
We then obtain the xyon mass by subtracting the values of the same 
diagrams with $M_{\lambda,i}$, $M_{\lambda,X}$ and $M_{\chi,X}^2$ 
set to zero.

We start by the gaugino-loop diagrams of Fig.~\ref{fig:xyon-mass}a and 
\ref{fig:xyon-mass}b.  We first note that the gaugino mass parameters 
on the IR brane are, in general, arbitrary complex parameters.  We take 
this into account by rewriting $M_{\lambda,i} \rightarrow M_{\lambda,i} 
e^{i\phi_i}$ and $M_{\lambda,X} \rightarrow M_{\lambda,X} e^{i\phi_X}$, 
so $M_{\lambda,i}$ and $M_{\lambda,X}$ are now real parameters. 
Calculating the diagrams in Fig.~\ref{fig:xyon-mass}a and 
\ref{fig:xyon-mass}b, we obtain the following contribution 
to the xyon mass:
\begin{equation}
  \delta \hat{m}_\varphi^2|_{\lambda} 
  = \delta \hat{m}_\varphi^2|_{\lambda}^{(1)} 
  + \sum_{i=1,2,3}\cos(\phi_i+\phi_X)\,
    \delta \hat{m}_\varphi^2|_{\lambda,i}^{(2)}.
\label{eq:xyon-gaugino}
\end{equation}
Here, the first term in the right-hand-side is the piece that does 
not depend on the phases of the gaugino mass parameters:
\begin{equation}
  \delta \hat{m}_\varphi^2|_{\lambda}^{(1)}
  = \delta m_\varphi^2|_{\lambda}^{(1)}
    - \delta m_\varphi^2|_{\lambda,M_{\lambda,i}=M_{\lambda,X}=0}^{(1)},
\label{eq:xyon-gaugino-1}
\end{equation}
where
\begin{eqnarray}
  \delta m_\varphi^2|_{\lambda}^{(1)}
  &=& -\frac{1}{2\pi^2} \sum_{i=1,2,3} \frac{C_i^\varphi}{(g_5^2 k)^2} 
  \int \frac{dz_1 dz_2}{z_1 z_2} {\cal F}(z_1) {\cal F}(z_2) 
  \int d|p|\, |p|^3 
\nonumber\\
  && {} \times \Biggl[ 
    |p|^2 f'^{321,i}(z_2,z_1;p) f^{\rm XY}(z_1,z_2;p)
  + |p|^2 f'^{\rm XY}(z_2,z_1;p) f^{321,i}(z_1,z_2;p)
\nonumber\\
  && \quad
  {} + 2 \biggl\{ \Bigl( \partial_{z_2}-\frac{1}{2z_2} \Bigr)
    f'^{321,i}(z_2,z_1;p) \biggr\}
    \biggl\{ \Bigl( \partial_{z_1}-\frac{1}{2z_1} \Bigr)
    f'^{\rm XY}(z_1,z_2;p) \biggr\} \Biggr].
\label{eq:xyon-gaugino-1-exp}
\end{eqnarray}
We here subtracted the supersymmetric piece in 
Eq.~(\ref{eq:xyon-gaugino-1}).  The functions $f^{321,i}$, $f^{\rm XY}$, 
$f'^{321,i}$ and $f'^{\rm XY}$ are given in Appendix~B.1.  The function 
${\cal F}(z)$ is the wavefunction of xyons in the $z$ coordinate
\begin{equation}
  {\cal F}(z) = \sqrt{\frac{2 g_5^2 k}{e^{2\pi kR}-1}} (kz)^2,
\label{eq:xyon-wavef-z}
\end{equation}
(see Eq.~(\ref{eq:xyon-wavef-y})), and $C_i^\varphi$ are the group 
theoretical factors: $(C_1^\varphi, C_2^\varphi, C_3^\varphi) = 
(5/12, 3/4, 4/3)$ for $SU(5)$ xyons.  The second term in the 
right-hand-side of Eq.~(\ref{eq:xyon-gaugino}) is the piece that 
depends on the phases:
\begin{equation}
  \delta \hat{m}_\varphi^2|_{\lambda,i}^{(2)}
  = \delta m_\varphi^2|_{\lambda,i}^{(2)}
    - \delta m_\varphi^2|_{\lambda,i,M_{\lambda,i}=M_{\lambda,X}=0}^{(2)},
\label{eq:xyon-gaugino-2}
\end{equation}
where
\begin{eqnarray}
  \delta m_\varphi^2|_{\lambda,i}^{(2)}
  &=& \frac{1}{2\pi^2} \frac{C_i^\varphi}{(g_5^2 k)^2} 
  \int \frac{dz_1 dz_2}{z_1 z_2} {\cal F}(z_1) {\cal F}(z_2) 
  \int d|p|\, |p|^3 
\nonumber\\
  && {} \times \Biggl[ 
    h'^{321,i}(z_2,z_1;p) h^{\rm XY}(z_1,z_2;p)
  + h'^{\rm XY}(z_2,z_1;p) h^{321,i}(z_1,z_2;p)
\nonumber\\
  && \quad
  {} + \frac{2}{|p|^2} \biggl\{ \Bigl( \partial_{z_2}-\frac{1}{2z_2} \Bigr)
    h'^{321,i}(z_2,z_1;p) \biggr\}
    \biggl\{ \Bigl( \partial_{z_1}-\frac{1}{2z_1} \Bigr)
    h'^{\rm XY}(z_1,z_2;p) \biggr\} \Biggr].
\label{eq:xyon-gaugino-2-exp}
\end{eqnarray}
The functions $h^{321,i}$, $h^{\rm XY}$, $h'^{321,i}$ and $h'^{\rm XY}$ 
are given in Appendix~B.1.

We next consider the contributions from gauge-scalar loops.  They are 
given by
\begin{equation}
  \delta \hat{m}_\varphi^2|_{\chi} 
  = \delta \hat{m}_\varphi^2|_{\chi}^{(1)} 
  + \delta \hat{m}_\varphi^2|_{\chi}^{(2)}.
\label{eq:xyon-scalar}
\end{equation}
The first term in the right-hand-side represents the contribution 
from the diagram of Fig.~\ref{fig:xyon-mass}c:
\begin{equation}
  \delta \hat{m}_\varphi^2|_{\chi}^{(1)}
  = \delta m_\varphi^2|_{\chi}^{(1)}
    - \delta m_\varphi^2|_{\chi,M_{\chi,X}^2=0}^{(1)},
\label{eq:xyon-scalar-1}
\end{equation}
where
\begin{eqnarray}
  && \delta m_\varphi^2|_{\chi}^{(1)}
  = \frac{C^\varphi}{8\pi^2 g_5^4} 
  \int dz_1 dz_2\, {\cal F}(z_1) {\cal F}(z_2) 
  \int d|p|\, |p|^3 
\nonumber\\
  && \quad {} \times \Biggl[ \partial_{z_1} \partial_{z_2}
    \Bigl( \hat{G}^{321}_{\chi\chi}(z_1,z_2;p) 
    \hat{G}^{\rm XY}_{\chi\chi}(z_1,z_2;p) \Bigr)
  - 4 \Bigl( \partial_{z_1} \hat{G}^{321}_{\chi\chi}(z_1,z_2;p) \Bigr) 
    \Bigl( \partial_{z_2} \hat{G}^{\rm XY}_{\chi\chi}(z_1,z_2;p) \Bigr) 
  \Biggr].
\label{eq:xyon-scalar-1-exp}
\end{eqnarray}
Here, $C^\varphi \equiv \sum_{i=1,2,3}C_i^\varphi = 5/2$, and 
$\hat{G}^{321}_{\chi\chi}(z,z';p)$ and $\hat{G}^{\rm XY}_{\chi\chi}(z,z';p)$ 
are the propagators for the (rescaled) gauge scalars, whose explicit forms 
are given in Appendix~B.2.  The second term in the right-hand-side of 
Eq.~(\ref{eq:xyon-scalar}) is the contribution from the diagram of 
Fig.~\ref{fig:xyon-mass}d:
\begin{equation}
  \delta \hat{m}_\varphi^2|_{\chi}^{(2)}
  = \delta m_\varphi^2|_{\chi}^{(2)}
    - \delta m_\varphi^2|_{\chi,M_{\chi,X}^2=0}^{(2)},
\label{eq:xyon-scalar-2}
\end{equation}
where
\begin{equation}
  \delta m_\varphi^2|_{\chi}^{(2)}
  = \frac{i C^\varphi}{8\pi^2 (g_5^2 k)} 
  \int \frac{dz}{z} \{ {\cal F}(z) \}^2
  \int d|p|\, |p|^3 \hat{G}^{\rm XY}_{\chi\chi}(z,z;p).
\label{eq:xyon-scalar-2-exp}
\end{equation}
Here we note that the 4D-momentum integral in $\delta 
\hat{m}_\varphi^2|_{\chi}^{(2)}$ are divergent even after 
subtracting the supersymmetric piece.  This contribution, however, 
is canceled by a part of the contribution arising from $\delta 
\hat{m}_\varphi^2|_{\chi}^{(1)}$ in the following way.  In general, 
the propagators of 5D fields take the form of $\hat{G}(z,z';p) = 
\theta(z-z'){\cal G}(z,z';p) + \theta(z'-z){\cal G}(z',z;p)$, where 
${\cal G}(z,z';p)$ is some function depending on the field of interest 
(this is implicit in the expressions of Appendix~B, which use the 
symbols $z_<$ and $z_>$ instead of the theta function).  This gives 
the contribution in Eq.~(\ref{eq:xyon-scalar-1-exp}) proportional to 
$\delta(z_1-z_2)$ in the integrand, which arises from the piece with 
$z$-derivatives hitting the theta functions.  We then find that the 
resulting contribution in $\delta \hat{m}_\varphi^2|_{\chi}^{(1)}$, 
which can be written in the form of a single $z$ integral, is 
exactly canceled with $\delta \hat{m}_\varphi^2|_{\chi}^{(2)}$ 
in Eq.~(\ref{eq:xyon-scalar}). 

To summarize, the mass of xyons is given by
\begin{equation}
  m_\varphi^2 
  = \delta \hat{m}_\varphi^2|_{\lambda}^{(1)} 
    + \sum_{i=1,2,3}\cos(\phi_i+\phi_X)\,
      \delta \hat{m}_\varphi^2|_{\lambda,i}^{(2)}
    + \delta \hat{m}_\varphi^2|_{\chi},
\label{eq:xyon-mass-fin}
\end{equation}
where $\delta \hat{m}_\varphi^2|_{\lambda}^{(1)}$, 
$\delta \hat{m}_\varphi^2|_{\lambda,i}^{(2)}$ and 
$\delta \hat{m}_\varphi^2|_{\chi}$ are given by 
Eqs.~(\ref{eq:xyon-gaugino-1},~\ref{eq:xyon-gaugino-1-exp}), 
(\ref{eq:xyon-gaugino-2},~\ref{eq:xyon-gaugino-2-exp}) and 
(\ref{eq:xyon-scalar}~--~\ref{eq:xyon-scalar-2-exp}), respectively. 
The quantities $\delta \hat{m}_\varphi^2|_{\lambda}^{(1)}$, 
$\delta \hat{m}_\varphi^2|_{\lambda,i}^{(2)}$ and $\delta 
\hat{m}_\varphi^2|_{\chi}$ are separately finite (up to 
exponentially suppressed contributions), reflecting the fact 
that the xyon obtains a mass only through the $SU(5)$ breaking 
on the UV brane while its wavefunction is strongly localized to 
the IR brane.

To see the relative importance of various contributions, we have 
listed values of $\delta \hat{m}_\varphi^2|_{\lambda}^{(1)}$, 
$\delta \hat{m}_\varphi^2|_{\lambda,i}^{(2)}$ and 
$\delta \hat{m}_\varphi^2|_{\chi}$ in Table~\ref{table:app-A} 
for several points in the parameter space.  Here, $r_{\lambda,i}$, 
$r_{\lambda,X}$ and $r_{\chi,X}$ are dimensionless parameters 
defined by $r_{\lambda,i} = M_{\lambda,i}/(M_*/16\pi^2)$, 
$r_{\lambda,X} = M_{\lambda,X}/(M_*/16\pi^2)$ and $r_{\chi,X} 
= M_{\chi,X}^2/(M_*^2/16\pi^2)$, which naturally take values 
of order one (see Eq.~(\ref{eq:susy-br-para-2})).  
\begin{table}
\begin{center}
\begin{tabular}{|c|c|ccccc|ccccc|c|}  \hline 
 $k'$ & $\frac{M_*}{\pi k}$ & \multicolumn{5}{c}{$\{ r_{\lambda,1}, 
 r_{\lambda,2}, r_{\lambda,3}, r_{\lambda,X}, r_{\chi,X} \}$} & 
 \multicolumn{5}{c}{$[\{ \delta \hat{m}_\varphi^2|_{\lambda}^{(1)}, 
 \delta \hat{m}_\varphi^2|_{\lambda,1}^{(2)}, 
 \delta \hat{m}_\varphi^2|_{\lambda,2}^{(2)}, 
 \delta \hat{m}_\varphi^2|_{\lambda,3}^{(2)}, 
 \delta \hat{m}_\varphi^2|_{\chi} \} ]^{1/2}$} & $[m_\varphi^2]^{1/2}$ 
\\ \hline
 $10$ &  $3$ &   $1$ &   $1$ &   $1$ &   $1$ &    $1$ & 
    \,\,\,\quad$3.7$\, & \quad$0.04$ &  \quad$0.1$ & \quad\,$0.3$ & $-1.5$ & 
     $3.4$ \\
 $10$ &  $3$ & $0.1$ & $0.1$ & $0.1$ &   $1$ &    $1$ & 
    \,\,\,\quad$3.2$\, & \quad$0.01$ & \quad$0.03$ & \quad\,$0.1$ & $-1.5$ & 
     $2.9$ \\
 $10$ &  $3$ &   $1$ &   $1$ &   $1$ & $0.1$ &    $1$ & 
    \,\,\,\quad$2.0$\, & \quad$0.04$ &  \quad$0.1$ & \quad\,$0.3$ & $-1.5$ & 
     $1.3$ \\
 $10$ &  $3$ &   $1$ &   $1$ &   $1$ &   $1$ & $0.05$ & 
    \,\,\,\quad$3.7$\, & \quad$0.04$ &  \quad$0.1$ & \quad\,$0.3$ & $-1.0$ & 
     $3.6$ \\
 $10$ & $3$ & $0.1$ & $0.1$ & $0.1$ & $0.1$ &    $1$ & 
    \,\,\,\quad$0.8$\, & \quad$0.01$ & \quad$0.03$ & \quad\,$0.1$ & $-1.5$ & 
    $-1.3$ 
\\ \hline
 $10$ &  $3$ & $2.5$ & $0.9$ & $1.0$ &   $1$ &    $1$ & 
    \,\,\,\quad$3.7$\, & \quad$0.05$ &  \quad$0.1$ & \quad\,$0.3$ & $-1.5$ & 
     $3.4$ \\
 $20$ &  $3$ & $2.5$ & $0.9$ & $1.0$ &   $1$ &    $1$ & 
    \,\,\,\quad$7.2$\, &  \quad$0.1$ &  \quad$0.2$ & \quad\,$0.6$ & $-3.0$ & 
     $6.5$ \\
 $40$ &  $3$ & $2.5$ & $0.9$ & $1.0$ &   $1$ &    $1$ & 
     \,\,\,\quad$14$\, &  \quad$0.2$ &  \quad$0.4$ & \quad\,$1.2$ & $-6.0$ & 
      $13$ \\
 $10$ &  $1$ & $2.5$ & $0.9$ & $1.0$ &   $1$ &    $1$ & 
    \,\,\,\quad$2.0$\, & \quad$0.05$ & \quad$0.07$ & \quad\,$0.2$ & $-1.2$ & 
     $1.7$ \\
 $10$ & $10$ & $2.5$ & $0.9$ & $1.0$ &   $1$ &    $1$ & 
    \,\,\,\quad$5.1$\, & \quad$0.03$ & \quad$0.08$ & \quad\,$0.2$ & $-1.6$ & 
     $4.8$ 
\\ \hline
\end{tabular}
\end{center}
\caption{Values of various contributions to the xyon mass squared 
 for several sample parameter points.  Here, $[X]^{n} \equiv 
 {\rm sgn}(X) \cdot |X|^{n}$, and all the masses are given in 
 units of TeV.}
\label{table:app-A}
\end{table}
\ From the table we find the following: 
\begin{itemize}
\item
The contributions to the xyon mass squared from the gaugino loop 
diagrams with the supersymmetric piece subtracted are positive, while 
that from the gauge-scalar loop diagrams with the supersymmetric 
piece subtracted is negative. 
\item
In most of natural parameter regions, the xyon mass squared, 
$m_\varphi^2$, is positive, which is crucial for the model to 
be viable.
\item
The sizes of phase-dependent gaugino contributions, 
$\delta \hat{m}_\varphi^2|_{\lambda,1}^{(2)}$, 
$\delta \hat{m}_\varphi^2|_{\lambda,2}^{(2)}$ and $\delta 
\hat{m}_\varphi^2|_{\lambda,3}^{(2)}$, are much smaller than that of 
the phase-independent one, $\delta \hat{m}_\varphi^2|_{\lambda}^{(1)}$. 
This implies that the size of the xyon mass does not depend much on 
the complex phases of the gaugino mass parameters on the IR brane. 
\item
The values of $\delta \hat{m}_\varphi^2|_{\lambda}^{(1)}$, 
$\delta \hat{m}_\varphi^2|_{\lambda,1}^{(2)}$, 
$\delta \hat{m}_\varphi^2|_{\lambda,2}^{(2)}$ and 
$\delta \hat{m}_\varphi^2|_{\lambda,3}^{(2)}$ 
($|\delta \hat{m}_\varphi^2|_{\chi}|$) decrease with decreasing 
values of $r_{\lambda,1}$, $r_{\lambda,2}$, $r_{\lambda,3}$ and 
$r_{\lambda,X}$ ($r_{\chi,X}$) in the parameter region considered. 
With fixed values of $k'$ and $M_*/\pi k'$, the quantity $\delta 
\hat{m}_\varphi^2|_{\lambda}^{(1)}$ depends only on $r_{\lambda,i}$ 
and $r_{\lambda,X}$, $\delta \hat{m}_\varphi^2|_{\lambda,i}^{(2)}$ 
on $r_{\lambda,i}$ (with a tiny dependence on $r_{\lambda,X}$), and 
$\delta \hat{m}_\varphi^2|_{\chi}$ on $r_{\chi,X}$.
\item
The xyon mass scales almost linearly with $k'$.
\item
The scaling of the xyon mass with $M_*/\pi k'$ is not very simple, 
but they are positively correlated. (Note that $M_*/\pi k' \rightarrow 
\alpha (M_*/\pi k')$ with the other parameters fixed is equivalent 
to $r_{\lambda,i} \rightarrow \alpha r_{\lambda,i}$, $r_{\lambda,X} 
\rightarrow \alpha r_{\lambda,X}$ and $r_{\chi,X} \rightarrow \alpha^2 
r_{\chi,X}$ with the other parameters fixed.)
\end{itemize}
These features are used in section~\ref{subsec:xyon-mass-2} in the 
discussion of the xyon mass in the holographic 5D theories.

\section*{Appendix B}

In this Appendix we present the propagators for the gauginos and gauge 
scalars, relevant for the calculation of the xyon mass.

\subsection*{B.1~~~Gaugino propagators}

Here we present the propagators for the $U(1)_Y$, $SU(2)_L$, $SU(3)_C$ 
and XY gauginos.  We present the propagators $\hat{G}$ for the rescaled 
fields, $\hat{\lambda}^A_\alpha = e^{-2k|y|}\lambda^A_\alpha$ and 
$\hat{\lambda}'^A_\alpha = e^{-2k|y|}\lambda'^A_\alpha$.  The propagators 
for the non-rescaled fields are given by $G = e^{2k(|y|+|y'|)}\hat{G}$.

The gaugino propagators take the form of $4 \times 4$ matrices in the 
space of $\{ \hat{\lambda}^A_\alpha, \hat{\lambda}'^{\dagger\dot{\alpha}A}, 
\hat{\lambda}^{\dagger\dot{\alpha}A}, \hat{\lambda}'^A_\alpha \}$:
\begin{equation}
  \hat{G}^S \equiv \left( \begin{array}{cc|cc} 
    \hat{G}^S_{\lambda \lambda^\dagger}(z,z';p)_{\alpha\dot{\beta}} & 
    \hat{G}^S_{\lambda \lambda'}(z,z';p)_\alpha^{~\beta} & 
    \hat{G}^S_{\lambda \lambda}(z,z';p)_\alpha^{~\beta} & 
    \hat{G}^S_{\lambda \lambda'^\dagger}(z,z';p)_{\alpha\dot{\beta}} \\
    \hat{G}^S_{\lambda'^\dagger 
               \lambda^\dagger}(z,z';p)^{\dot{\alpha}}_{~\dot{\beta}} & 
    \hat{G}^S_{\lambda'^\dagger \lambda'}(z,z';p)^{\dot{\alpha}\beta} & 
    \hat{G}^S_{\lambda'^\dagger \lambda}(z,z';p)^{\dot{\alpha}\beta} & 
    \hat{G}^S_{\lambda'^\dagger 
               \lambda'^\dagger}(z,z';p)^{\dot{\alpha}}_{~\dot{\beta}} \\ \hline
    \hat{G}^S_{\lambda^\dagger 
               \lambda^\dagger}(z,z';p)^{\dot{\alpha}}_{~\dot{\beta}} & 
    \hat{G}^S_{\lambda^\dagger \lambda'}(z,z';p)^{\dot{\alpha}\beta} & 
    \hat{G}^S_{\lambda^\dagger \lambda}(z,z';p)^{\dot{\alpha}\beta} & 
    \hat{G}^S_{\lambda^\dagger 
               \lambda'^\dagger}(z,z';p)^{\dot{\alpha}}_{~\dot{\beta}} \\
    \hat{G}^S_{\lambda' \lambda^\dagger}(z,z';p)_{\alpha\dot{\beta}} & 
    \hat{G}^S_{\lambda' \lambda'}(z,z';p)_\alpha^{~\beta} & 
    \hat{G}^S_{\lambda' \lambda}(z,z';p)_\alpha^{~\beta} & 
    \hat{G}^S_{\lambda' \lambda'^\dagger}(z,z';p)_{\alpha\dot{\beta}} 
  \end{array} \right),
\label{eq:gaugino-prop-1}
\end{equation}
where the superscript $S$ takes either ``$321,i$'' (with $i=1,2,3$ for 
$U(1)_Y$, $SU(2)_L$ and $SU(3)_C$) or ``XY'' (for $SU(5)/321$).  Here, 
we have given the propagators in the mixed position-momentum space: 
$z \equiv e^{k|y|}/k$ (and $z'$) represents the coordinate for the fifth 
dimension and $p$ the 4D momentum. 

The propagators are written in the following form~\cite{Nomura:2003qb}:
\begin{equation}
  \hat{G}^S = \left( \begin{array}{cc|cc} 
    i \sigma^\mu_{\alpha\dot{\beta}} p_\mu f^S &
    i \delta_\alpha^{~\beta} (-\partial_z+\frac{1}{2z}) f'^S &
    i \delta_\alpha^{~\beta} h^S &
    \frac{i \sigma^\mu_{\alpha\dot{\beta}} p_\mu}{p^2}
      (\partial_z-\frac{1}{2z}) h'^S \\
    i \delta^{\dot{\alpha}}_{~\dot{\beta}} (\partial_z+\frac{1}{2z}) f^S &
    i \bar{\sigma}^{\mu \dot{\alpha}\beta} p_\mu f'^S &
    \frac{i\bar{\sigma}^{\mu \dot{\alpha}\beta} p_\mu}{p^2}
      (-\partial_z-\frac{1}{2z}) h^S &
    i \delta^{\dot{\alpha}}_{~\dot{\beta}} h'^S \\ \hline
    i \delta^{\dot{\alpha}}_{~\dot{\beta}} h^S &
    \frac{i\bar{\sigma}^{\mu \dot{\alpha}\beta} p_\mu}{p^2}
      (\partial_z-\frac{1}{2z}) h'^S &
    i \bar{\sigma}^{\mu \dot{\alpha}\beta} p_\mu f^S &
    i \delta^{\dot{\alpha}}_{~\dot{\beta}} (-\partial_z+\frac{1}{2z}) f'^S \\
    \frac{i\sigma^\mu_{\alpha\dot{\beta}} p_\mu}{p^2}
      (-\partial_z-\frac{1}{2z}) h^S &
    i \delta_\alpha^{~\beta} h'^S &
    i \delta_\alpha^{~\beta} (\partial_z+\frac{1}{2z}) f^S &
    i \sigma^\mu_{\alpha\dot{\beta}} p_\mu f'^S
  \end{array} \right).
\label{eq:gaugino-prop-2}
\end{equation}
Here, $f^S$, $f'^S$, $h^S$ and $h'^S$ are the functions of $z$, $z'$ and $p$:
\begin{eqnarray}
  f^S(z,z';p)  &=& \frac{g_5^2 \sqrt{z_< z_>}}{(C^S-A^S)^2+(B^S)^2} 
    \biggl( I_1(|p|z_<)+C^S K_1(|p|z_<) \biggr) \nonumber\\
  && {} \times \biggl( (C^S-A^S) \Bigl\{ I_1(|p|z_>)+A^S K_1(|p|z_>) \Bigr\}
    - (B^S)^2 K_1(|p|z_>) \biggr),
\label{eq:fz} \\
  f'^S(z,z';p) &=& -\frac{g_5^2 \sqrt{z_< z_>}}{(C^S-A^S)^2+(B^S)^2} 
    \biggl( I_0(|p|z_<)-C^S K_0(|p|z_<) \biggr) \nonumber\\
  && {} \times \biggl( (C^S-A^S) \Bigl\{ I_0(|p|z_>)-A^S K_0(|p|z_>) \Bigr\}
    + (B^S)^2 K_0(|p|z_>) \biggr),
\label{eq:fpz} \\
  h^S(z,z';p)  &=& -\frac{g_5^2 |p| \sqrt{z_< z_>}}{(C^S-A^S)^2+(B^S)^2} 
    \biggl( I_1(|p|z_<)+C^S K_1(|p|z_<) \biggr) \nonumber\\
  && {} \times B^S \biggl( I_1(|p|z_>)+C^S K_1(|p|z_>) \biggr),
\label{eq:hz} \\
  h'^S(z,z';p) &=& \frac{g_5^2 |p| \sqrt{z_< z_>}}{(C^S-A^S)^2+(B^S)^2} 
    \biggl( I_0(|p|z_<)-C^S K_0(|p|z_<) \biggr) \nonumber\\
  && {} \times B^S \biggl( I_0(|p|z_>)-C^S K_0(|p|z_>) \biggr),
\label{eq:hpz}
\end{eqnarray}
where $|p| \equiv \sqrt{p^2}$ and $z_<$ ($z_>$) is the lesser (greater) 
of $z$ and $z'$.

The coefficients $A^S$, $B^S$ and $C^S$ are given by
\begin{equation}
  A^S = \frac{X_I^S X_K^S - Y_I^S Y_K^S}{(X_K^S)^2 + (Y_K^S)^2}, \qquad
  B^S = \frac{X_I^S Y_K^S + X_K^S Y_I^S}{(X_K^S)^2 + (Y_K^S)^2}, \qquad
  C^S = \frac{Z_I^S}{Z_K^S}.
\label{eq:ABC}
\end{equation}
For the 321 gauginos ($S = 321,i$),
\begin{equation}
  \left\{ \begin{array}{l}
    X_I^{321,i} = \frac{1}{g_5^2} I_0(\frac{|p|}{k'}), \\
    X_K^{321,i} = \frac{1}{g_5^2} K_0(\frac{|p|}{k'}),
  \end{array} \right. \quad
  \left\{ \begin{array}{l}
    Y_I^{321,i} = M_{\lambda,i} I_1(\frac{|p|}{k'}), \\
    Y_K^{321,i} = M_{\lambda,i} K_1(\frac{|p|}{k'}),
  \end{array} \right. \quad
  \left\{ \begin{array}{l}
    Z_I^{321,i} = \frac{1}{g_5^2} I_0(\frac{|p|}{k}) 
        - \frac{|p|}{\tilde{g}_{0,i}^2} I_1(\frac{|p|}{k}), \\
    Z_K^{321,i} = \frac{1}{g_5^2} K_0(\frac{|p|}{k}) 
        + \frac{|p|}{\tilde{g}_{0,i}^2} K_1(\frac{|p|}{k}),
  \end{array} \right.
\label{eq:XYZ321}
\end{equation}
where we have neglected the contributions from the IR-brane gauge kinetic 
terms (see the text).  For the XY gauginos ($S = {\rm XY}$),
\begin{equation}
  \left\{ \begin{array}{l}
    X_I^{\rm XY} = \frac{1}{g_5^2} I_1(\frac{|p|}{k'}), \\
    X_K^{\rm XY} = -\frac{1}{g_5^2} K_1(\frac{|p|}{k'}),
  \end{array} \right. \quad
  \left\{ \begin{array}{l}
    Y_I^{\rm XY} = M_{\lambda,X} I_0(\frac{|p|}{k'}), \\
    Y_K^{\rm XY} = -M_{\lambda,X} K_0(\frac{|p|}{k'}),
  \end{array} \right. \quad
  \left\{ \begin{array}{l}
    Z_I^{\rm XY} = I_1(\frac{|p|}{k}), \\
    Z_K^{\rm XY} = -K_1(\frac{|p|}{k}).
  \end{array} \right.
\label{eq:XYZXY}
\end{equation}

\subsection*{B.2~~~Gauge scalar propagators}

Here we present the propagators for the gauge scalars, $\chi^{321}$ 
and $\chi^{\rm XY}$.  For gauge scalars, differences among the $U(1)_Y$, 
$SU(2)_L$ and $SU(3)_C$ components are not important, as these fields 
are strongly localized towards the IR brane and do not feel logarithmically 
enhanced UV-brane terms.  We again present the propagators $\hat{G}$ for 
the rescaled fields, $\hat{\chi}^A = e^{-2k|y|}\chi^A$. 

In the mixed position-momentum space, the propagators are given by
\begin{equation}
  \hat{G}^S_{\chi\chi}(z,z';p)
  = \frac{i g_5^2}{k} \frac{1}{D^S-E^S} 
    \biggl( I_0(|p|z_<)+D^S K_0(|p|z_<) \biggr) 
    \biggl( I_0(|p|z_>)+E^S K_0(|p|z_>) \biggr),
\label{eq:scalar-prop}
\end{equation}
where the superscript $S$ takes either 321 or XY.  The coefficients 
$D^S$ and $E^S$ are given for $\chi^{321}$ by
\begin{equation}
  D^{321} = -\frac{I_0(\frac{|p|}{k})}{K_0(\frac{|p|}{k})}, \qquad
  E^{321} = -\frac{I_0(\frac{|p|}{k'})}{K_0(\frac{|p|}{k'})},
\label{eq:DE321}
\end{equation}
and for $\chi^{\rm XY}$ by
\begin{equation}
  D^{\rm XY} = \frac{I_1(\frac{|p|}{k})}{K_1(\frac{|p|}{k})}, \qquad
  E^{\rm XY} = \frac{\frac{|p|}{k'}I_1(\frac{|p|}{k'}) 
      + \frac{g_5^2 M_{\chi,X}^2}{k}I_0(\frac{|p|}{k'})}
    {\frac{|p|}{k'}K_1(\frac{|p|}{k'}) 
      - \frac{g_5^2 M_{\chi,X}^2}{k}K_0(\frac{|p|}{k'})}.
\label{eq:DEXY}
\end{equation}

\end{document}